\documentclass[useAMS,usenatbib]{mn2e}
\usepackage[dvips]{graphicx}
\usepackage{aas_macros}

\title[GMRT 325-MHz observations of ELAIS-N1]
  {325-MHz observations of the ELAIS-N1 field using the Giant Metrewave Radio Telescope}
\author[S. K. Sirothia et al.]
  {S.~K.~Sirothia,$^1$\thanks{Email: sirothia@ncra.tifr.res.in}
  M~.Dennefeld,$^2$ D.~J.~Saikia,$^1$ H.~Dole,$^3$
\newauthor F.~Ricquebourg,$^2$ and J.~Roland$^2$ \\
  $^1$ National Centre for Radio Astrophysics, Tata Institute of Fundamental Research, 
       Post Bag 3, Ganeshkhind, Pune 411007, India\\
  $^2$ Institut d'Astrophysique de Paris, 98bis Boulevard Arago, F-75014, Paris, France \\
  $^3$ Institut d'Astrophysique Spatiale, B\^at. 121, Universit\'e Paris-Sud 11 and CNRS (UMR 8617), 
       F-91405 Orsay Cedex, France\\
}
\date{Accepeted for publication in MNRAS}

\pagerange{\pageref{firstpage}--\pageref{lastpage}} \pubyear{2002}

\def\LaTeX{L\kern-.36em\raise.3ex\hbox{a}\kern-.15em
    T\kern-.1667em\lower.7ex\hbox{E}\kern-.125emX}

\newcounter{saveeqn}

\pubyear{2008}

\begin{document}
\label{firstpage}

\maketitle

\begin{abstract}
We present observations of the European Large-Area {\it ISO} Survey-North 1 (ELAIS-N1) at 325 MHz using the Giant Metrewave Radio Telescope (GMRT), with the ultimate objective of identifying active galactic nuclei and starburst galaxies and examining their evolution with cosmic epoch. After combining the data from two different days we have achieved a median rms noise of  $\approx40 \mu$Jy beam$^{-1}$, which is the lowest that has been achieved at this frequency.  We detect 1286 sources with a total flux density above $\approx270 \mu$Jy. In this paper, we use our deep radio image to examine the spectral indices of these sources by comparing our flux density estimates with those of Garn et al. at 610 MHz with the GMRT, and surveys with the Very Large Array at 1400 MHz. We attempt to identify very steep spectrum sources which are likely to be either relic sources or high-redshift objects as well as inverted-spectra objects which could be Giga-Hertz Peaked Spectrum objects. We present the source counts, and report the possibility of a flattening in the normalized differential counts at low flux densities which has so far been reported at higher radio frequencies. 
\end{abstract}


\section{INTRODUCTION}

One of the key questions in extragalactic studies is to understand the 
evolution of galaxies, namely their star formation history, formation of 
active galactic nuclei (AGN), building-up of large disks 
and bulges and the formation and evolution of supermassive black holes over 
the redshift range from $z\approx5$ to the present 
epoch. This is the range where the global star-formation rate has passed 
through a maximum between a redshift of 1 and 2 
(\citeauthor{1996MNRAS.283.1388M}~\citeyear{1996MNRAS.283.1388M}; \citeauthor*{1998ApJ...498..106M}~\citeyear{1998ApJ...498..106M}; see the recent compilation by \citeauthor{2006ApJ...651..142H}~\citeyear{2006ApJ...651..142H}).  It 
is conjectured that merging of galaxies has occurred significantly at these epochs, 
triggering collapse of molecular clouds and star formation, often in dusty 
environments. Many of these galaxies may also harbour an AGN and are copious
emitters in the infrared region of the spectrum. It has become clear over the
last decade that a population of galaxies which radiate most of their power
in the infrared constitutes an important and significant component of the 
Universe (\citeauthor*{2005ARA+A..43..727L}~\citeyear{2005ARA+A..43..727L} for a review).

Multiwavelength observations of high-redshift infrared galaxies could provide
valuable insights into issues of galaxy formation and evolution.
The spectral energy distribution (SED) of the cosmic infrared 
(IR) background (CIB, which is the emission at wavelengths larger than a few
microns) is due to the formation and evolution of both AGN and starburst galaxies
(e.g. \citeauthor{1996A+A...308L...5P}~\citeyear{1996A+A...308L...5P};
\citeauthor{1998ApJ...508...25H}~\citeyear{1998ApJ...508...25H};
\citeauthor{1999A+A...344..322L}~\citeyear{1999A+A...344..322L};
\citeauthor{2000A+A...360....1G}~\citeyear{2000A+A...360....1G};
\citeauthor{2005PhR...409..361K}~\citeyear{2005PhR...409..361K};
\citeauthor{2006A+A...451..417D}~\citeyear{2006A+A...451..417D}). The SED of the CIB  
peaks around 150 $\mu$m, accounting for about half the total energy in the 
optical/infrared extragalactic background light (cf. \citeauthor{2001ARA+A..39..249H}~\citeyear{2001ARA+A..39..249H}; 
\citeauthor{2004NewAR..48..465W}~\citeyear{2004NewAR..48..465W};
\citeauthor{2006Natur.440.1018A}~\citeyear{2006Natur.440.1018A}). 
It has been shown recently that the mid-infrared (MIR) 24~$\mu$m selected sources 
contribute more than 70 per cent of the CIB at 70~and~160~$\mu$m.
Galaxies contributing the most to the total CIB are z$\sim$1 luminous 
infrared galaxies, which have intermediate stellar masses \citep{2006A+A...451..417D}.

To explore these issues related to galaxy formation and evolution and possible
relationships between AGN and starburst activity 
require large samples over wide enough areas, to avoid biases 
due to large-scale structure variations, as well as multi-wavelength coverages to 
have a global appraisal of the energy output and of its origin. This means, 
in particular, a far-IR coverage to access the ``dusty'' objects 
to get the bolometric energy output; an 
optical coverage as this is the range where the usual spectroscopic diagnostics 
are available; 
the ultraviolet (UV) range as this is where the massive stars have their impact and where 
the energy is seen  in the absence of dust; the radio range to distinguish 
thermal from non-thermal sources and obtain the best spatial resolution; and 
possibly a high-energy coverage in X-rays to detect the most powerful AGNs. 
Amongst the different fields which are being studied extensively, a well-known one
selected at infrared wavelengths is the European Large Area ISO (Infrared Space Observatory)
Survey (ELAIS), from which we have selected one of the northern fields, the ELAIS-N1, observable with GMRT.   

The ELAIS-N1 field has been chosen in a region of the sky with low-IR 
foreground emission, to allow detection of fainter (and presumably more distant) 
galaxies; also, because of this advantage, it  has already  been 
covered at far-IR wavelengths by the ISO-FIRBACK (Far-Infrared Background) survey at 170~$\mu$m, to find 
dusty, high-z galaxies expected to significantly contribute to the far-IR cosmic 
background and is covered by Spitzer as well, as part of the SWIRE 
(Spitzer Wide-area Infrared Extragalactic)
legacy survey \citep{2004ApJS..154...54L}. The details on the 
first far-IR data can be found in \cite{2001A+A...372..364D}. Our team has been studying
this field in detail, and has particularly made some of the optical spectroscopic 
observations. Most of the  sources identified up to now, essentially the IR-brighter ones, 
are local, cold, moderately 
star-forming galaxies rather than the violent starbursters expected at larger 
distances \citep{2005A+A...440....5D}. It is now understood that this is partly due 
to the sensitivity limit of ISO, so that the higher z sources, needed to 
reproduce the number counts, have to be found among the fainter IR sources detected by Spitzer. 
Due to its interest in 
studies of galaxy evolution, it has also been included in the Galex UV survey. 
It has already optical imaging coverage from LaPalma as part of the 
INT (Isaac Newton Telescope) Wide Field Survey \citep{2001NewAR..45...97M} and a band-merged catalogue 
is available \citep{2004yCat..73511290R}. Redshift surveys of the region are underway \citep[e.g.][]{2008MNRAS.386..697R} and it is also
partly covered by the Sloan Digital Sky Survey \citep[SDSS,][]{2008ApJS..175..297A}. 

Radio surveys provide important constraints in our understanding of
the evolution of the Universe. Traditionally counts of the number of sources
as a function of the radio flux density have provided information on the 
evolution of radio source populations with cosmic epoch 
(e.g. \citeauthor{1966MNRAS.133..421L}~\citeyear{1966MNRAS.133..421L};
\citeauthor{1993MNRAS.263..123R}~\citeyear{1993MNRAS.263..123R}). 
More recently,
deep radio surveys have shown a flattening of the source counts at about a
mJy at both 610 and 1400 MHz (e.g. \citeauthor*{1990ASPC...10..389W}~\citeyear{1990ASPC...10..389W};
\citeauthor{2006A+A...457..517P}~\citeyear{2006A+A...457..517P} and references therein;
\citeauthor{2007MNRAS.378..995M}~\citeyear{2007MNRAS.378..995M} and references therein). 
The change of slope is believed to be due to a new population of radio sources, 
the so-called sub-mJy population, consisting largely of low-luminosity AGN
and starburst galaxies \citep[e.g.][]{2007ASPC..380..205P, 2008ApJS..177...14S}.
The source counts at higher flux densities are 
dominated by the classical double radio galaxies and quasars \citep[e.g.][]{1989ApJ...338...13C}.

Concerning the ELAIS-N1 field,
a Very Large Array (VLA) radio survey has been made at $\lambda$20 cm by 
\citet{1999MNRAS.302..222C}, reaching a brightness limit of 
$\approx$1 mJy beam$^{-1}$  (5 sigma) 
over the 1.54 square degrees of coverage, and deeper in 
smaller areas. This field has been studied at $\lambda$50 cm using the Giant
Metrewave Radio Telescope (GMRT) by \citet{2008MNRAS.383...75G}, hereafter 
referred to as G2008. They have covered a
total area of $\approx$~9~deg$^2$ with a resolution of 6~$\times$~5~arcsec$^2$ with
a total of 19 pointings. In 4 of their pointings they reach an rms of
 $\approx$~40~$\mu$Jy~beam$^{-1}$ and an rms of $\approx$~70~$\mu$Jy~beam$^{-1}$ in the rest of the pointings.
They have catalogued 2500 sources and present a mosaic of maps from their survey.
Preliminary results from a study of polarised compact sources
 \citep{2007ApJ...666..201T} at 1420 MHz using the Dominion Radio Astrophysical
 Observatory Synthesis Telescope (DRAO ST) are also available. They currently present
30\% of observations from a survey of 7.4~deg$^2$ centered on $16^h11^m,~+55^\circ00^\prime$.
They present maps in Stokes I, Q and U with a maximum sensitivity of 78~$\mu$Jy~beam$^{-1}$,
 with a resolution of $\approx$~1~arcmin$^2$.

A GMRT coverage of this field, at $\lambda=$~90 and 200~cm, provides
an ideal complement to get accurate spectral shapes and indexes 
over a large frequency range and help establish the nature of the objects. 
In particular it will help to separate the respective contribution of AGNs 
and starbursts, a long standing problem for the interpretation of the far-IR 
emission. It is also relevant to note here that there is a tight correlation 
between the far-IR and radio luminosities of galaxies (e.g. 
\citeauthor*{1985ApJ...298L...7H}~\citeyear{1985ApJ...298L...7H};
\citeauthor{1992ARA+A..30..575C}~\citeyear{1992ARA+A..30..575C})
which applies to both local and global properties of
galaxies in the nearby and distant Universe (\citeauthor{2002A+A...384L..19G}~\citeyear{2002A+A...384L..19G};
\citeauthor{2005ChJAA...5..448L}~\citeyear{2005ChJAA...5..448L};
\citeauthor{2006MNRAS.370..363H}~\citeyear{2006MNRAS.370..363H};
\citeauthor{2006ApJ...638..157M}~\citeyear{2006ApJ...638..157M};
\citeauthor{2005ApJ...622..772C}~\citeyear{2005ApJ...622..772C}).

As part of our program to explore the issues outlined earlier, we are 
making deep observations of the ELAIS-N1 field with the GMRT at low frequencies.
In this paper we present the results of our GMRT observations at 325 MHz
pointed towards the centre of the field ($16^h10^m, +54^\circ36^\prime$). We have made the
observations on two different days and have made images of the first day
separately to check for consistency.  The final image made by combining
the data of both the days has a median rms
noise figure of $\approx$40 $\mu$Jy beam$^{-1}$ towards the centre of the field, and is
amongst the deepest images made at this frequency. It is more sensitive
than the low-frequency survey of the XMM-LSS (Large Scale Structure) field at 325 MHz with the VLA
which has a 5$\sigma$ brightness limit of $\sim$4 mJy beam$^{-1}$  
\citep{2006A+A...456..791T}. For comparison, the 5$\sigma$ limiting brightness of 
the Westerbork Northern Sky Survey (WENSS; \citeauthor{1997A+AS..124..259R}~\citeyear{1997A+AS..124..259R})
 is $\sim$18 mJy beam$^{-1}$, while for the deeper surveys of selected fields with the Westerbork telescope
the corresponding values range from 2.4 to 3.5 mJy beam$^{-1}$ \citep{1992A+A...256..331W}.
The details of our observations and analysis are described in Section~\ref{ef0:sec:obs}, while the results
are presented in Section~\ref{ef0:sec:result}. The discussion and conclusions are summarised 
in Section~\ref{ef0:sec:discussions}. 

\begin{table}
\begin{center}
\caption{Observation summary}
\label{ef0:table:obs_sum}
\begin{tabular}{| l | l |} \hline
Date & 2005 June 26, 2005 June 27  \\
Working antennas & 25 \\
Centre Frequency & 325 MHz \\
Bandwidth & 32 MHz \\
Visibility integration time & 4.19 s \\
Total observation time & 10.3 hrs $\times$ 2 \\ \hline
\multicolumn{2}{l}{~~~~~~~~ Flux Calibrator}\\ \hline
Source & 3C286\\
Time & 0.3 hrs $\times$ 2\\
Flux density & 25.97 Jy\\
Scale & Perley-Taylor 99\\ \hline
\multicolumn{2}{l}{~~~~~~~~ Phase Calibrator}\\ \hline
Source & J1459+7140 \\
Time & 1.8 hrs $\times$ 2\\
Flux density & 20.17 Jy\\ \hline
\multicolumn{2}{l}{~~~~~~~~ Target Field}\\ \hline
Phase centre & J1610+5440 \\
Time & 8.2 hrs $\times$ 2\\
Fraction ($f_d$) of data used & $\approx$ 40\% \\
\hline
\end{tabular}
\end{center}
\end{table}

\begin{table}
\begin{center}
\caption{Imaging summary}
\label{ef0:table:imag_sum}
\begin{tabular}{| l | l |} \hline
Package & AIPS++ (version: 1.9, build \#1556)\\
Image size & $5000\times5000$\\
Pixel size & $2^{\prime\prime}\times2^{\prime\prime}$ \\
Clean Algorithm & Wide-field Clark clean\\
Imaging Weight & Briggs (rmode=`norm', robust=0)\\ \hline
\multicolumn{2}{l}{Widefield Imaging Parameters}\\ \hline
Number of facets & 9 with 1.5 overlap\\
Wproject planes & 256\\
Primary Beam correction & Gaussian, FWHM=84.4$^{\prime}$\\
Rms noise & $\approx 40 \mu$Jy beam$^{-1}$\\
Synthesised beam & $9.35^{\prime\prime}\times7.38^{\prime\prime}$,\ PA=73.25$^\circ$ \\
\hline
\end{tabular}
\end{center}
\end{table}

\section{OBSERVATIONS AND DATA REDUCTION}
\label{ef0:sec:obs}
The observation summary along with the calibrators used is 
presented in Table~\ref{ef0:table:obs_sum}, while some of the
imaging details are listed in Table~\ref{ef0:table:imag_sum}, both of
which are self explanatory. The data were acquired with a small
visibility integration time from the correlator to help in the 
identification of radio frequency interference (RFI) and minimize 
smearing effects. The data reduction was mainly done 
using AIPS++, and with 3C286 as the primary flux density and 
bandpass calibrator. After applying bandpass corrections on the
phase calibrator (J1459+7140), gain and phase variations were estimated,
and then flux density, bandpass, gain and phase calibration from 
3C286 and J1459+7140 were applied on the target field (J1610+5440).

While calibrating the data, bad data were flagged at various stages. 
The data for antennas with high errors in antenna-based solutions
were examined and flagged over certain time ranges. Some baselines
were flagged based on closure errors on the bandpass calibrator.
Channel and time-based flagging of data points corrupted by 
RFI were done using a median filter 
with a $6\sigma$ threshold. Residual errors above $5\sigma$ were 
also flagged after a few rounds of imaging and self calibration.
The system temperature ($T_{sys}$) was found to vary with antenna, 
the ambient temperature which has a diurnal variation, and elevation
\citep{sks2008}. 
In the absence of regular $T_{sys}$ measurements for GMRT antennas, 
this correction was estimated from the residuals of corrected data 
with respect to the model data. The corrections were then applied 
to the data. The final image was made after several rounds of phase
self calibration, and one round of amplitude self calibration, where
the data were normalized by the median gain for all the data. This
image was made by combining the data from both the sidebands, giving
a total bandwidth of 32 MHz.  The
final image was also primary beam corrected using the Gaussian 
parameter as mentioned in Table~\ref{ef0:table:imag_sum}.

\begin{figure}
\begin{center}
\includegraphics[angle=0, totalheight=2.8in, width=4in]{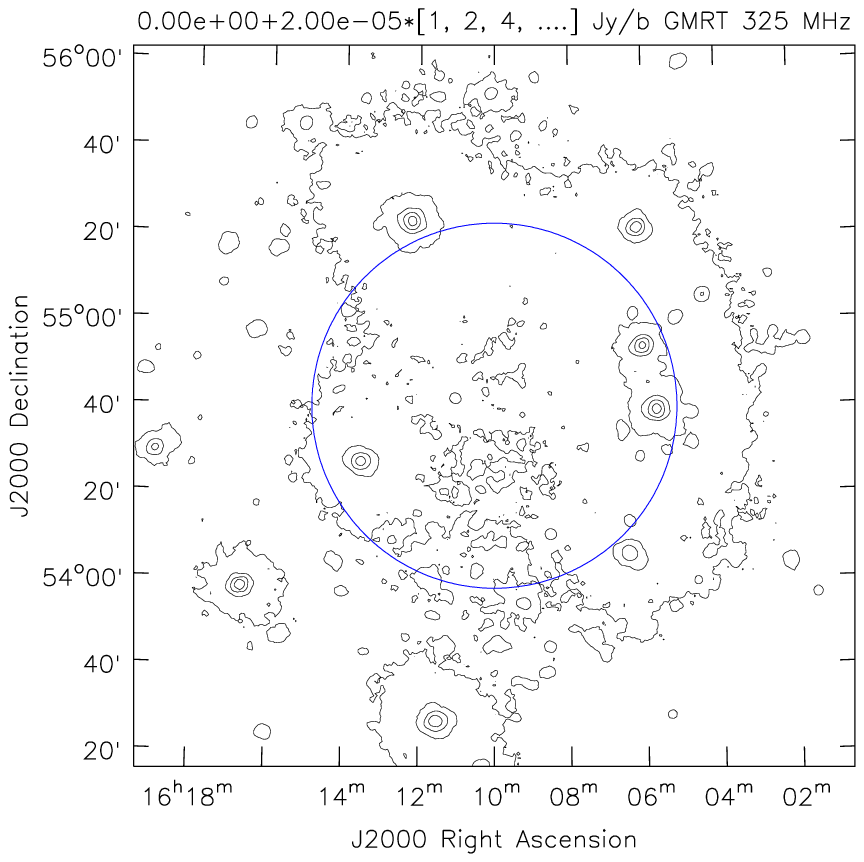}
\caption{The variation of rms noise across the image before primary beam correction. In this
figure and in all the images presented here the contour levels in units of Jy beam$^{-1}$ 
are represented by mean$+$rms$\times$(n) where n is the multiplication factor. These levels
are shown above each image. All negative contours appear as dashed lines.}
\label{ef0:fig:rms_noise}
\end{center}
\end{figure}

The final image has a median rms noise of $\approx$40 $\mu$Jy beam$^{-1}$ near the
phase centre and 30 $\mu$Jy beam$^{-1}$ near the half power of the 
beam before primary beam correction. The noise variation across the
beam is shown in Fig.~\ref{ef0:fig:rms_noise}.
Issues related to dynamic range affect the quality of the image near the
bright sources where the noise is greater, as can be seen in Fig.~\ref{ef0:fig:rms_noise}. 
An image of a strong source
in the field, and the corresponding image at 610 MHz from 
G2008 is discussed later in the paper. 
The theoretically expected noise near the phase centre,
 is given by the following equation, 
\begin{equation}
\sigma=\frac{\sqrt{2} T_{sys}}{G\sqrt{n(n-1)f_d\Delta\nu\tau}}
\end{equation}
where, $T_{sys}\approx108$K, the antenna gain
$G\approx.32$K Jy$^{-1}$, n is the number of working antennas, $\Delta\nu$
is the bandwidth and $\tau$ is the total duration of the observations,
$f_d$ is the total fraction of data which has been used for making the final
image; the last four parameters being listed in Table~\ref{ef0:table:obs_sum}.
For our data set the theoretically calculated value is $\sim$22 $\mu$Jy beam$^{-1}$.

\begin{figure*}
\vbox{
  \hbox{
   \includegraphics[angle=0, totalheight=2.8in, viewport=19 212 573 627, clip]{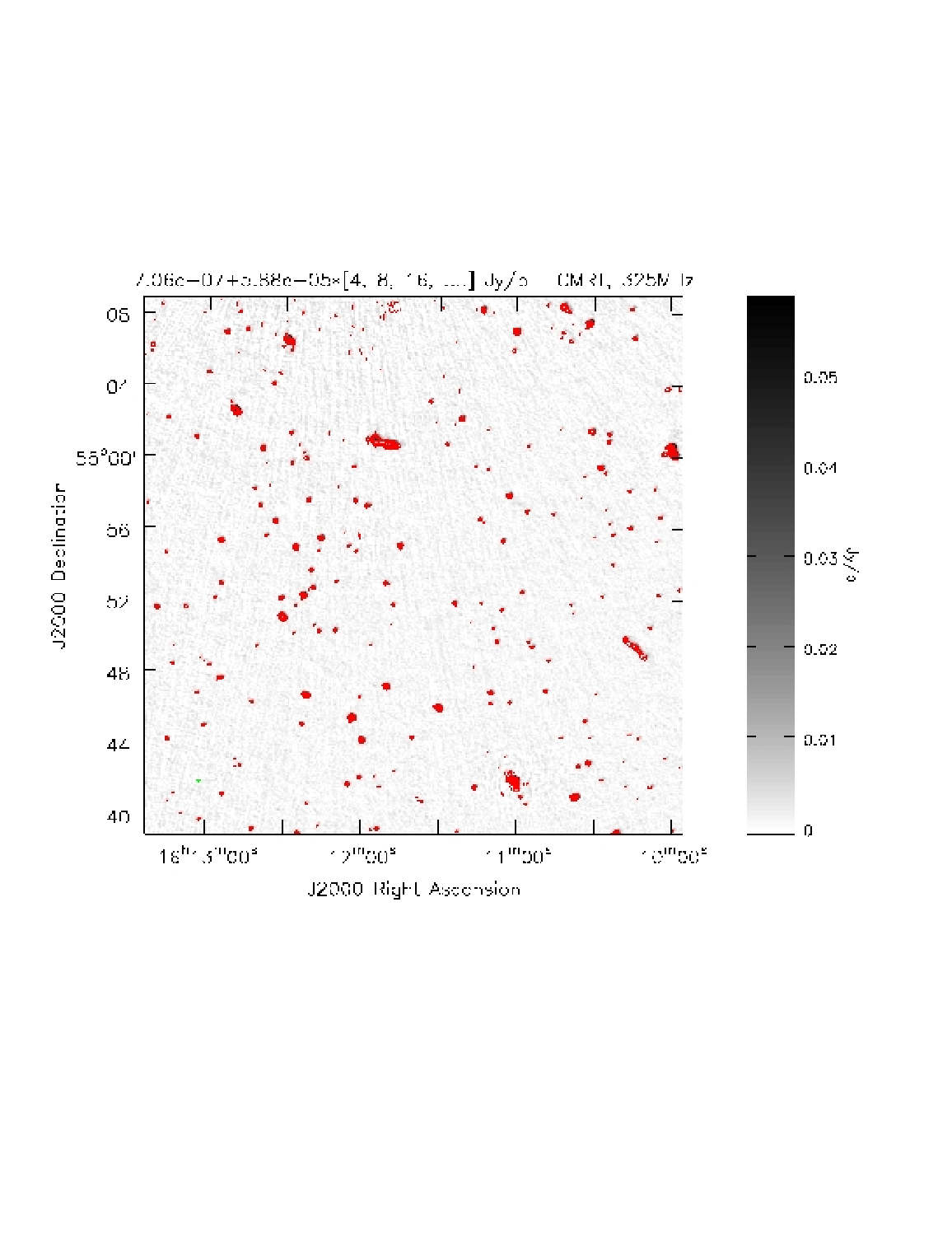}
   \includegraphics[angle=0, totalheight=2.8in, viewport=19 212 573 627, clip]{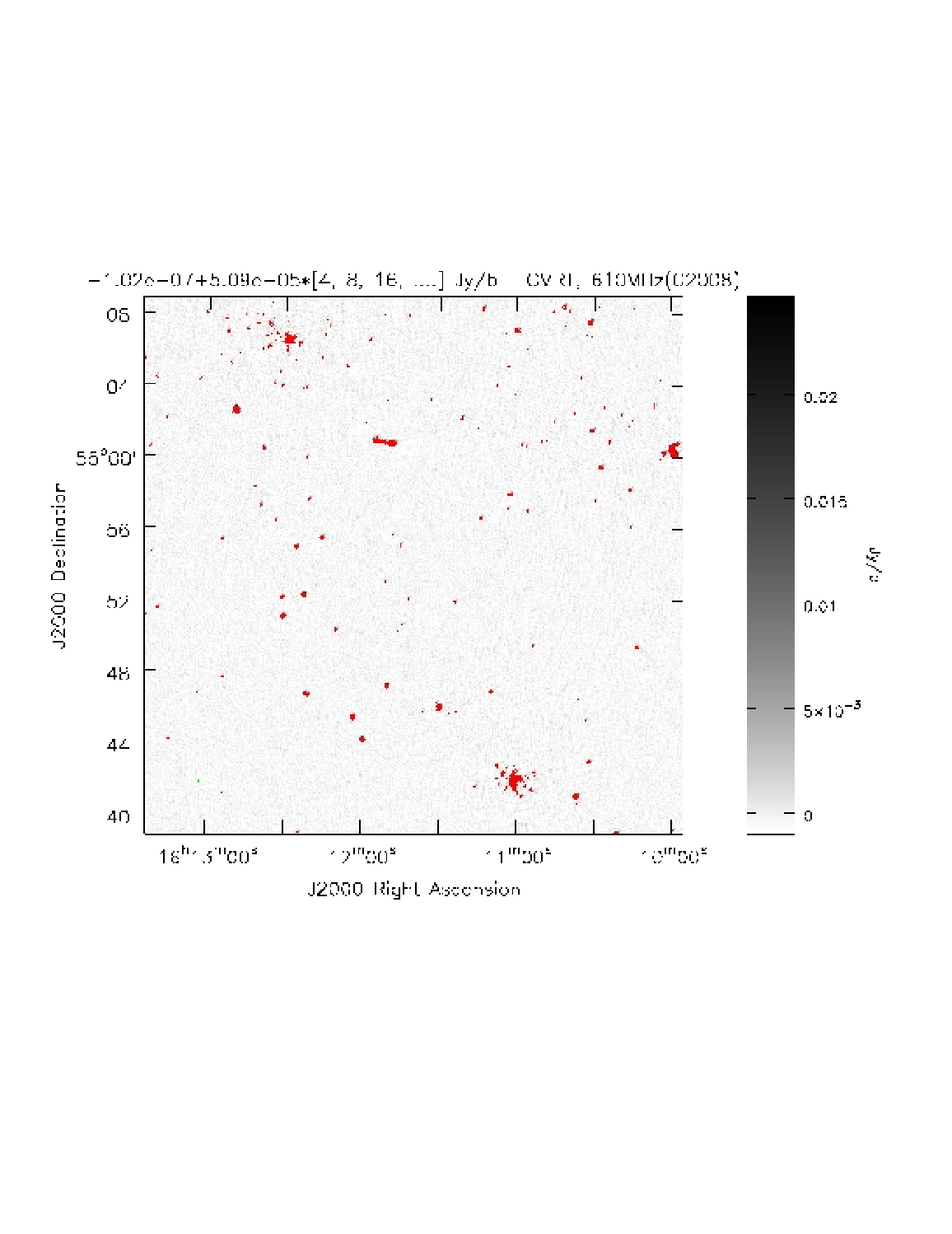}
       }
  \hbox{
   \includegraphics[angle=0, totalheight=2.8in, viewport=19 212 573 627, clip]{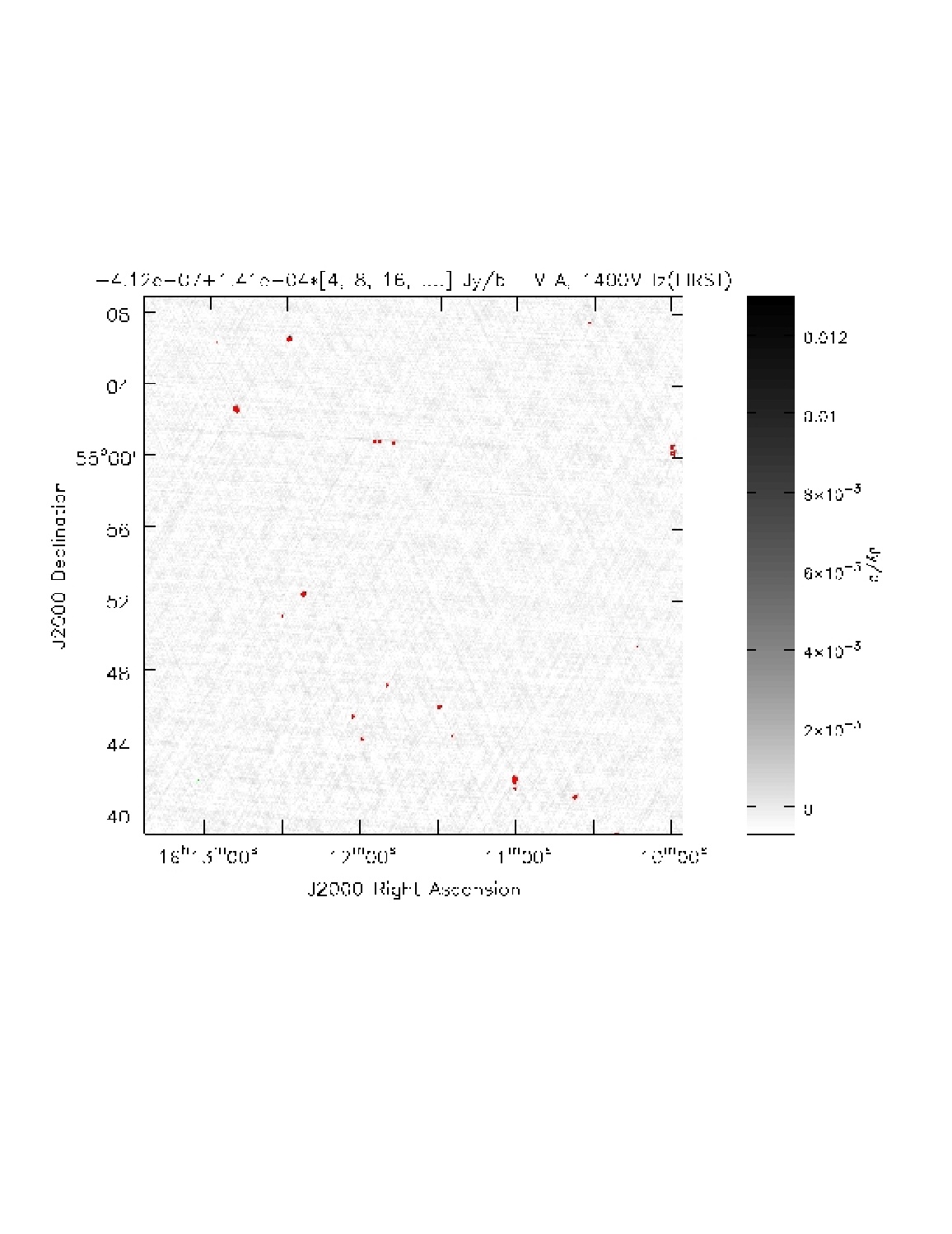}
   \includegraphics[angle=0, totalheight=2.8in, viewport=19 212 573 627, clip]{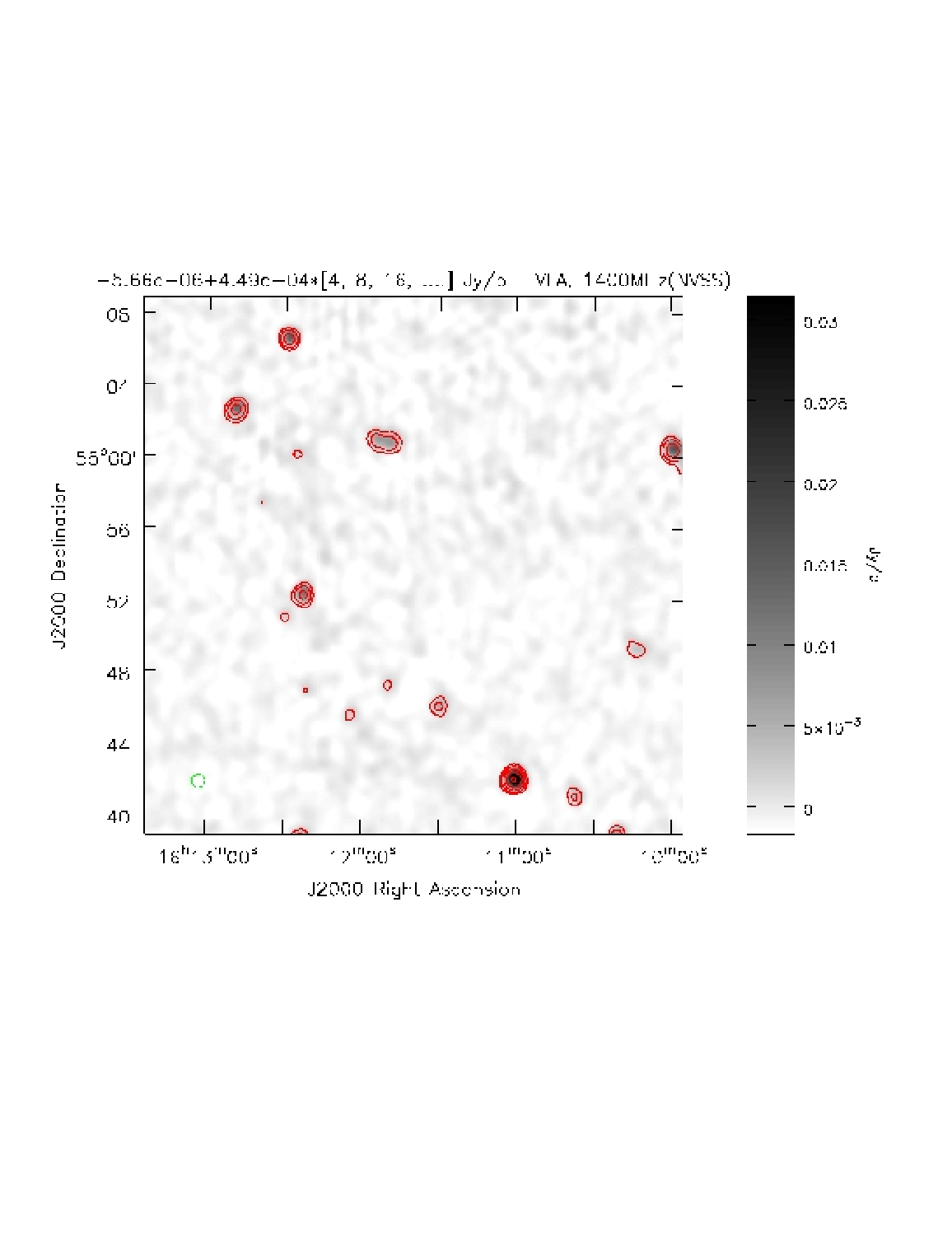}
       }
}
\caption{GMRT image of an area of 30$\times$30 arcmin$^2$ from our observations at 325MHz (top left), and
the corresponding images of the same area at 610 MHz with the GMRT from G2008 (top right), 
and from FIRST (bottom left) and NVSS (bottom right) at 1400 MHz.}
\label{ef0:fig:gmrt_mrao_first_nvss}
\end{figure*}

\section{SOURCE CATALOGUE AT 325 MHz}
\label{ef0:sec:result}
As an illustration, the GMRT 325-MHz image of a region of about 
30$\times$30 arcmin$^2$ along with the
Faint Images of the Radio Sky at Twenty-centimeters \citep*[FIRST,][]{1995ApJ...450..559B},
and NRAO VLA Sly Survey \citep[NVSS,][]{1998AJ....115.1693C} images at 1400 MHz and the GMRT 
image at 610 MHz (G2008)
are shown in Fig.~\ref{ef0:fig:gmrt_mrao_first_nvss}.  The brighter 
sources are seen clearly in all the images. One of the
first tasks for the entire field is to extract and list all
the sources along with their positions and flux densities.

\subsection{Source extraction criterion} 
\label{ef0:sec:source_ext}
A catalogue of sources within 1.1$^\circ$ radius of the phase centre, which is
18.3 per cent of the primary beam peak value, was created with the peak 
source brightness greater than 6 times the local rms noise value. A comparison
of our sources with those detected in WENSS 
\citep{1997A+AS..124..259R} for different distances from our phase centre showed no
systematic effects and suggests this to be a reasonable value for this
catalogue. The rms noise has been evaluated
over an area of approximately 63$\times$63 pixels excluding the
source pixels. Since the local noise
varies with distance from the phase centre and also in the 
vicinity of bright sources, this approach has helped minimize
the detection of spurious sources. 

\begin{figure}
\begin{center}
\vbox{
   \includegraphics[angle=0, totalheight=2.0in, viewport=19 212 573 627, clip]{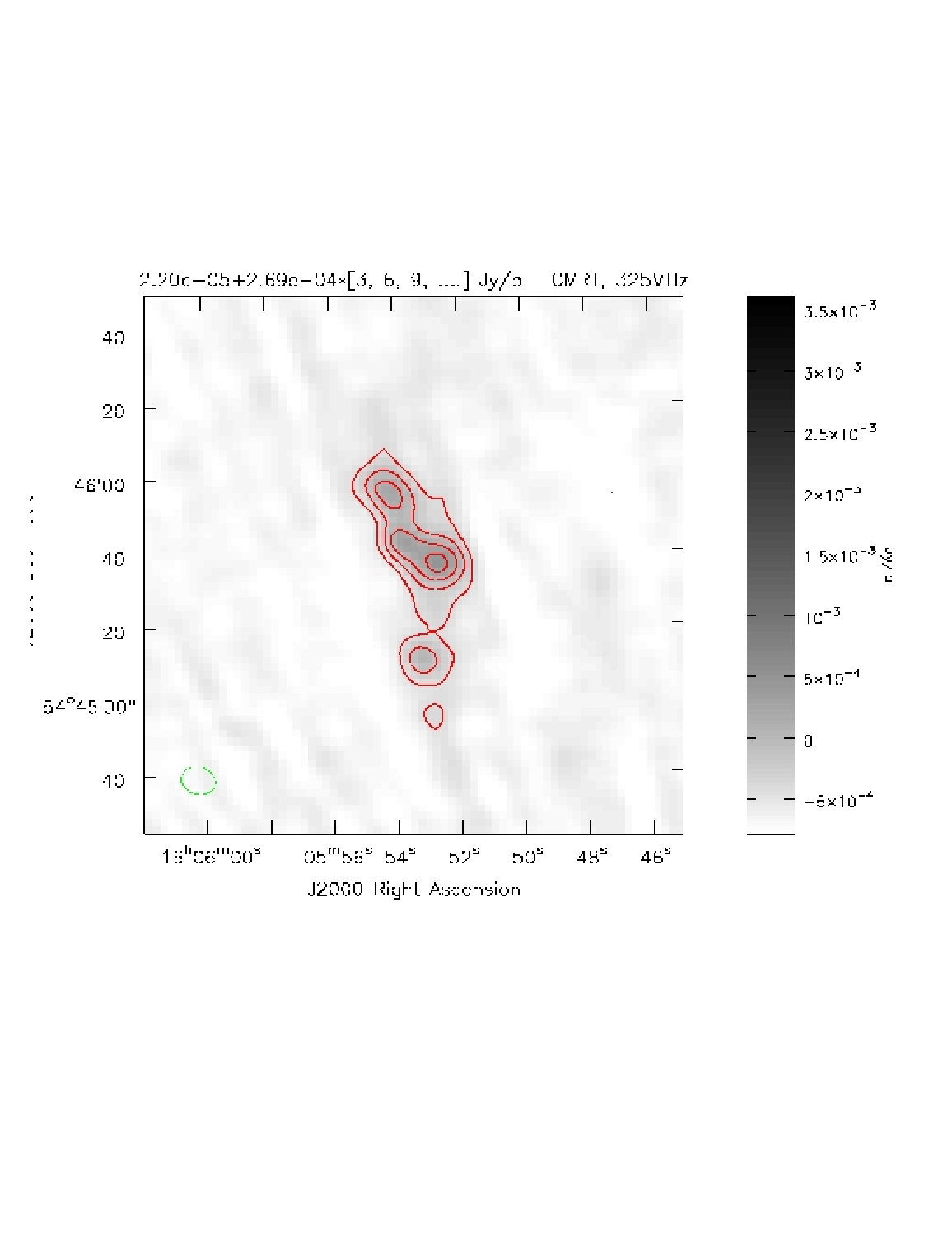}
   \includegraphics[angle=0, totalheight=2.0in, viewport=19 212 573 627, clip]{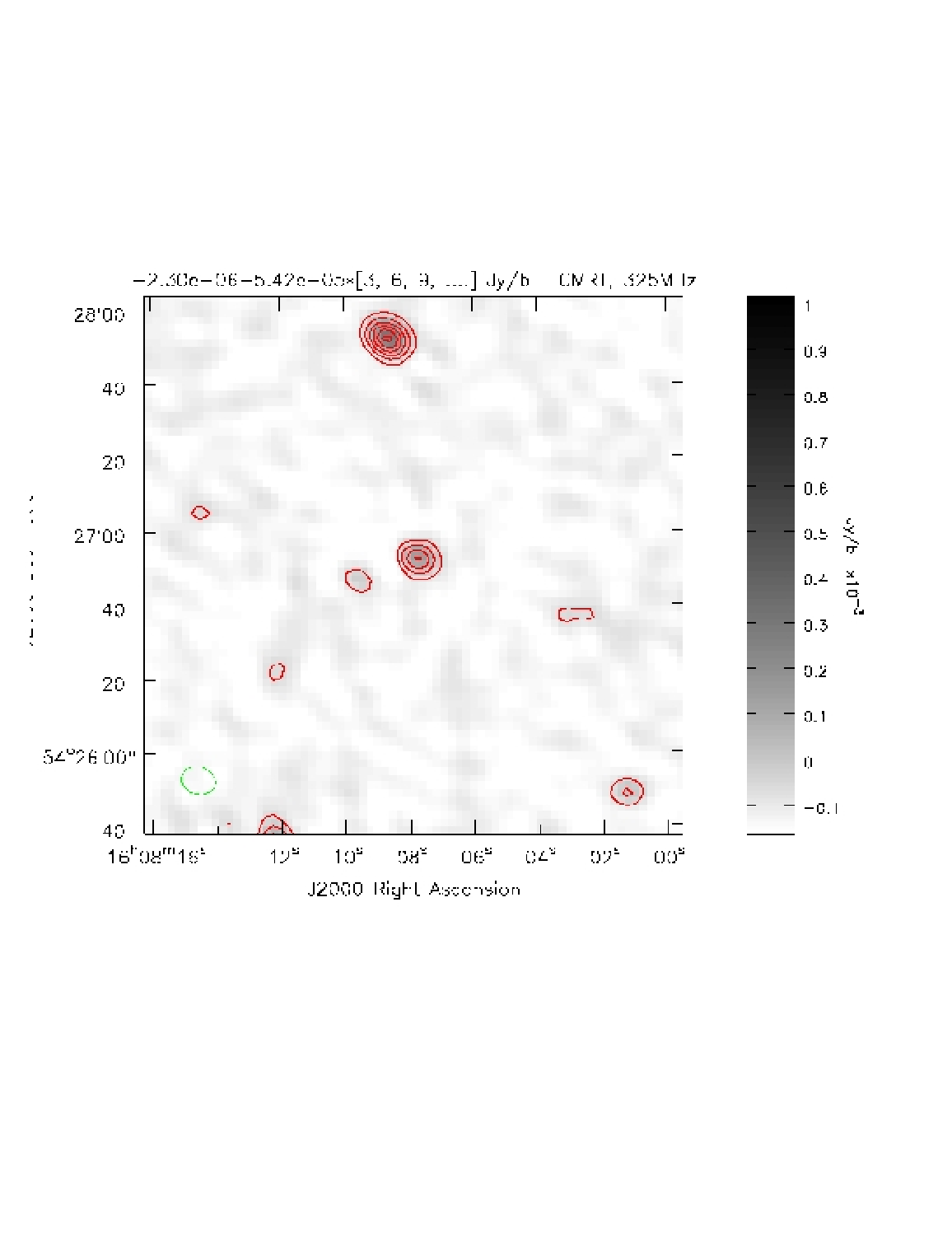}
   \includegraphics[angle=0, totalheight=2.0in, viewport=19 212 573 627, clip]{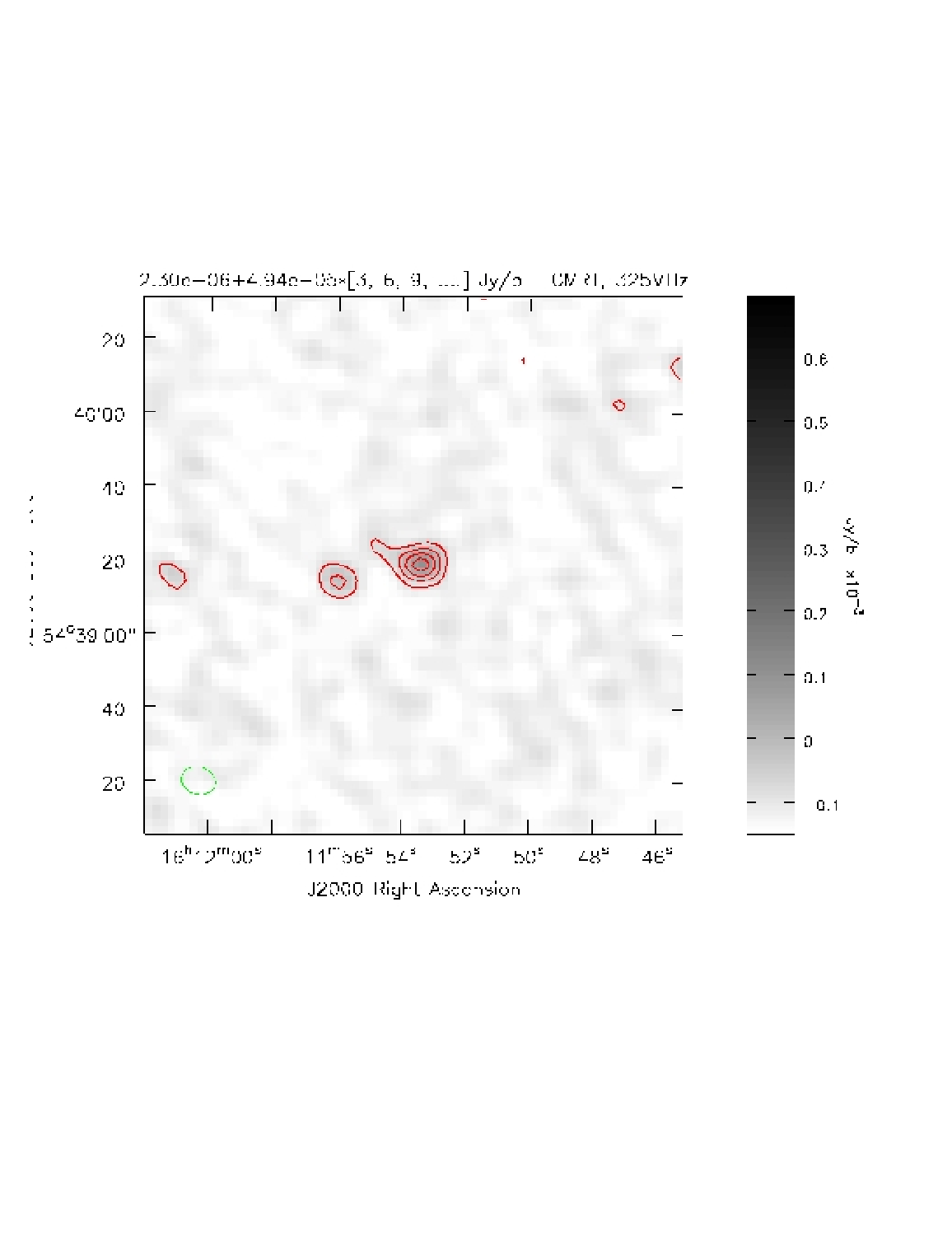}
      
}
\caption{The figure illustrates the 3-$\sigma$ contours used for estimating the corresponding flux densities 
of a few selected sources from our GMRT observations at 325 MHz. The upper panel shows a resolved source
where the total flux density is the value within the 3-$\sigma$ contour, while the middle and lower panels
show two unresolved sources where S$_{\rm total}$ has been taken equal to S${\rm peak}$.}
\label{ef0:fig:diff_1s_3s}
\end{center}
\end{figure}

An extract of sources from the full Table are listed in 
Table~\ref{ef0:table:srclist},
along with some of their observed
properties. The full Table consisting of 1286 sources,
is available in the on-line version  
of the paper. Column 1: source name in J2000 co-ordinates where
hhmmss represents the hours, minutes and seconds of right ascension and 
ddmmss represents the degrees, arcmin and arcsec of declination; 
columns 2 and 3: the right ascension and declination of the source 
which is the flux-density weighted centroid of all the emission enclosed by
the 3$\sigma$ contour. We have estimated the errors in the 
positions using the formalism outlined by \cite{1998AJ....115.1693C}.
The typical error in the positions of the sources is about 1.4 arcsec,
consistent with the comparisons of our positions with those of G2008 and
FIRST, as discussed in Section~\ref{ef0:sec:spi}.
Column 4: distance of the centroid from
the phase centre in degrees; column 5: the local rms noise value in units
of $\mu$Jy beam$^{-1}$. The peak and total flux densities within the 3-$\sigma$ contour are listed 
in columns 6 and 7 respectively.
The flux densities of the different sources were estimated as follows. 
For unresolved sources, the total flux density S$_{\rm total}$ in units of
mJy has been taken equal to the peak value S$_{\rm peak}$ in our analysis,
although the flux density within the 3-$\sigma$ contour is also listed.
For the 
extended sources the total flux density is the value within the 3-$\sigma$ 
contour. Fig.~\ref{ef0:fig:diff_1s_3s} shows the 3-$\sigma$ contours for
one resolved source and two unresolved sources. 
For ease of readers the values of total flux densities (S$_{\rm total}$) thus estimated are mentioned in column 8. 
For resolved sources we list the largest angular size in 
column 9, while the unresolved sources have been marked as U in this
column. The largest angular size has been estimated by fitting Gaussians
to the resolved sources. For single sources it is the deconvolved size
of a single Gaussian, while for extended sources it is       
the sum of the distance between the furthest Gaussian components
and their semi-major axes.
The categorization of unresolved sources and a discussion of the flux densities 
are presented in Section~\ref{ef0:sec:source_flux}. 

\begin{table*}
\caption{The source catalogue. This is a sample source catalogue; the complete catalogue appears in the on-line version.}
\label{ef0:table:srclist}
\begin{tabular}{l r r r r r r r r r r r}
\hline
\multicolumn{1}{c}{Source name} & \multicolumn{1}{c}{RA} & \multicolumn{1}{c}{DEC} & \multicolumn{1}{c}{Dist} & \multicolumn{1}{c}{$\sigma_{rms}$} & \multicolumn{1}{c}{S$_{peak}$} & \multicolumn{1}{c}{S$_{total~3\sigma}$} & \multicolumn{1}{c}{S$_{total}$} & \multicolumn{1}{c}{Size} & \multicolumn{1}{c}{Notes} \\ 
                    & \multicolumn{1}{c}{hh:mm:ss.s} & \multicolumn{1}{c}{dd:mm:ss.s} & \multicolumn{1}{c}{deg} & \multicolumn{1}{c}{$\mu$Jy b$^{-1}$}  & \multicolumn{1}{c}{mJy b$^{-1}$} &  \multicolumn{1}{c}{mJy} &  \multicolumn{1}{c}{mJy}  & \multicolumn{1}{c}{$^{\prime\prime}$} &        \\ 
\multicolumn{1}{c}{(1)} & \multicolumn{1}{c}{(2)} & \multicolumn{1}{c}{(3)} & \multicolumn{1}{c}{(4)} &  \multicolumn{1}{c}{(5)} & \multicolumn{1}{c}{(6)} & \multicolumn{1}{c}{(7)} & \multicolumn{1}{c}{(8)} & \multicolumn{1}{c}{(9)} & \multicolumn{1}{c}{(10)} \\
\hline
GMRT160515+0544536  & 16:05:15.2 & +54:45:38.0 &     0.69 &      231 &     6.56 &    13.22 &    13.22 &     10.7 \\
GMRT160522+0542924  & 16:05:22.4 & +54:29:26.7 &     0.69 &      197 &    13.56 &    28.63 &    28.63 &     33.4 \\
GMRT160524+0544710  & 16:05:24.3 & +54:47:13.7 &     0.67 &      149 &     1.30 &     1.49 &     1.30 &        U \\
GMRT160538+0543919  & 16:05:38.5 & +54:39:26.4 &     0.63 &     1692 &   435.71 &  1022.63 &  1022.63 &     77.1 \\
GMRT160538+0544130  & 16:05:38.3 & +54:41:32.3 &     0.63 &      416 &     2.58 &     5.48 &     2.58 &        U \\
\\
GMRT160539+0544544  & 16:05:39.8 & +54:45:45.5 &     0.63 &      236 &     3.72 &     3.21 &     3.72 &        U \\
GMRT160542+0544640  & 16:05:42.6 & +54:46:45.6 &     0.63 &      245 &    25.15 &    30.74 &    30.74 &     13.0 \\
GMRT160551+0543842  & 16:05:52.0 & +54:38:49.1 &     0.60 &      478 &     4.96 &    11.57 &    11.57 &     12.4 \\
GMRT160553+0542223  & 16:05:53.6 & +54:22:26.6 &     0.66 &       96 &     1.30 &     1.45 &     1.30 &        U \\
GMRT160553+0544537  & 16:05:53.3 & +54:45:42.0 &     0.60 &      268 &     3.62 &    26.68 &    26.68 &     57.5 \\
\\
GMRT160556+0543043  & 16:05:56.8 & +54:30:52.1 &     0.61 &      106 &     0.95 &     0.92 &     0.95 &        U \\
GMRT160600+0545402  & 16:06:00.1 & +54:54:08.8 &     0.62 &      918 &   327.65 &  1076.82 &  1076.82 &    109.7 \\
GMRT160602+0543937  & 16:06:02.2 & +54:39:37.3 &     0.57 &      254 &     2.33 &     3.54 &     2.33 &        U \\
GMRT160603+0545930  & 16:06:03.6 & +54:59:37.9 &     0.65 &      140 &     1.40 &     1.42 &     1.40 &        U \\
GMRT160605+0542812  & 16:06:05.8 & +54:28:13.9 &     0.60 &      187 &    36.93 &    63.43 &    63.43 &     44.0 \\
\\
GMRT160606+0542857  & 16:06:06.4 & +54:28:62.9 &     0.59 &      160 &     1.24 &     1.25 &     1.24 &        U \\
GMRT160609+0542708  & 16:06:09.1 & +54:27:11.9 &     0.60 &      120 &     1.45 &     2.71 &     2.71 &      6.4 \\
GMRT160613+0543501  & 16:06:13.6 & +54:35:07.1 &     0.55 &      131 &     0.92 &     1.58 &     0.92 &        U \\
GMRT160613+0545322  & 16:06:13.4 & +54:53:28.8 &     0.59 &      398 &     4.45 &    14.50 &    14.50 &     14.8 \\
GMRT160613+0550149  & 16:06:13.6 & +55:01:50.4 &     0.65 &      200 &     2.70 &     7.90 &     7.90 &     22.4 \\
\\
GMRT160621+0545636  & 16:06:21.3 & +54:56:37.9 &     0.59 &      202 &     4.10 &    22.44 &    22.44 &     48.5 \\
GMRT160622+0541332  & 16:06:22.4 & +54:13:41.4 &     0.69 &      225 &    64.81 &    83.41 &    83.41 &     45.4 \\
GMRT160623+0544129  & 16:06:23.3 & +54:41:36.5 &     0.52 &      119 &     0.83 &     0.89 &     0.83 &        U \\
GMRT160624+0545437  & 16:06:24.5 & +54:54:38.4 &     0.57 &      199 &     1.23 &     0.86 &     1.23 &        U \\
GMRT160626+0542113  & 16:06:26.9 & +54:21:19.0 &     0.60 &       76 &     0.81 &     0.63 &     0.81 &        U \\
\\
GMRT160626+0544906  & 16:06:26.3 & +54:49:07.6 &     0.54 &      178 &     2.27 &     6.09 &     6.09 &     15.0 \\
GMRT160632+0542453  & 16:06:32.3 & +54:24:56.8 &     0.56 &       77 &     0.61 &     0.48 &     0.61 &        U \\
GMRT160633+0544322  & 16:06:33.6 & +54:43:23.6 &     0.50 &       98 &     0.66 &     0.48 &     0.66 &        U \\
GMRT160633+0545525  & 16:06:33.1 & +54:55:31.1 &     0.56 &      146 &     0.92 &     1.28 &     0.92 &        U \\
GMRT160634+0543451  & 16:06:34.8 & +54:34:57.9 &     0.50 &      160 &    10.58 &    70.94 &    70.94 &    108.5 \\
\\
GMRT160634+0544542  & 16:06:34.4 & +54:45:49.4 &     0.50 &      108 &     2.77 &     3.82 &     3.82 &      7.8 \\
GMRT160636+0543247  & 16:06:36.4 & +54:32:53.5 &     0.51 &      137 &    17.34 &    45.50 &    45.50 &     18.0 \\
GMRT160638+0545918  & 16:06:38.5 & +54:59:25.1 &     0.58 &      119 &     1.53 &     2.47 &     1.53 &        U \\
GMRT160639+0542313  & 16:06:39.1 & +54:23:13.2 &     0.56 &       77 &     1.14 &     1.40 &     1.14 &        U \\
GMRT160642+0542636  & 16:06:42.3 & +54:26:45.9 &     0.53 &       79 &     3.16 &     3.50 &     3.50 &      9.4 \\
\\
GMRT160642+0542711  & 16:06:42.8 & +54:27:17.4 &     0.52 &       81 &     0.79 &     0.68 &     0.79 &        U \\
GMRT160644+0545341  & 16:06:44.0 & +54:53:49.9 &     0.52 &      126 &     1.21 &     2.74 &     1.21 &        U \\
GMRT160647+0541508  & 16:06:47.2 & +54:15:16.6 &     0.62 &      102 &     5.50 &    10.74 &    10.74 &     34.2 \\
GMRT160650+0542511  & 16:06:50.7 & +54:25:16.1 &     0.52 &       75 &     0.68 &     1.32 &     0.68 &        U \\
GMRT160651+0541449  & 16:06:51.3 & +54:14:49.1 &     0.62 &      103 &     1.50 &     4.90 &     4.90 &     24.4 \\
\\
GMRT160655+0542753  & 16:06:55.2 & +54:27:60.0 &     0.49 &       74 &     0.55 &     0.52 &     0.55 &        U \\
GMRT160656+0541457  & 16:06:56.1 & +54:14:63.0 &     0.61 &       91 &     1.14 &     1.11 &     1.14 &        U \\
GMRT160657+0550156  & 16:06:57.5 & +55:01:57.0 &     0.57 &      102 &     1.35 &     4.32 &     4.32 &     21.7 \\
GMRT160657+0550350  & 16:06:57.8 & +55:03:54.9 &     0.59 &       92 &     1.47 &     1.74 &     1.47 &        U \\
GMRT160658+0544328  & 16:06:58.0 & +54:43:30.0 &     0.44 &       91 &    13.00 &    22.90 &    22.90 &     35.7 \\
\\
GMRT160659+0541135  & 16:06:59.6 & +54:11:44.0 &     0.64 &       94 &     0.63 &     0.48 &     0.63 &        U \\
GMRT160700+0541343  & 16:07:00.1 & +54:13:48.0 &     0.62 &       84 &     0.65 &     0.42 &     0.65 &        U \\
GMRT160702+0542348  & 16:07:02.6 & +54:23:54.6 &     0.51 &       61 &     0.58 &     0.53 &     0.58 &        U \\
GMRT160703+0541626  & 16:07:03.9 & +54:16:27.1 &     0.58 &       82 &     0.53 &     0.60 &     0.53 &        U \\
GMRT160703+0543414  & 16:07:03.2 & +54:34:20.6 &     0.44 &       71 &     1.35 &     1.94 &     1.35 &        U \\
\hline \hline
\end{tabular}

\end{table*}
 
\subsubsection{Source size and flux densities} 
\label{ef0:sec:source_flux}
For low signal-to-noise ratio detections, Gaussian fitting 
routines may be significantly affected by noise spikes leading 
to errors in estimating both the width and flux density of the
source. The ratio 

\begin{figure}
\begin{center}
\includegraphics[angle=270, totalheight=2.5in, width=3.5in]{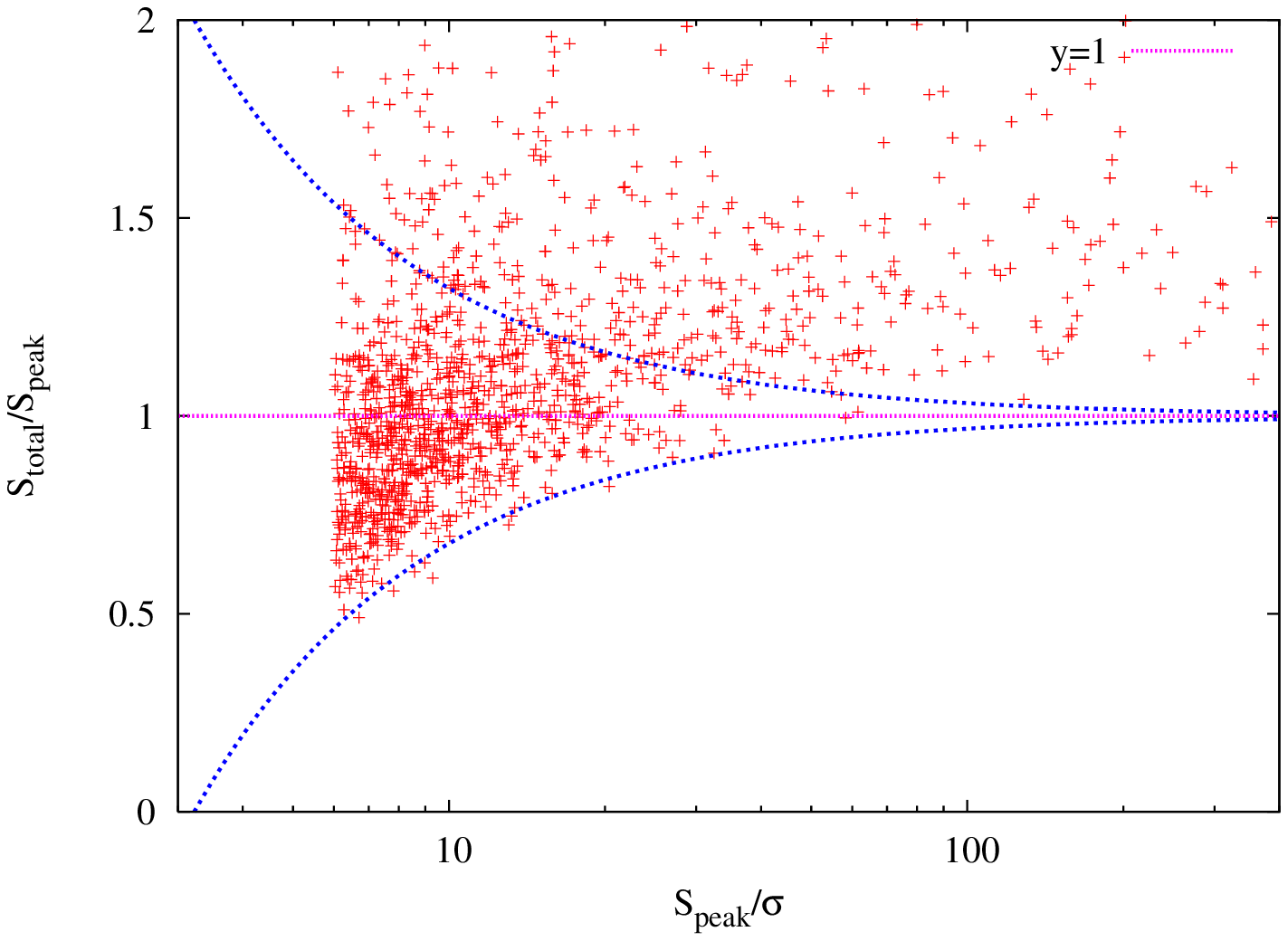}
\caption{The ratio of the total flux density, S$_{\rm total}$, 
to the peak value S$_{\rm peak}$
as a function of the source S$_{\rm peak}$-to-rms noise ($\sigma$) ratio. The 6$\sigma$ cut-off is 
adopted for the catalogue definition and the horizontal line gives
S$_{\rm total}$ = S$_{\rm peak}$ locus. Also shown are the lower and upper envelopes 
(dashed lines) of the flux density ratio distribution containing  almost all the unresolved
sources.}
\label{ef0:fig:src_res}
\end{center}
\end{figure}

S$_{\rm total}$/S$_{\rm peak}$ = $\theta_{\rm min}$ $\theta_{\rm maj}$ /($b_{\rm min}$ $b_{\rm maj}$)

where $\theta_{\rm min}$ and $\theta_{\rm maj}$ are the major and minor
axes of the detected source and $b_{\rm min}$ and $b_{\rm maj}$ are 
the major and minor axes of the restoring beam. The flux density 
ratio may be used to discriminate between unresolved sources and
those which are much larger than the beam \citep[see][]{2006A+A...457..517P}.
In Fig.~\ref{ef0:fig:src_res}, we plot the above ratio of the flux densities to the 
signal-to-noise ratio (S$_{\rm peak}$/$\sigma_{\rm local}$) for all
sources above the 6$\sigma_{\rm local}$ threshold. Clearly sources
for which S$_{\rm total}$/S$_{\rm peak}$ $<$1 have been affected by
noise. We first fit a functional form of the curve $f(x)=1.0\pm\frac{3.22}{x}$ such that almost all
of the points with S$_{\rm total}$/S$_{\rm peak}$ $<$1 lie above the curve. Reflecting
this curve about S$_{\rm total}$/S$_{\rm peak}$ $=$1 gives a list of 
all the sources which lie between the two curves and can be considered to be unresolved. 
This analysis shows that about 73 per cent of the sources (943/1286) are 
considered to be unresolved and these have been listed as 
`unresolved' sources (U) in the Table.

\subsection{Comparison with WENSS} 
\label{ef0:sec:source_wenss}
\begin{figure}
\begin{center}
\includegraphics[angle=270, totalheight=2.5in, width=3.5in]{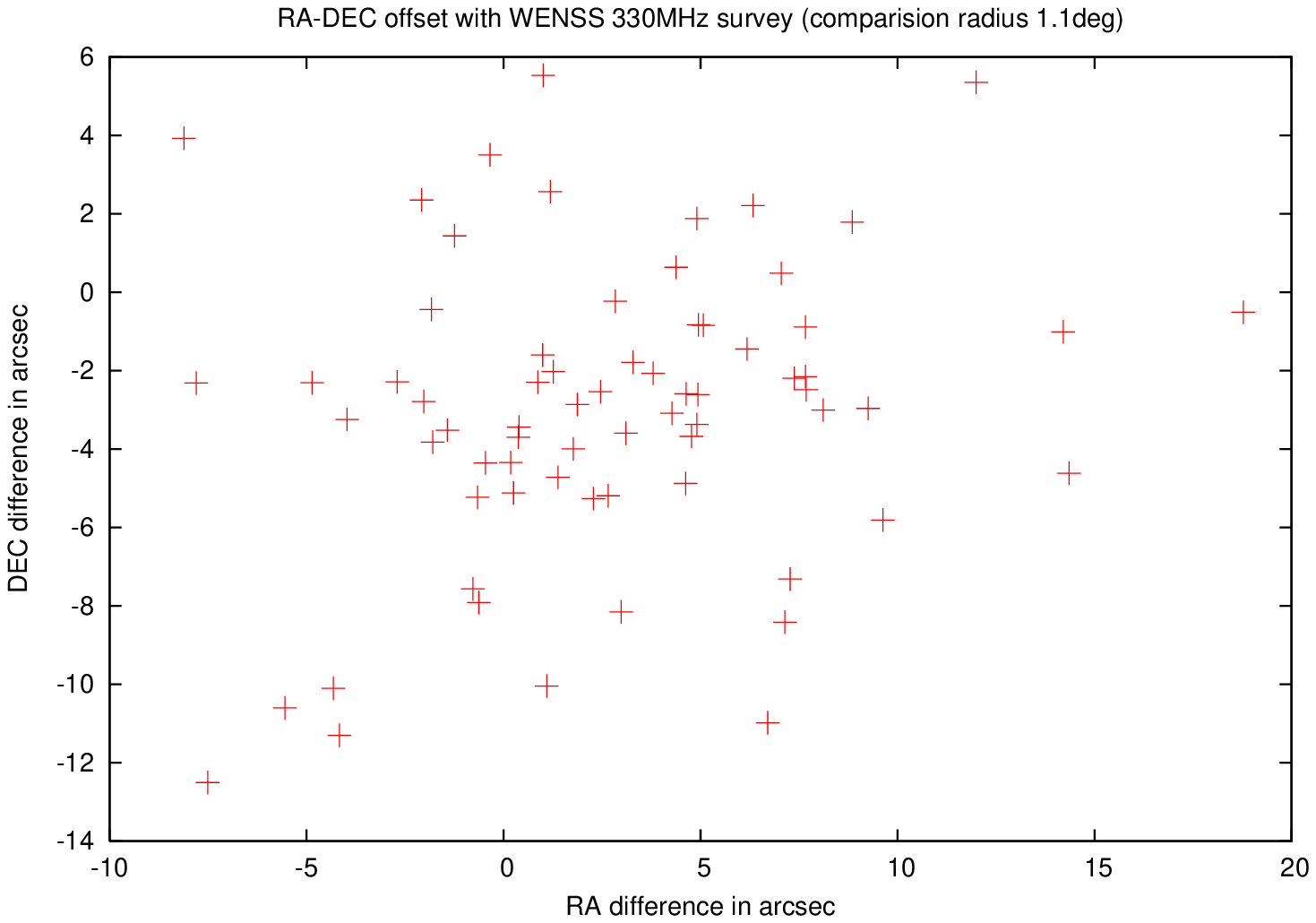}
\caption{Position difference of the corresponding sources observed by GMRT and WENSS for the 
unresolved WENSS sources.}
\label{ef0:fig:wenss_gmrt_radec}
\end{center}
\end{figure}

To compare our estimates of the flux densities with those of the WENSS catalogue we have
identified all WENSS sources which lie within 10 arcsec of our 3-$\sigma$ centroid position
from the catalog available via Vizier \citep{2000A+AS..143...23O}. We have compared the positions of
the centroids of emission for the unresolved WENSS sources with our sources 
(as shown in Fig.~\ref{ef0:fig:wenss_gmrt_radec}) and
find the mean and rms displacement to be 2.45 and 6.61 arcsec in right ascension and $-$3.45
and 5.78 arcsec in declination. The value of 10 arcsec is the half power synthesized beamwidth 
of our GMRT observations, while the quoted error in the WENSS positions is 
$\sim$1.5 arcsec \citep{2000yCat.8062....0D}. Within the uncertainties, the GMRT positions are 
consistent with the WENSS ones. 

\begin{figure}
\begin{center}
\hbox{
 \vbox{
  \includegraphics[angle=270, totalheight=2.5in, width=3.5in]{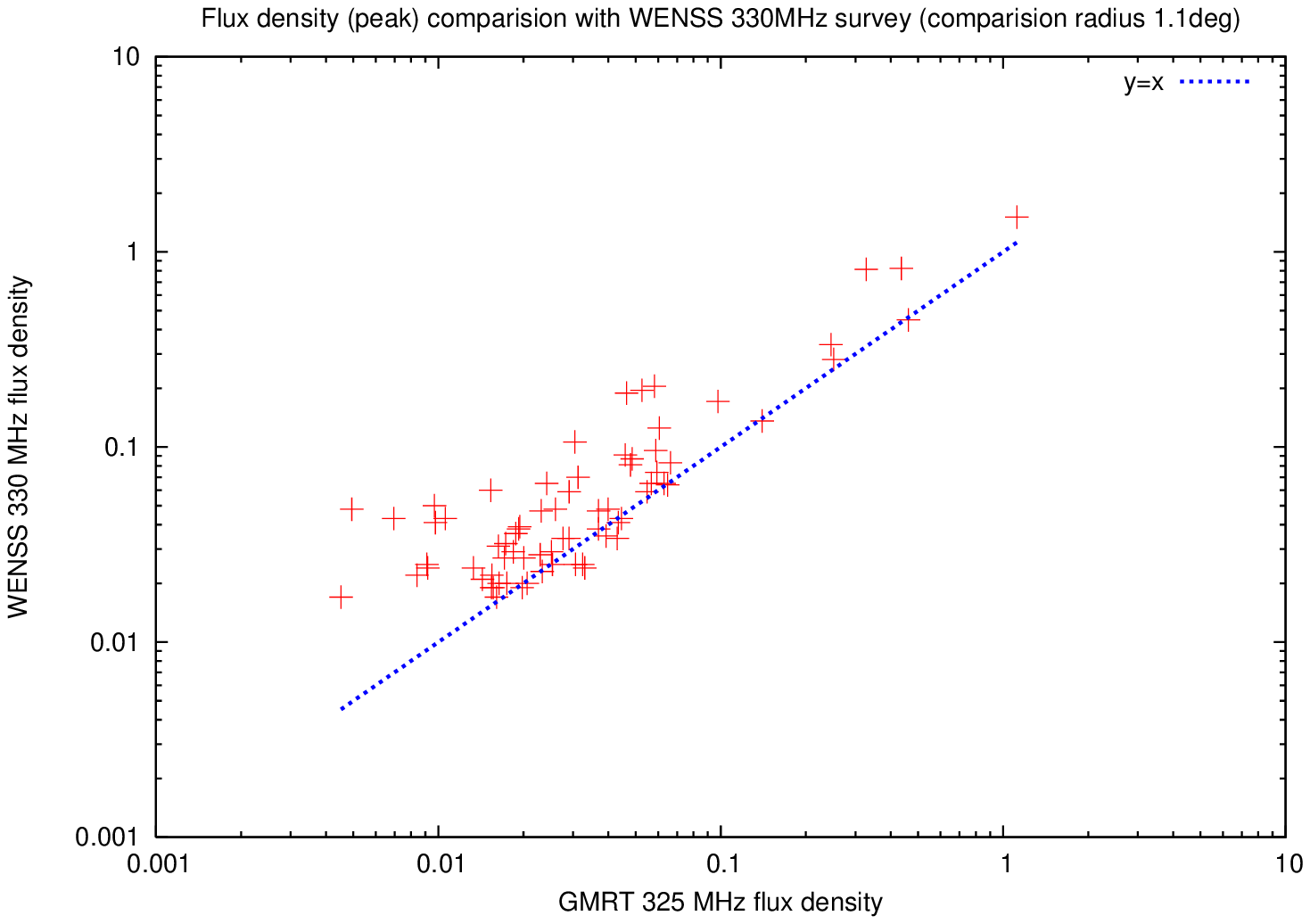}
  \includegraphics[angle=270, totalheight=2.5in, width=3.5in]{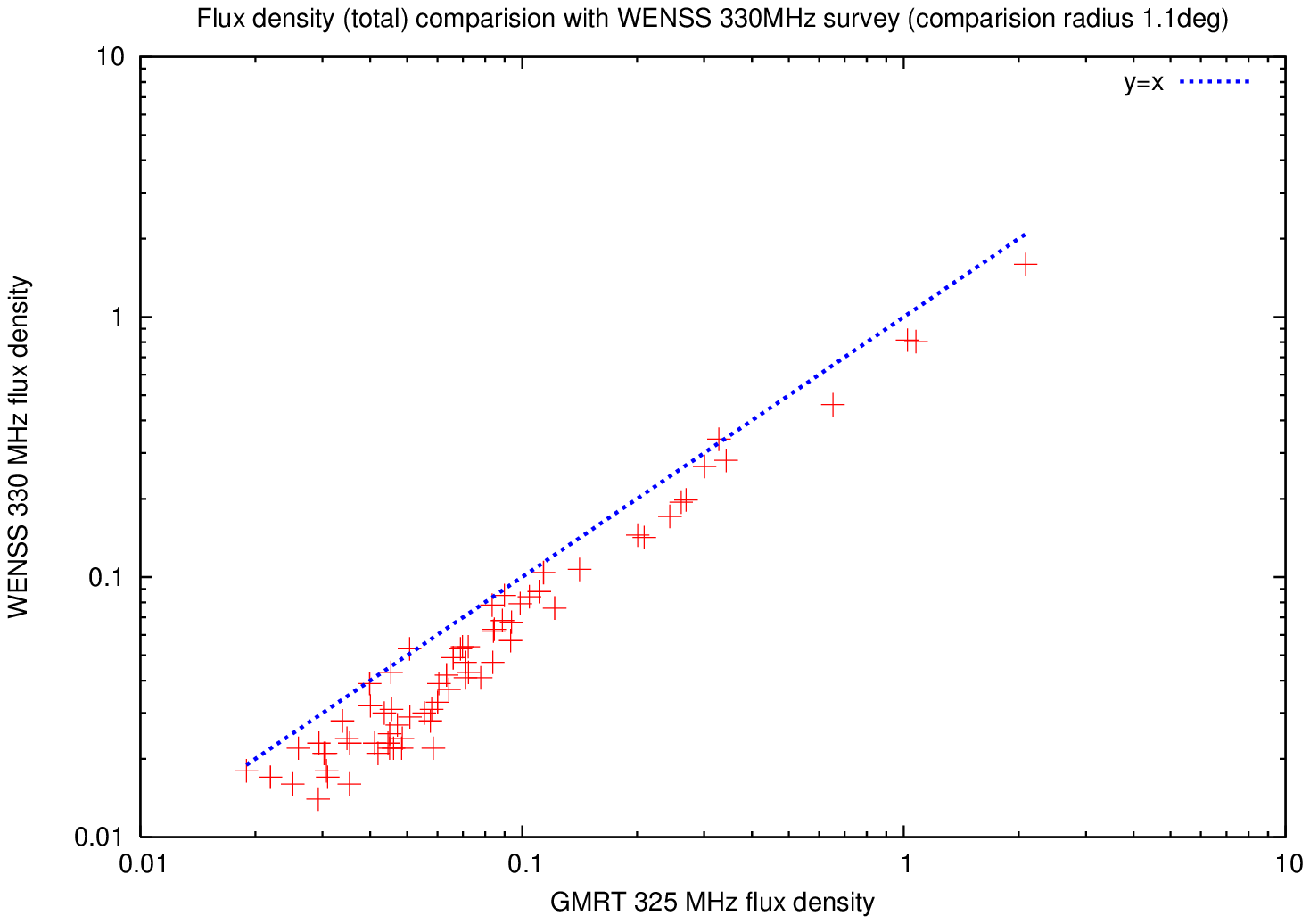}
 }
}
\caption{The peak (upper panel) and total (lower panel) flux densities from our observations plotted against the
corresponding values from WENSS. }
\label{ef0:fig:wenss_gmrt}
\end{center}
\end{figure}

Figure~\ref{ef0:fig:wenss_gmrt} shows a comparison of the peak and integrated flux densities
estimated from our GMRT observations at 325 MHz with those of the WENSS survey at 330 MHz.
There are a total of 72 sources. It is clear that the peak flux densities obtained from the
WENSS survey tend to be systematically larger than the GMRT ones by approximately 34 per
cent on average. This would be expected in the presence of extended emission because of the
larger beamwidth in the WENSS survey of 54$\times$54 cosec$\delta$ arcsec$^2$ compared with 
$\sim$10 arcsec for the GMRT image. However, while comparing the integrated flux densities,
the WENSS flux densities are $\approx$76\% of the GMRT values on average. This is perhaps a
reflection of the fact that the GMRT image being significantly deeper than the WENSS ones,
and thereby more extended emission contributes to the flux density within the 3-$\sigma$
contour level.

\subsubsection{Selected extended sources} 
\label{ef0:sec:source_double}
\begin{figure}
\begin{center}
   \includegraphics[angle=0, totalheight=2.1in, viewport=19 212 573 627, clip]{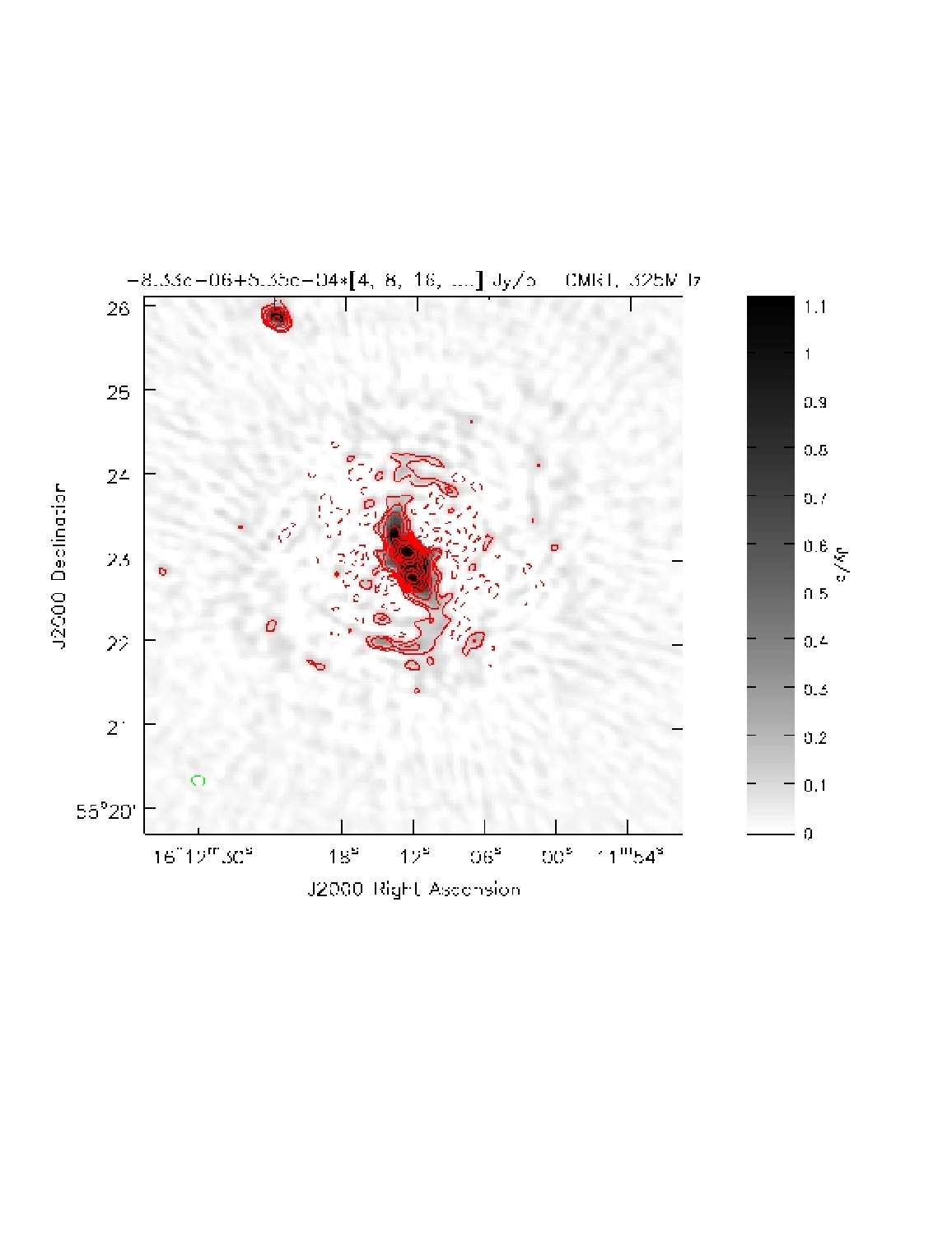}
   \includegraphics[angle=0, totalheight=2.1in, viewport=19 212 573 627, clip]{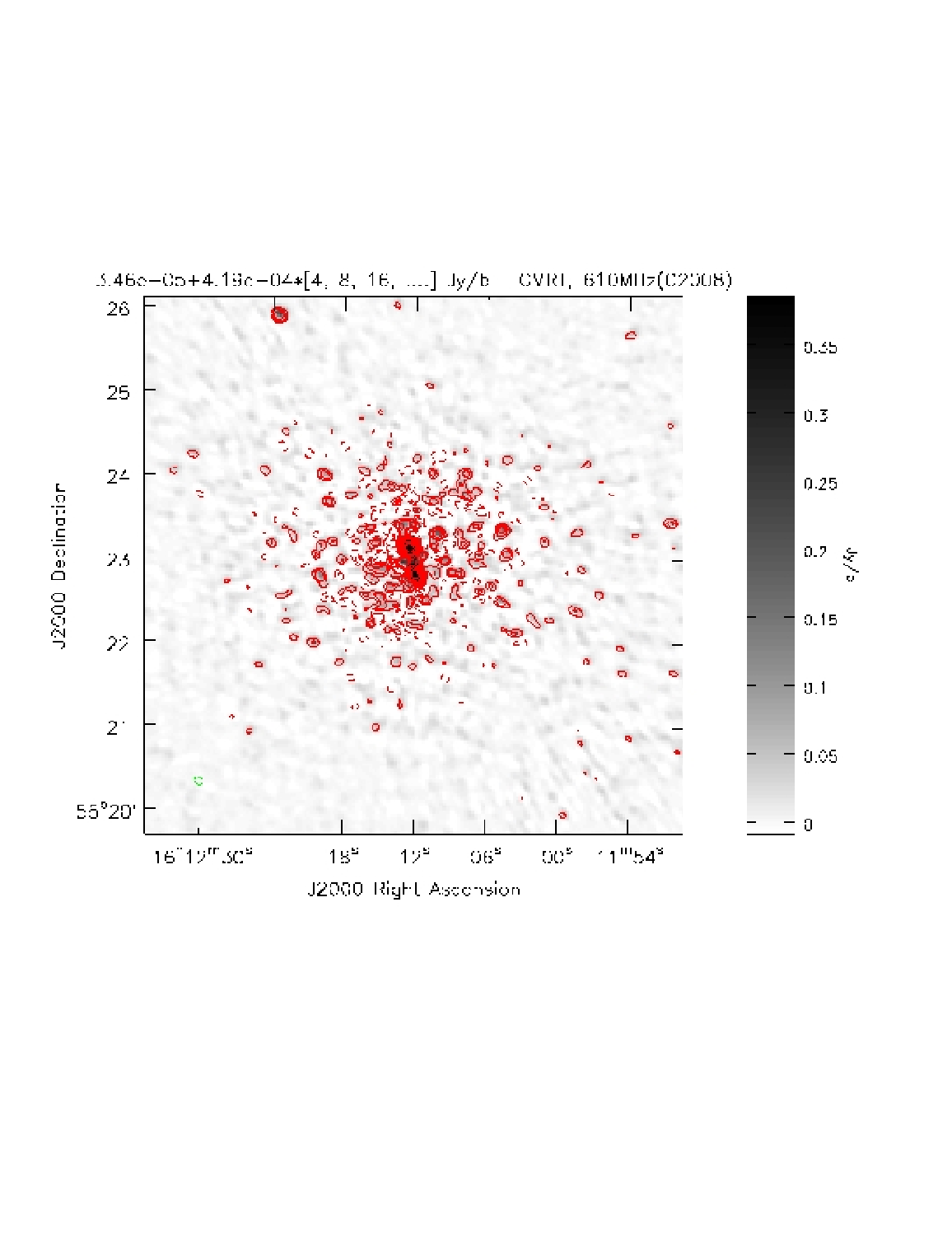}
   \includegraphics[angle=0, totalheight=2.1in, viewport=19 212 573 627, clip]{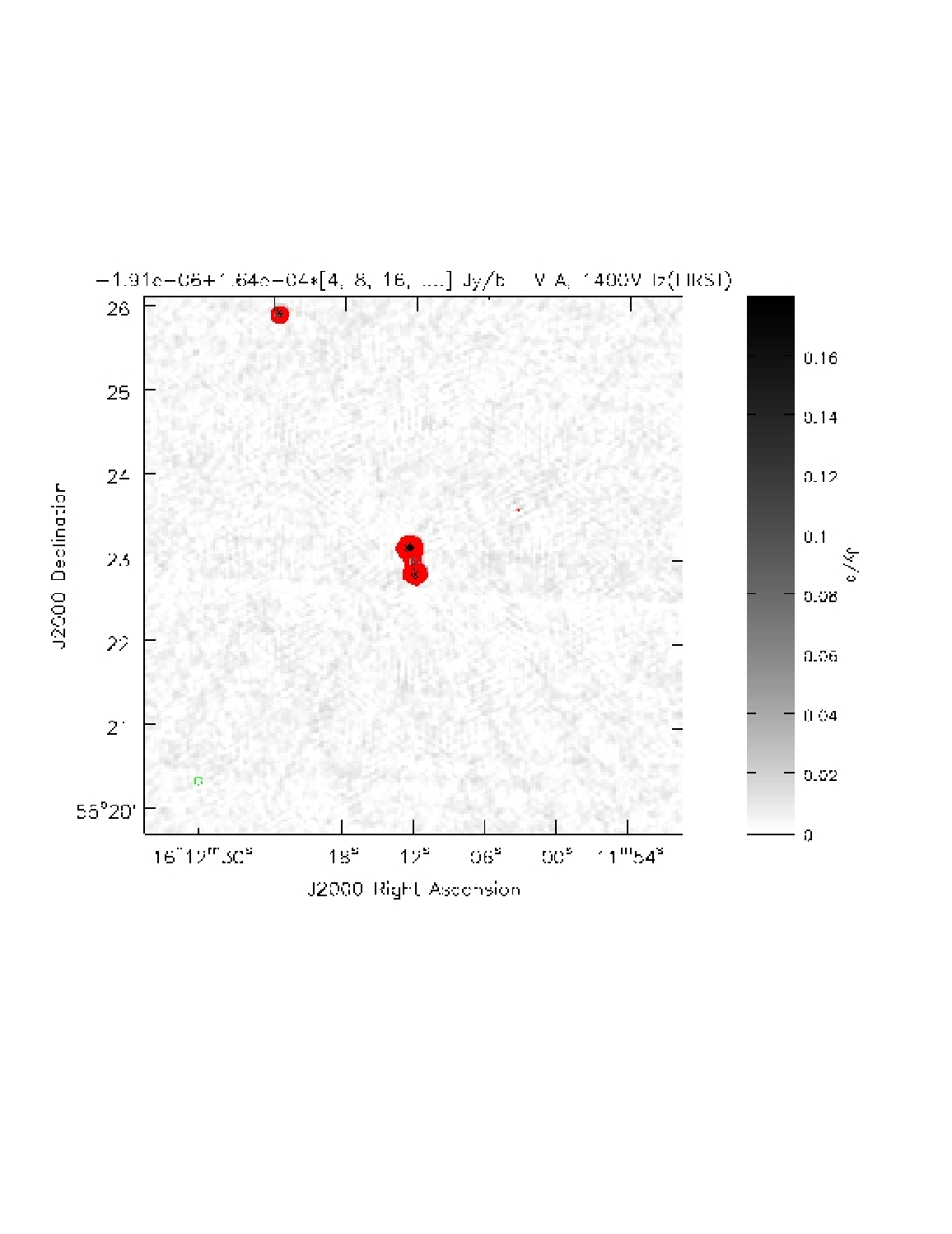}
   \includegraphics[angle=0, totalheight=2.1in, viewport=19 212 573 627, clip]{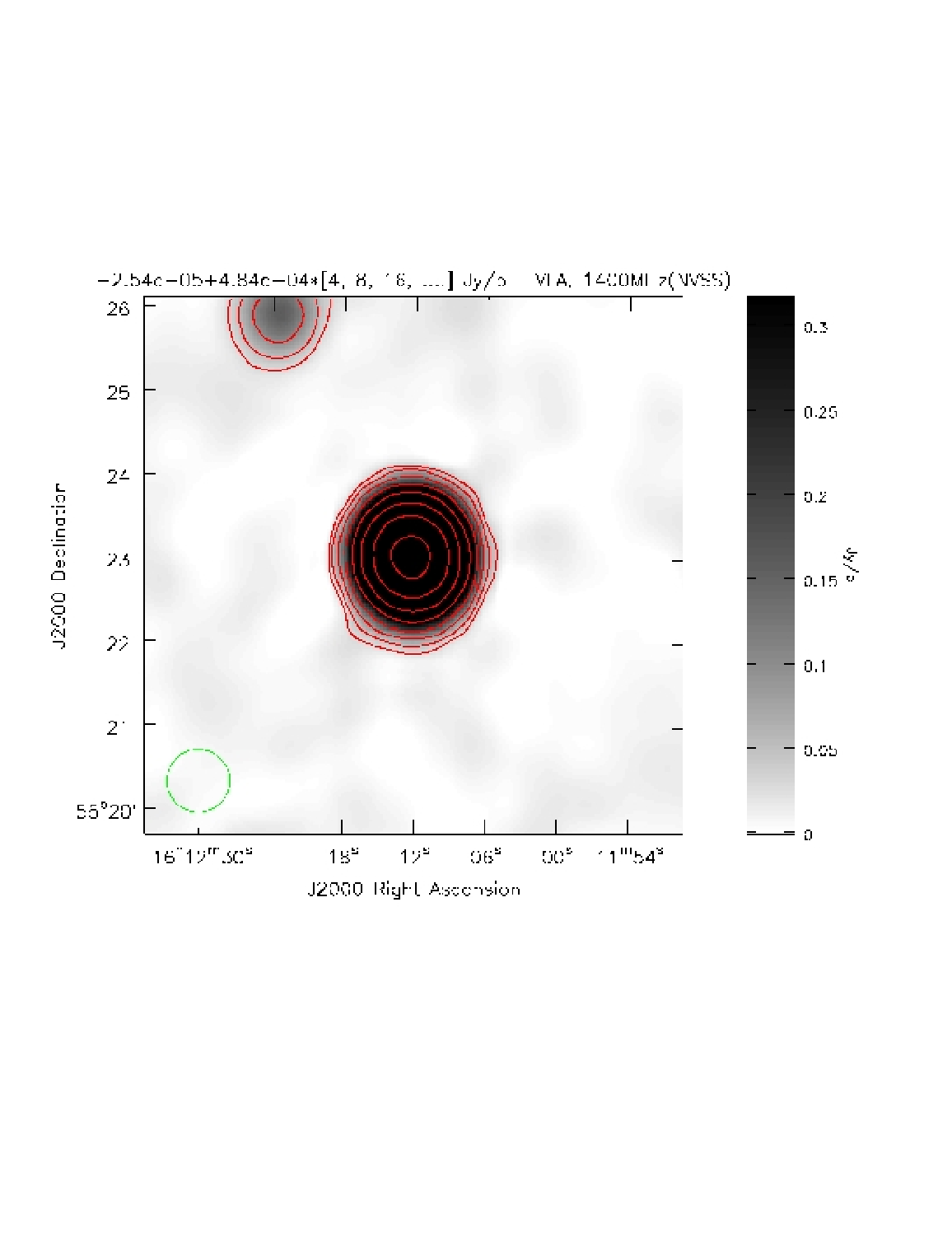}
\caption{GMRT image at 325 MHz of a strong source in the field to illustrate the effects of dynamic
range (top), and the corresponding images of the same region at 610 MHz with the GMRT from G2008, 
and from FIRST and NVSS at 1400 MHz in descending order.}
\label{ef0:fig:dynrng_1}
\end{center}
\end{figure}

\begin{figure}
\begin{center}
   \includegraphics[angle=0, totalheight=2.1in, viewport=19 212 573 627, clip]{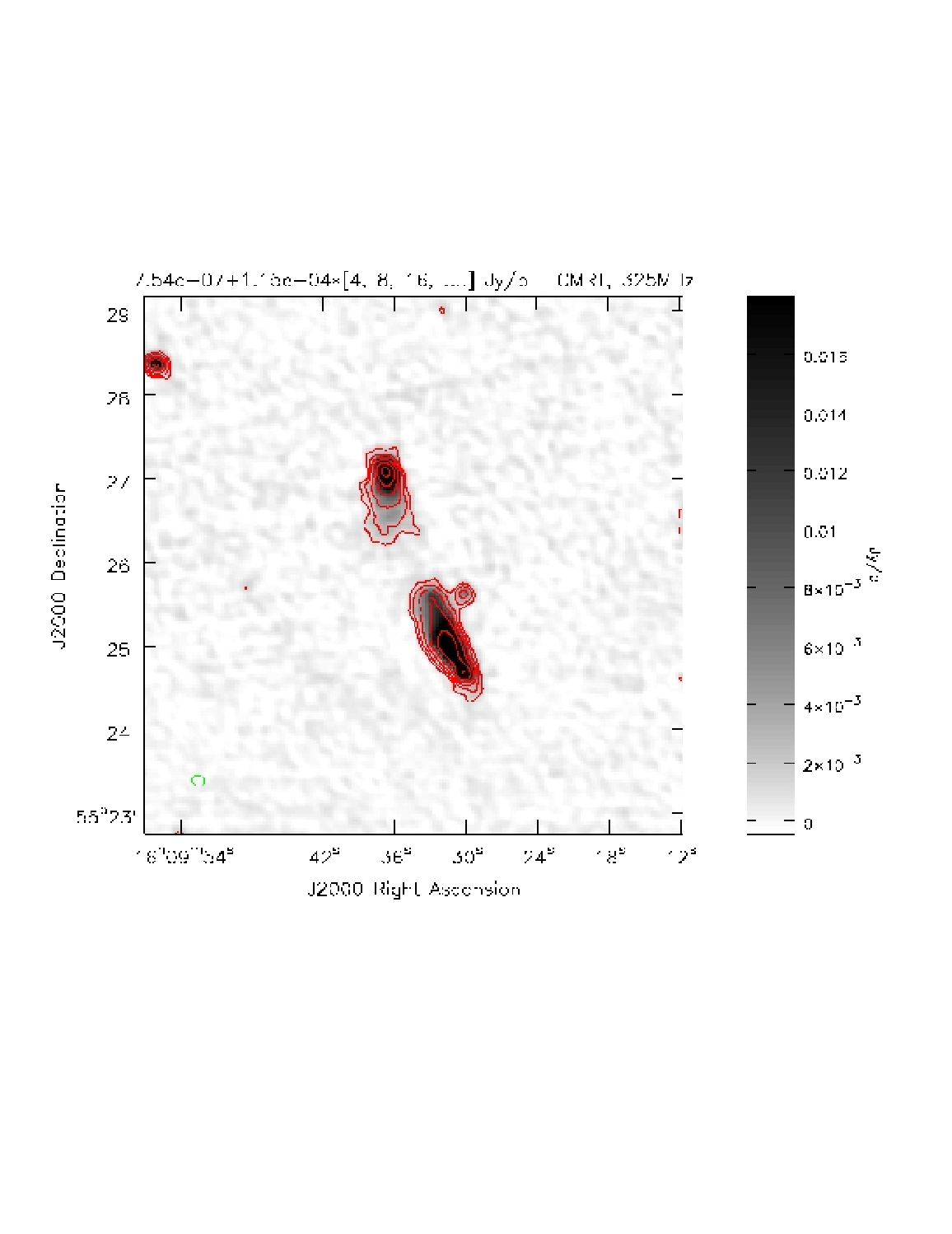}
   \includegraphics[angle=0, totalheight=2.1in, viewport=19 212 573 627, clip]{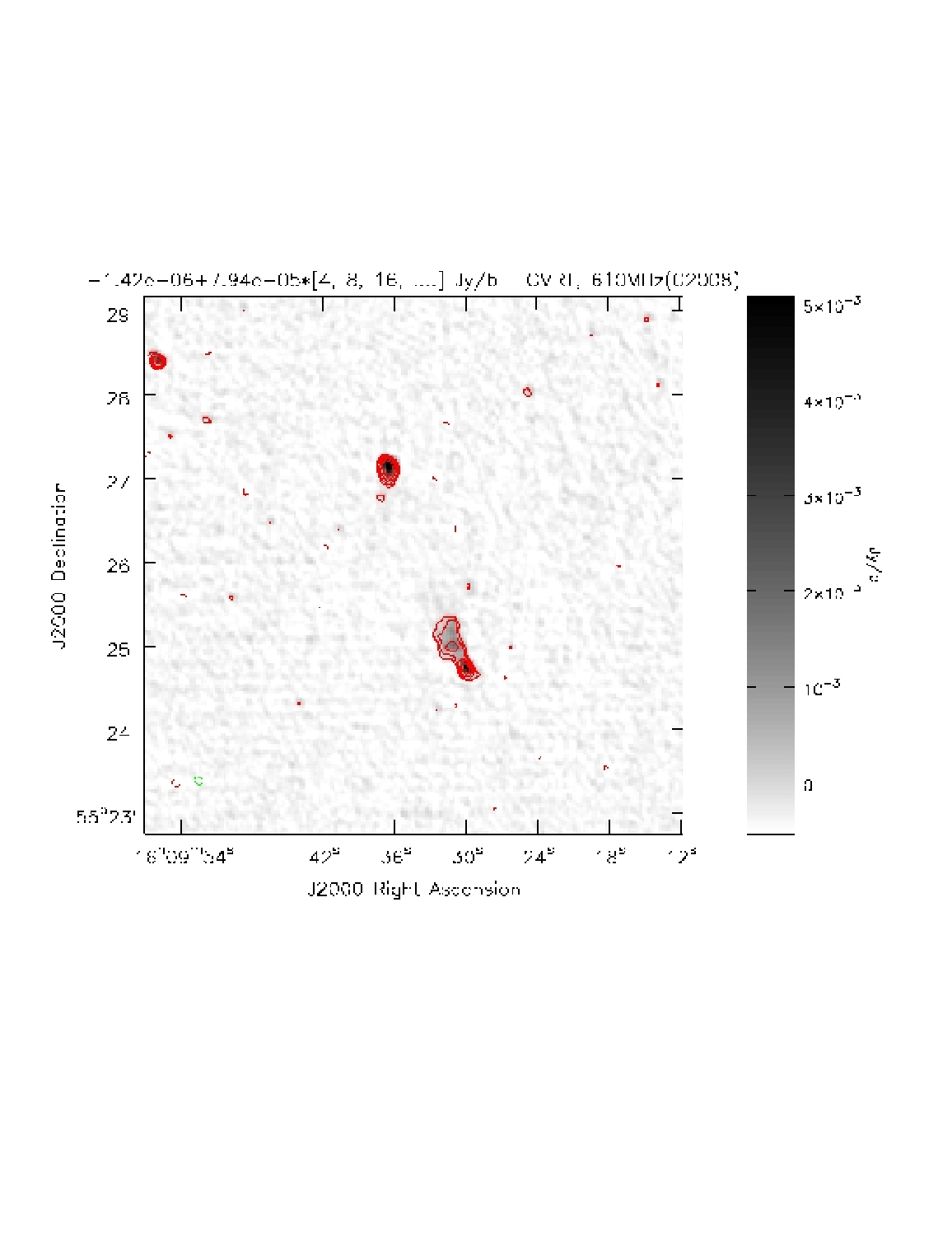}
   \includegraphics[angle=0, totalheight=2.1in, viewport=19 212 573 627, clip]{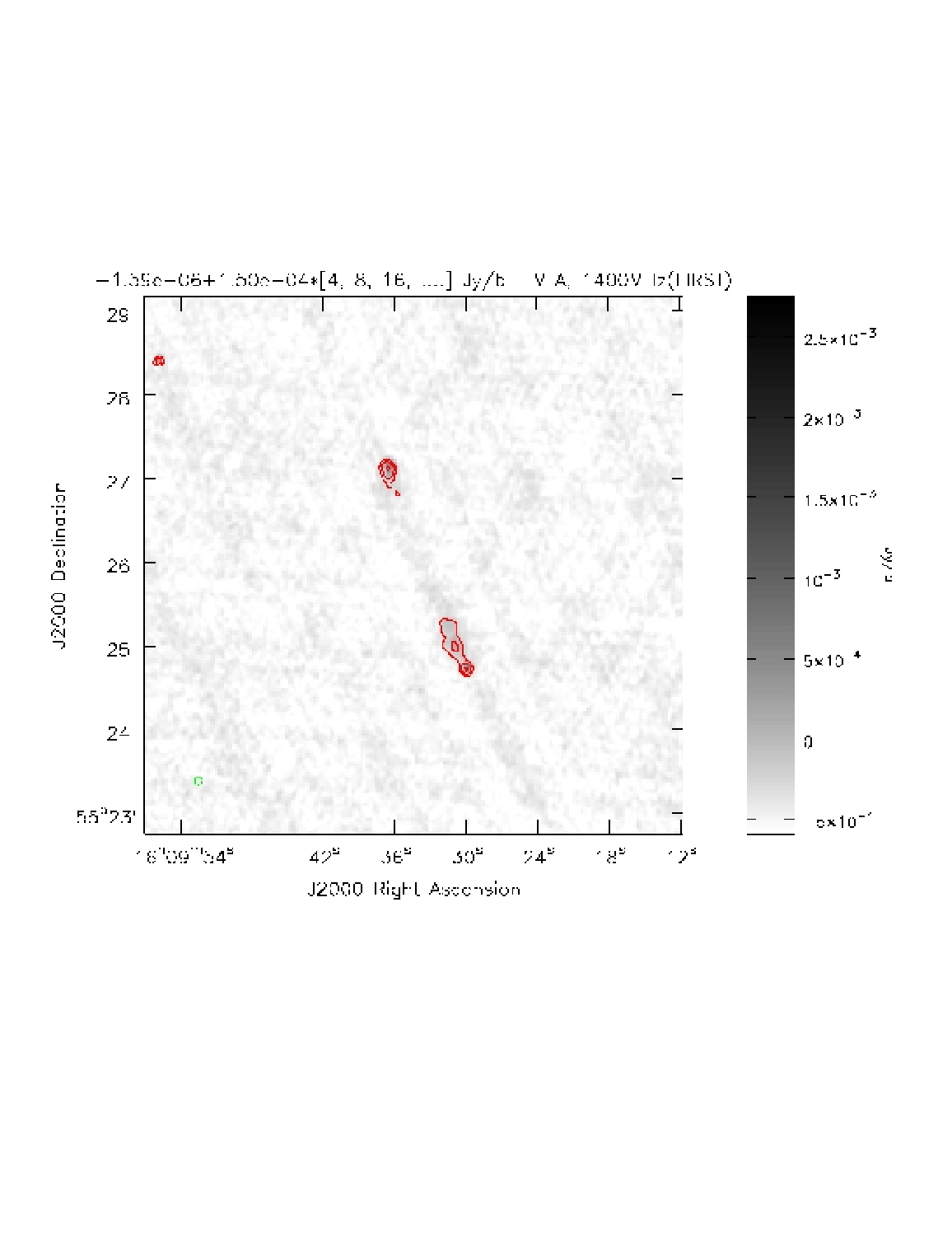}
   \includegraphics[angle=0, totalheight=2.1in, viewport=19 212 573 627, clip]{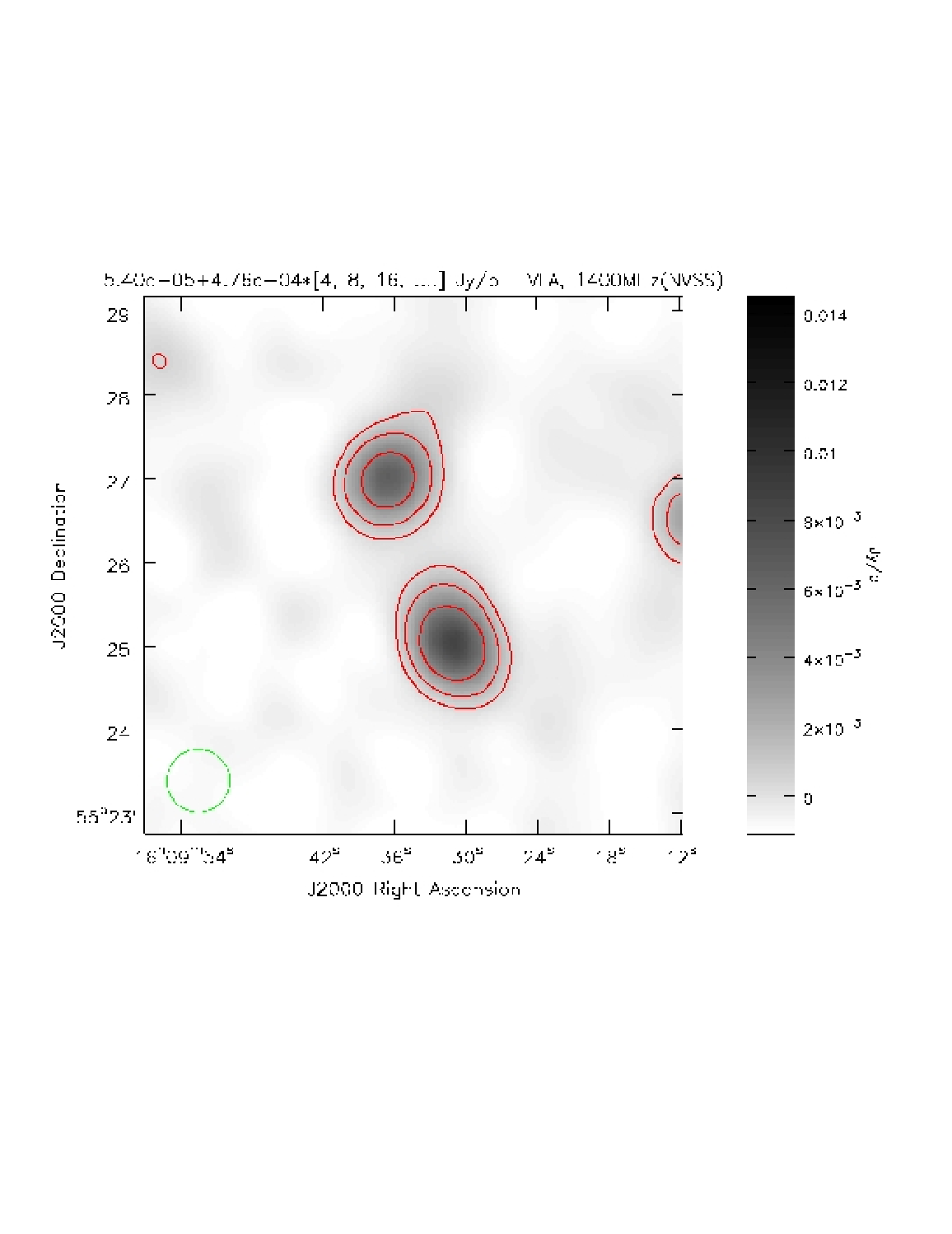}
\caption{GMRT image at 325 MHz of a source with extended lobes of emission (top), and
the corresponding images of the same region at 610 MHz with the GMRT from G2008, 
and from FIRST and NVSS  at 1400 MHz in descending order.}
\label{ef0:fig:diffemm_2}
\end{center}
\end{figure}

\begin{figure}
\begin{center}
   \includegraphics[angle=0, totalheight=2.1in, viewport=19 212 573 627, clip]{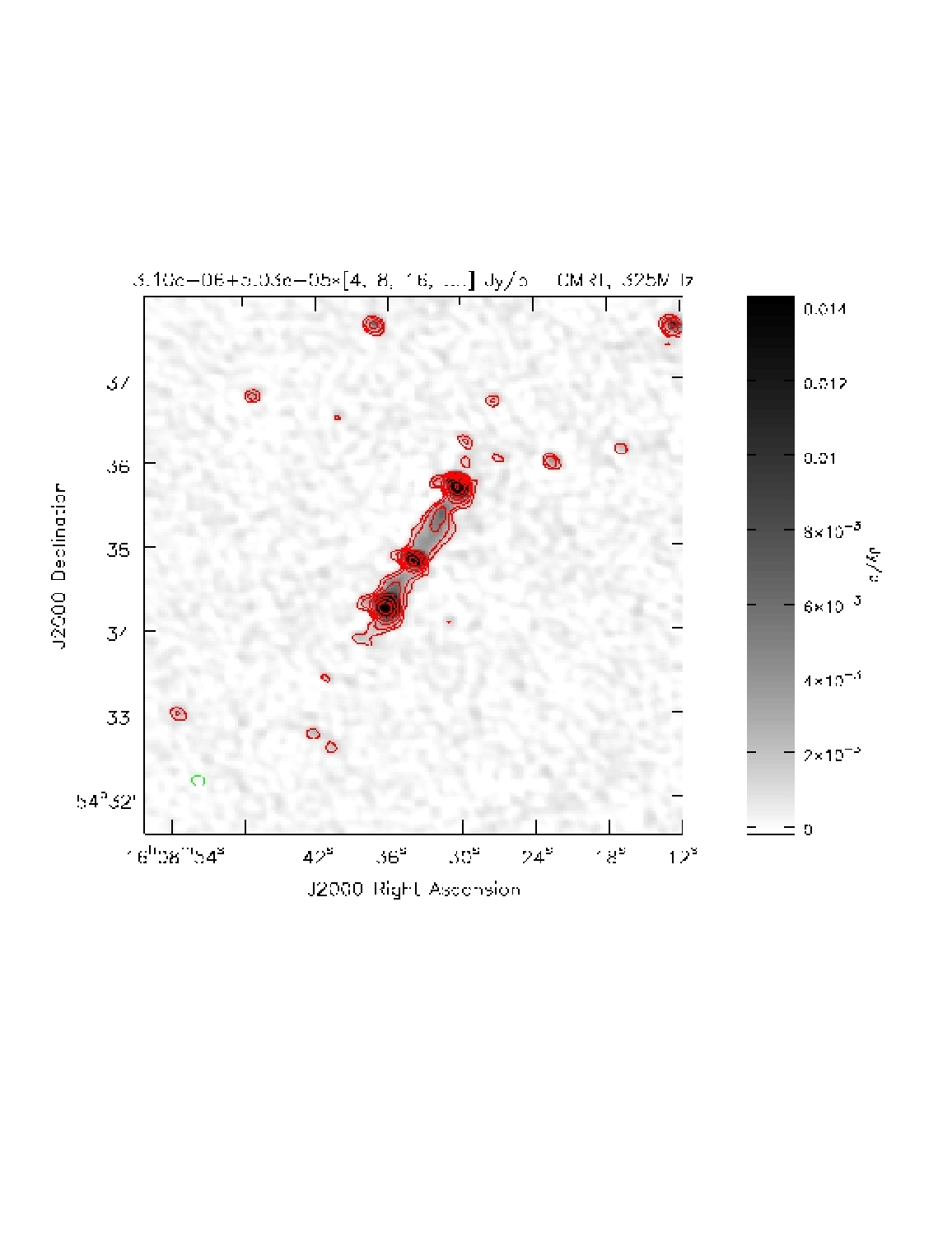}
   \includegraphics[angle=0, totalheight=2.1in, viewport=19 212 573 627, clip]{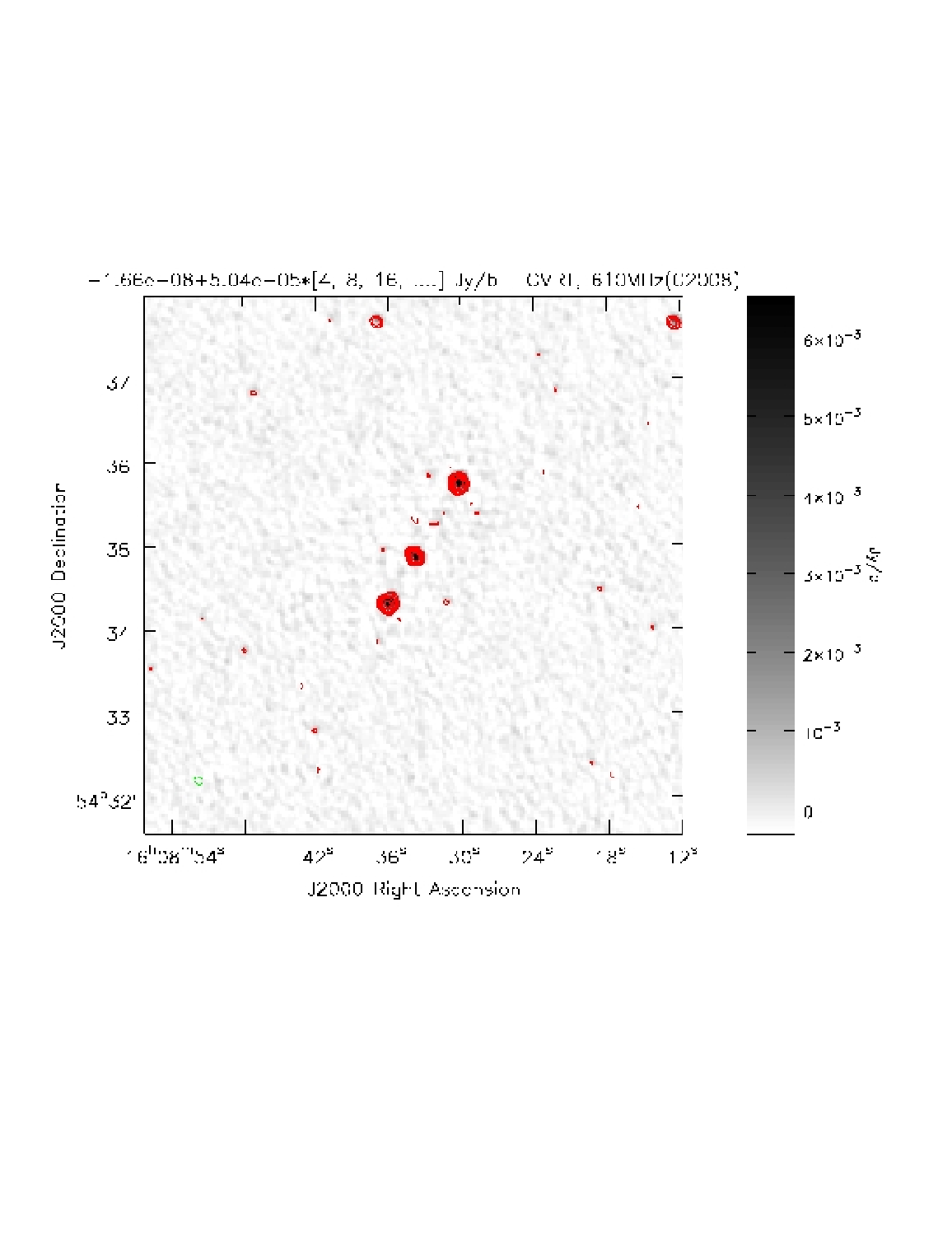}
   \includegraphics[angle=0, totalheight=2.1in, viewport=19 212 573 627, clip]{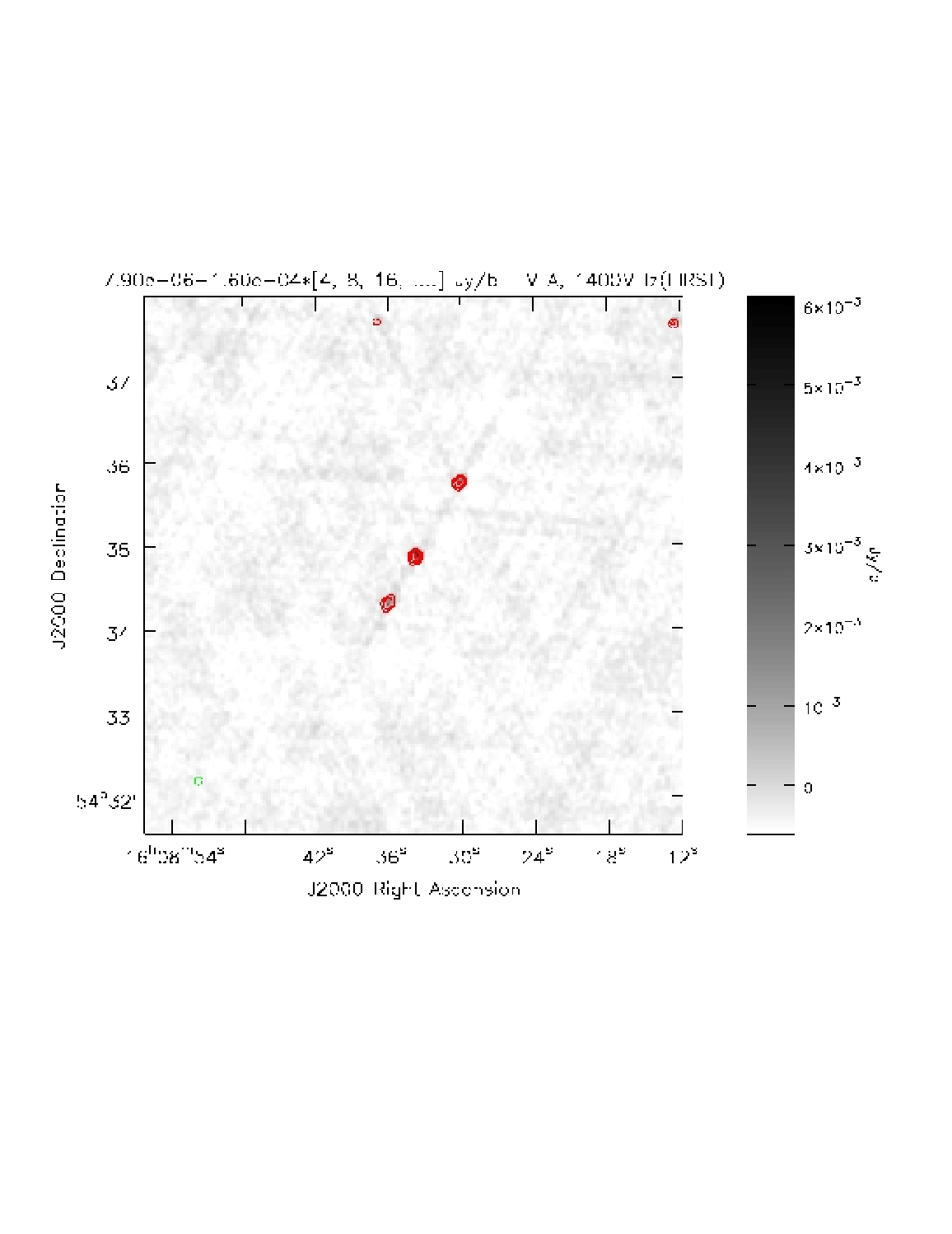}
   \includegraphics[angle=0, totalheight=2.1in, viewport=19 212 573 627, clip]{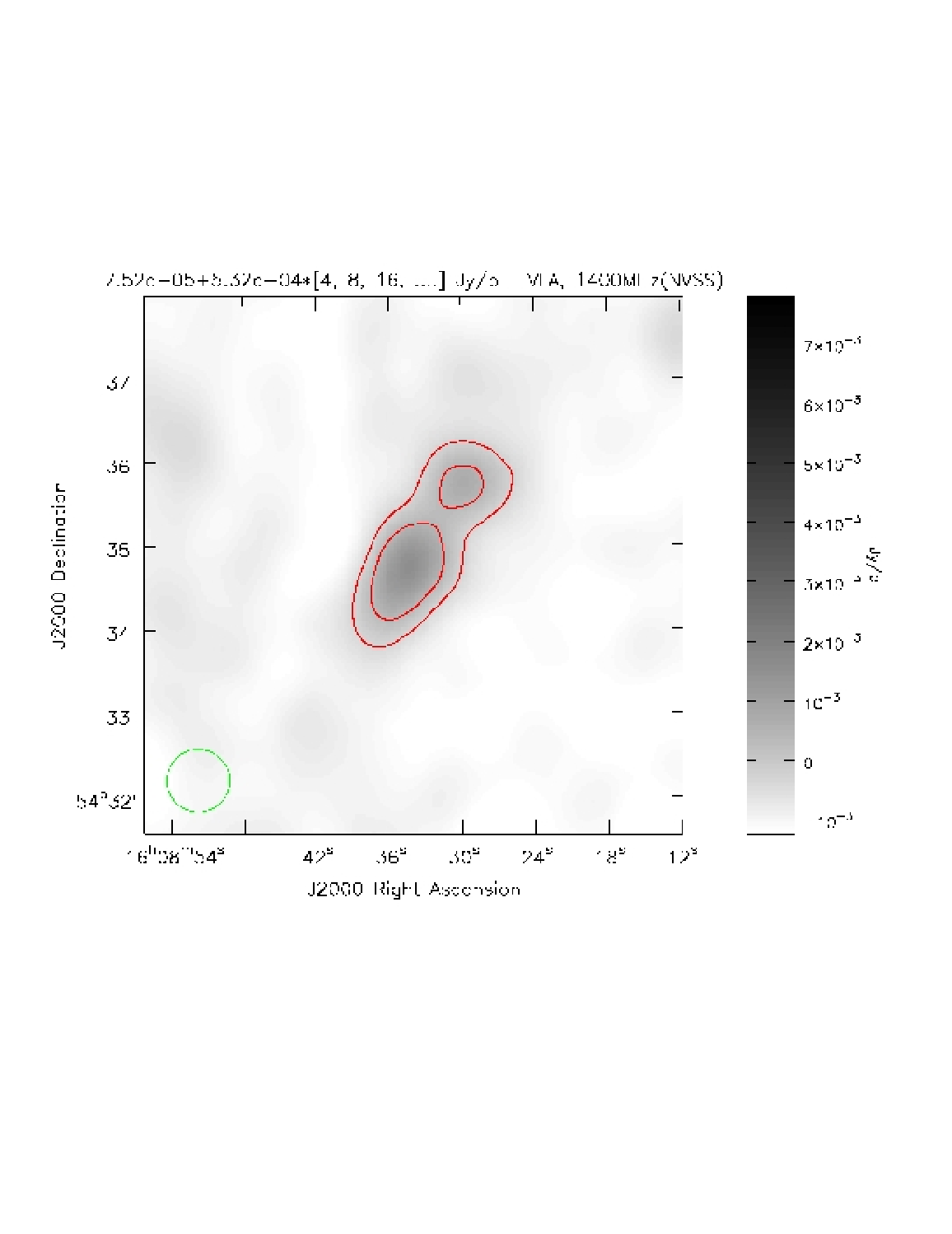}
\caption{GMRT image at 325 MHz of a source with a bridge of emission (top), and
the corresponding images of the same region at 610 MHz with the GMRT from G2008, 
and from FIRST and NVSS at 1400 MHz in descending order.}
\label{ef0:fig:diffemm_3}
\end{center}
\end{figure}

\begin{figure}
\begin{center}
   \includegraphics[angle=0, totalheight=2.1in, viewport=19 212 573 627, clip]{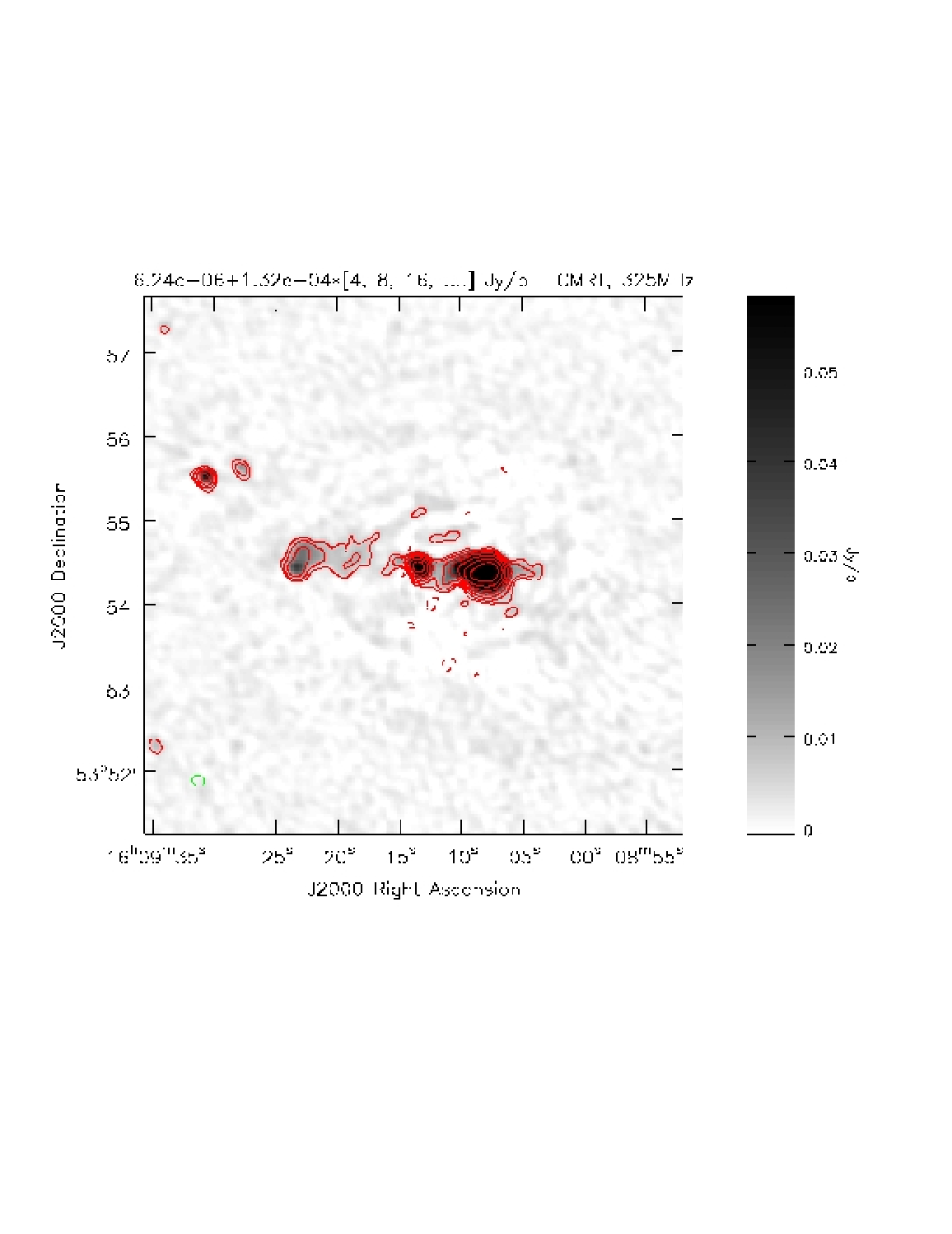}
   \includegraphics[angle=0, totalheight=2.1in, viewport=19 212 573 627, clip]{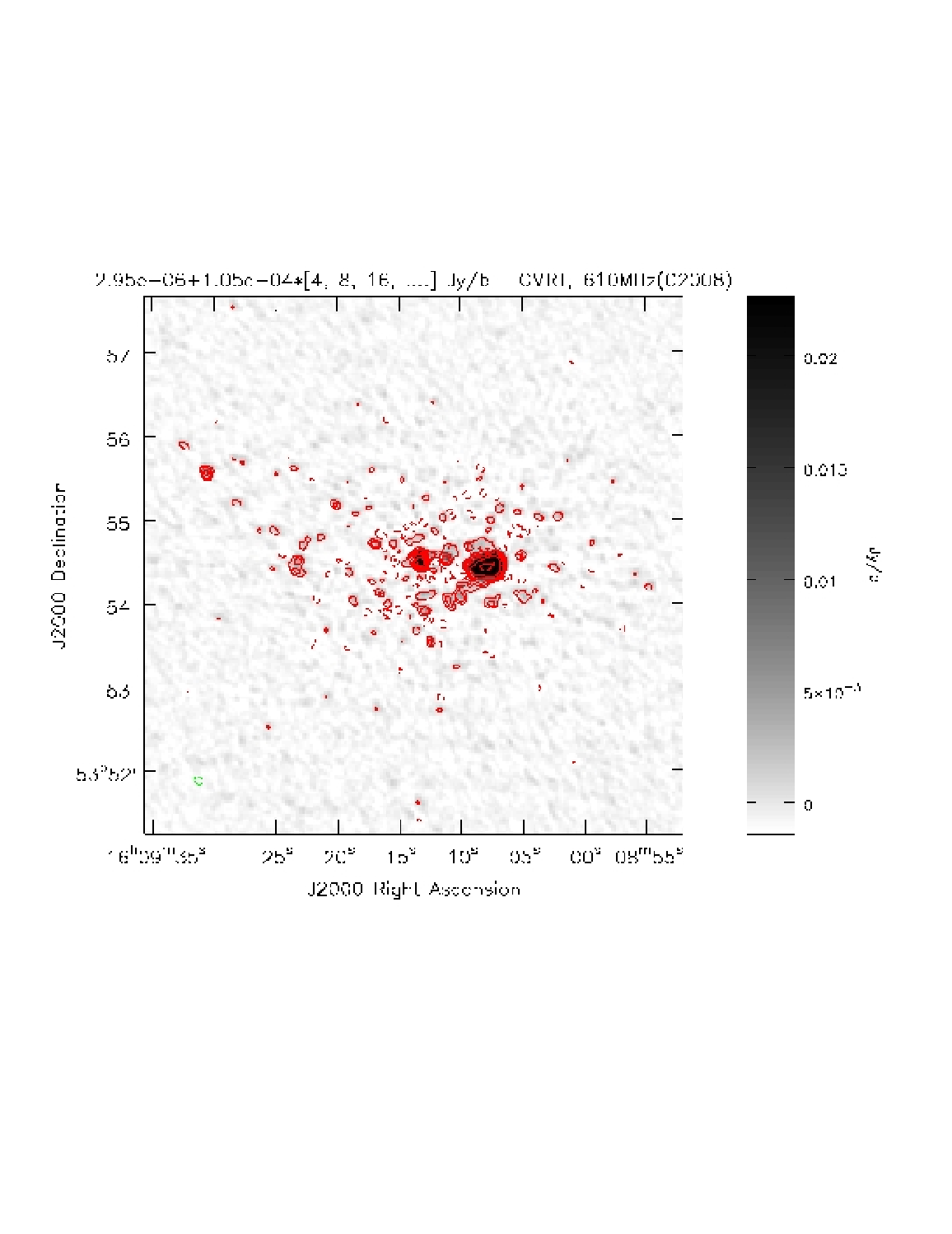}
   \includegraphics[angle=0, totalheight=2.1in, viewport=19 212 573 627, clip]{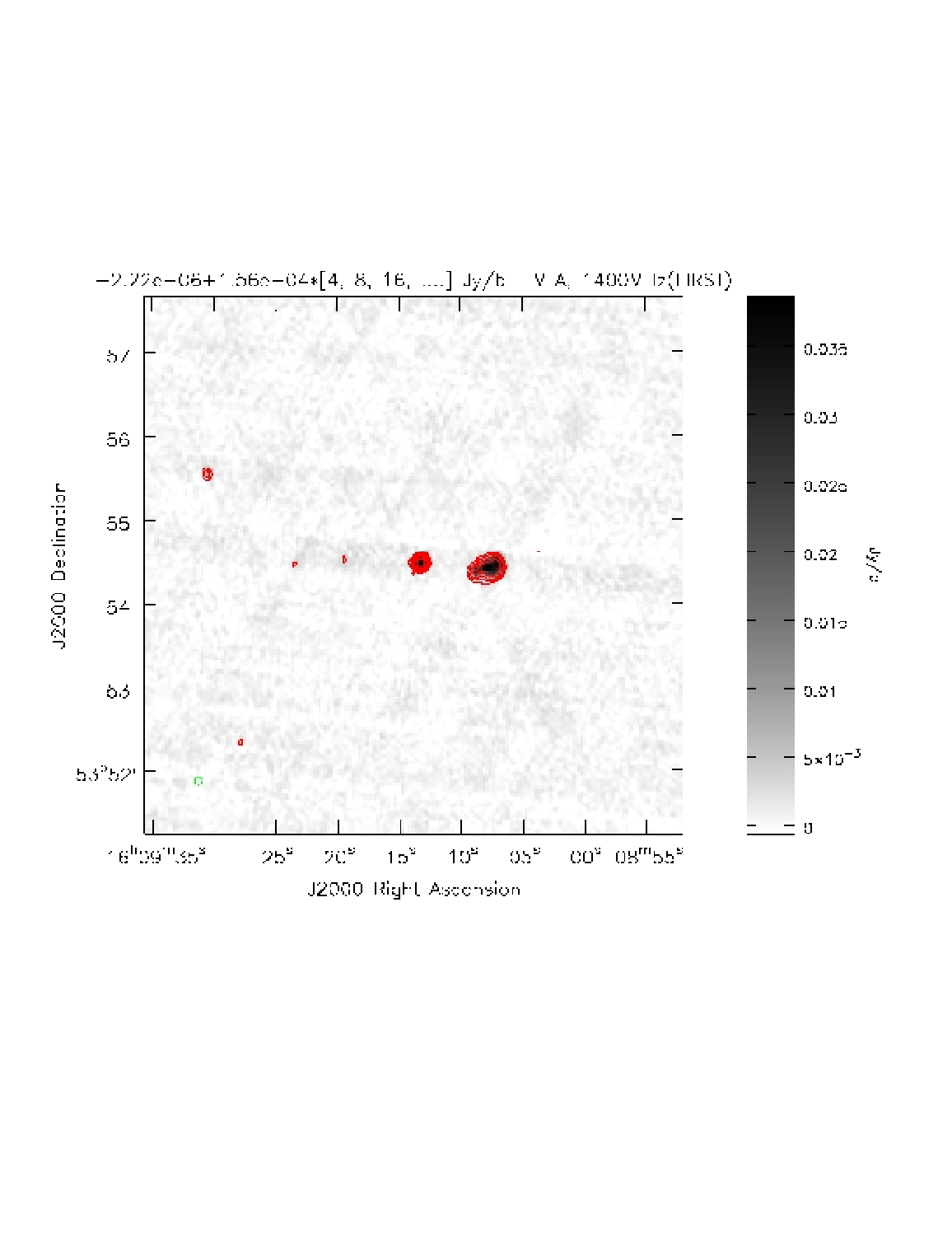}
   \includegraphics[angle=0, totalheight=2.1in, viewport=19 212 573 627, clip]{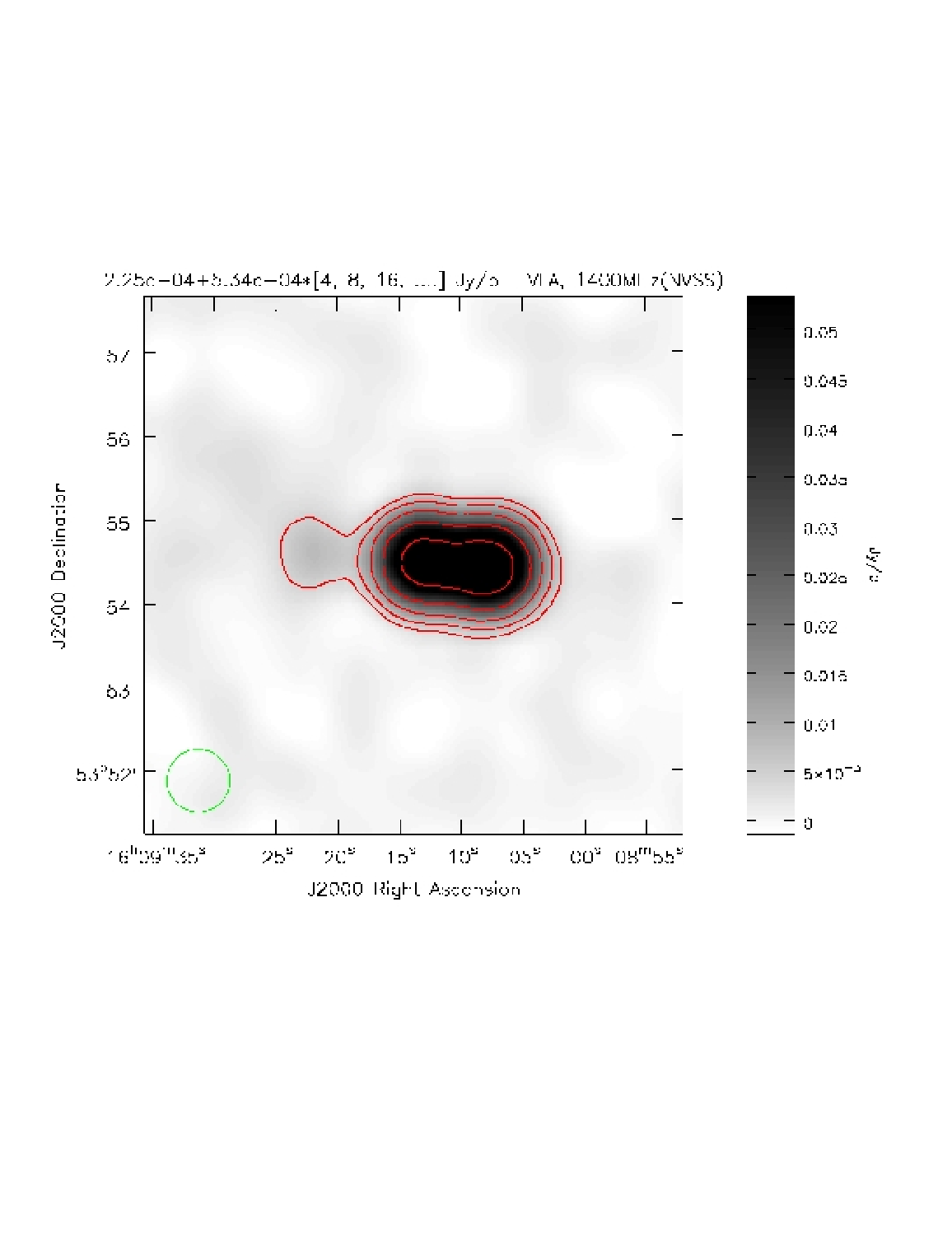}
\caption{GMRT image at 325 MHz of a highly asymmetric double-lobed source with a possible core (top), and 
the corresponding images of the same region at 610 MHz with the GMRT from G2008, 
and from FIRST and NVSS at 1400 MHz in descending order.}
\label{ef0:fig:diffemm_4}
\end{center}
\end{figure}

Figures~\ref{ef0:fig:dynrng_1},~\ref{ef0:fig:diffemm_2},~\ref{ef0:fig:diffemm_3}~and~\ref{ef0:fig:diffemm_4} show
selected images at 325 MHz of extended sources, alongwith their GMRT 610 MHz, FIRST and NVSS images.
Figure~\ref{ef0:fig:dynrng_1} shows the image of the source with a high peak brightness of 1.12~Jy~beam$^{-1}$
at 325 MHz. Although artefacts from the sidelobes are still visible in the 325-MHz
image, these are less than in the 610-MHz one, which has been made from  a number of `snap-shots'
with the GMRT (G2008). Figures~\ref{ef0:fig:diffemm_2},~\ref{ef0:fig:diffemm_3}~and~\ref{ef0:fig:diffemm_4}
show that diffuse extended emission from the lobes as well as bridges of emission are better
represented in the 325-MHz images compared with those at the higher frequencies.

\section{DISCUSSIONS}
\label{ef0:sec:discussions}

\subsection{Spectral Index}
\label{ef0:sec:spi}
\subsubsection{Source matching methodology}
The spectral indices for the sources which are listed
in our catalogue and which also appear either in the catalogues of 
G2008 or FIRST have been estimated. Ideally the resolutions of the
observations at the different frequencies should be identical. We have
chosen the two catalogues whose resolutions are within a factor of 2 of
our observations. For spectral index comparisons,
we presently restrict our sample to those sources within 0.7 deg of the phase
centre, the half power point of the primary beam. For a distance of 0.7 deg
from the phase centre the number of sources in our sample is 844 which 
is approximately 65 per cent of the total number of sources we list. 
The corresponding numbers in 
the G2008 and FIRST catalogues within this distance are 553 and 134 
respectively. 

To identify the sources in G2008 and FIRST which correspond to the
sources in our sample, 
we compared the positions of the unresolved sources in the
different catalogues and determined the mean shifts in both right
ascension and declination relative to one another. The mean shift between 
325$-$610 MHz is \mbox{RA: 3.27$\pm$3.60} arcsec and 
\mbox{DEC: $-$2.79$\pm$1.74} arcsec and between 325 MHz-FIRST is \mbox{RA: 3.11$\pm$1.65} arcsec 
and \mbox{DEC: $-$3.00$\pm$1.03} arcsec. After removing the mean shift, a source
is considered to be a match if it lies within 
7.5 arcsec of the centroid of emission for point sources and within
the sum of half the largest angular size and 7.5 arcsec for an extended
source. The value of 7.5 arcsec is approximately the sum of half the 
half power synthesized beamwidths of the observations at the two frequencies.
Since the components of a double or triple source are often listed as two or
more different components when no bridge emission is detected at 610 and 1400 MHz, 
the total flux densities at these frequencies have been estimated by summing the 
flux densities within the emission areas at 325 MHz for estimating the spectral indices.  
With these criteria 334 sources seen at 325 MHz have matches in the 610-MHz 
catalogue, while 102 have matches in the 1400-MHz FIRST catalogue. 
We have checked the number of matches by increasing 7.5 arcsec to 15 arcsec
and find that the number of matches remains the same. Since the observations at
610 and 1400 MHz sometimes do not detect the bridge emission and list the 
components as different sources, the number of matches for these catalogues
naturally increases. For instance, 360 of the 553 sources at 610 MHz (G2008) have a 
match at 325 MHz, while 126 of the 134 sources at 1400 MHz have a match at 325 MHz.
Sources may not have a 
match in the catalogues due to a combination of both sensitivity and spectral 
indices of the sources, as seen in our recent work on a deep survey of the sky
towards a couple of clusters of galaxies at 153, 244, 610 and 1260 MHz 
\citep{2008arXiv0809.4565S}. 
 
\subsubsection{Very steep-spectrum and GPS sources} 
\begin{figure}
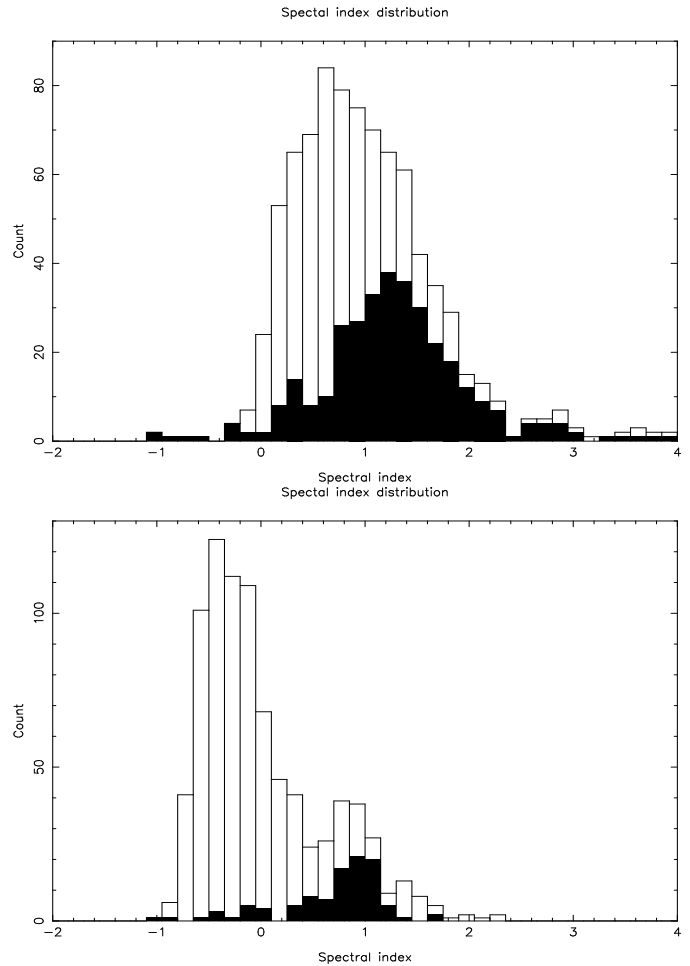

\begin{center}
\includegraphics[angle=270, totalheight=2.5in, width=3.5in]{./spdis_nu325_610_limit.ps}
\includegraphics[angle=270, totalheight=2.5in, width=3.5in]{./spdis_nu325_1400_limit.ps}
\caption{The spectral index distributions between 325 and 610 MHz (top panel) and between
325 and 1400 MHz (lower panel). The sources with estimated values are shown shaded while 
those with lower limits to the spectral indices are shown unshaded. }
\label{ef0:fig:spdis}
\end{center}
\end{figure}

We have attempted to identify very steep spectrum radio sources
with spectral index $\alpha$ $\geq$1.3 
($\alpha$ defined as S$\propto\nu^{-\alpha}$), 
as well as GPS (Giga-Hertz Peaked Spectrum) candidates which 
are either weak or not detected at 325 MHz and have $\alpha\leq -0.5$. 
The distributions of the spectral indices between the different frequencies
are presented in Fig.~\ref{ef0:fig:spdis}. The distribution of spectral
indices between 610 and 1400 MHz has been published by G2008 and is not
reproduced here.  The median values of $\alpha_{325}^{610}$ and 
$\alpha_{325}^{1400}$ are 1.28 and 0.83 
respectively. The distribution of $\alpha_{610}^{1400}$ 
has a broad peak, with the median value of 0.51  
being smaller due to a significant population of flat-spectrum objects (G2008). 
The median value of $\alpha_{325}^{610}$ appears to be on the higher side, and
this could be due to more diffuse emission being detected at the lower
frequency image. 

There are four candidate GPS sources which satsify the above criteria.
These are J1606+5427, J1613+5433, J1613+5422 and J1614+5437.
J1606+5427 has a flux density of 2.96 mJy at 1400 MHz in the FIRST catalogue,
while its total flux densities at 325 and 610 MHz are 0.68 and 1.3 mJy
respectively. The spectral index between 325 and 1400 MHz is $-$1.01.
J1613+5433, J1613+5422 and J1614+5437 have flux densities of 1.83,
3.49 and 4.54 mJy respectively in the FIRST catalog but have not been
detected at either 325 MHz or at 610 MHz by G2008. These sources are also
not visible in NVSS and require further confirmation. 

\begin{figure*}
\begin{center}
\vbox{
  \hbox{
   \includegraphics[angle=0, totalheight=1.8in, viewport=19 212 573 627, clip]{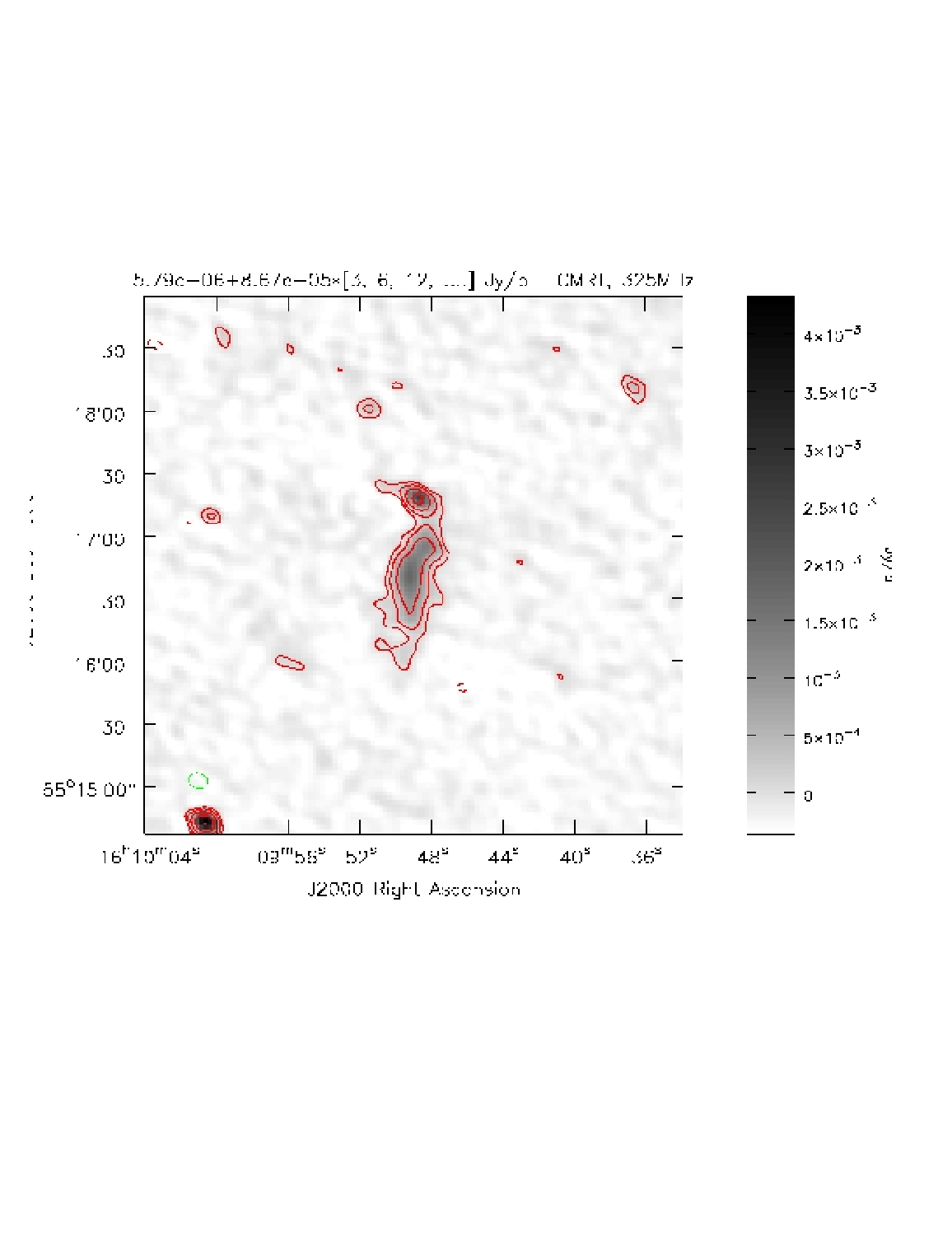}
   \includegraphics[angle=0, totalheight=1.8in, viewport=19 212 573 627, clip]{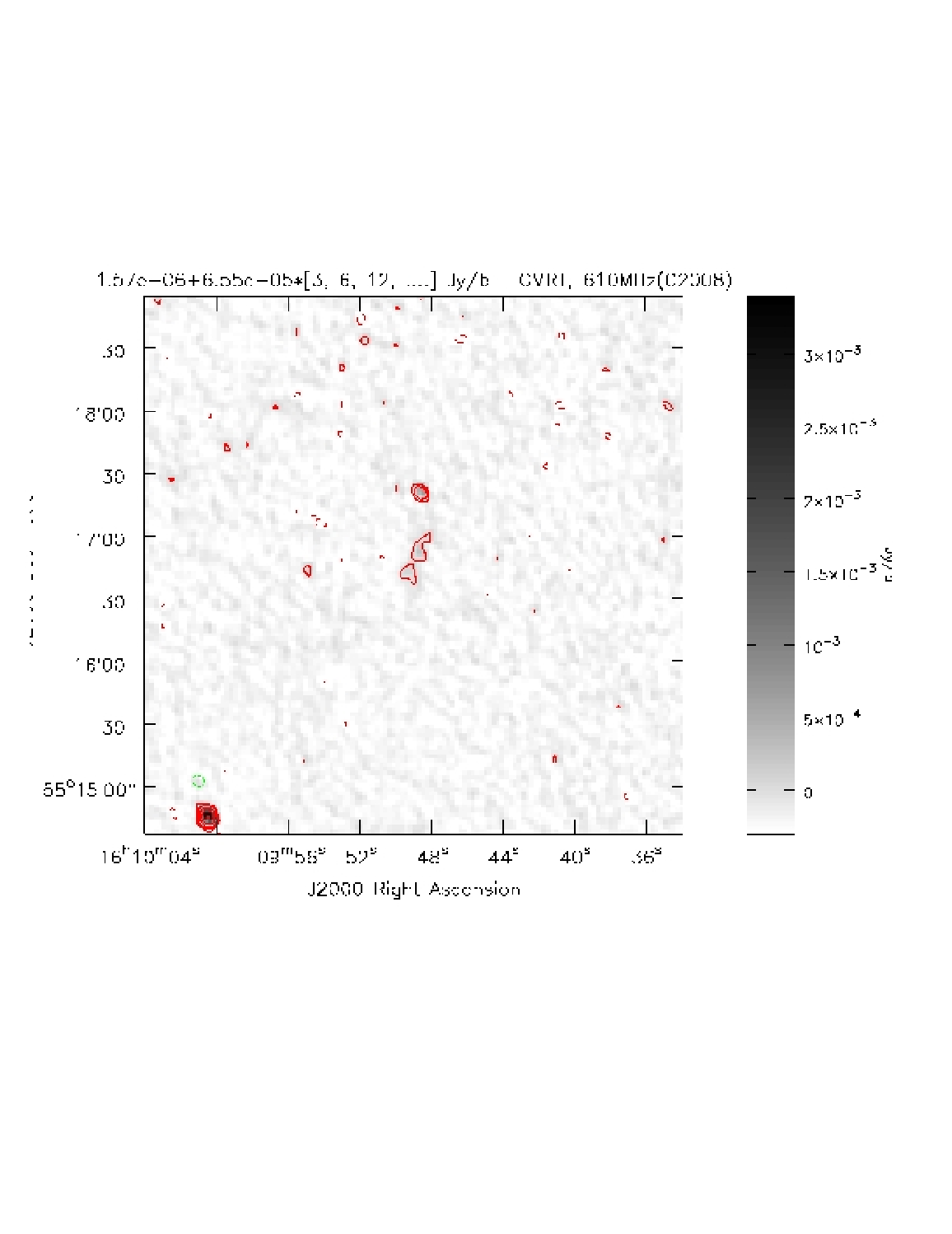}
   \includegraphics[angle=0, totalheight=1.8in, viewport=19 212 573 627, clip]{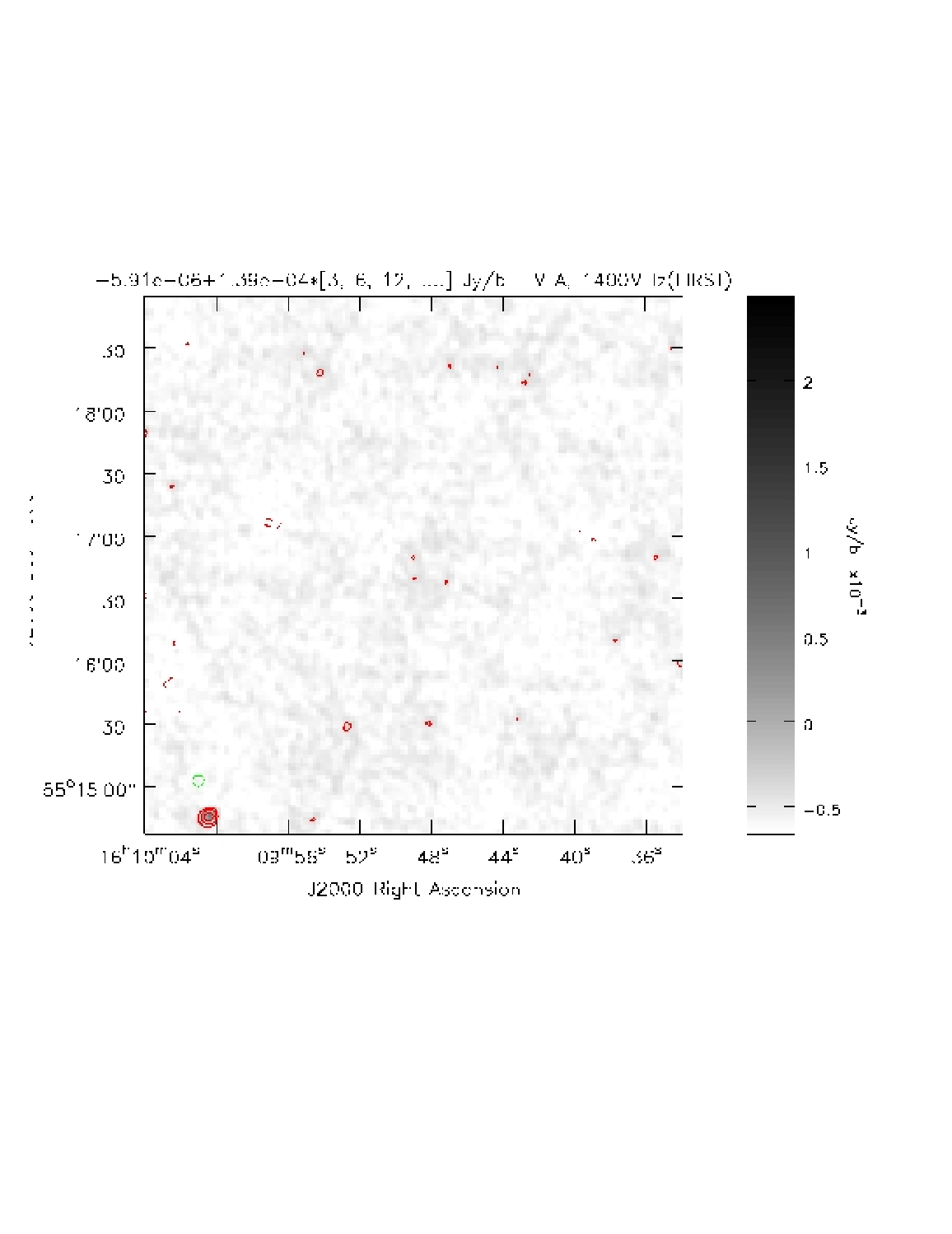}
  }
  \hbox{
   \includegraphics[angle=0, totalheight=1.8in, viewport=19 212 573 627, clip]{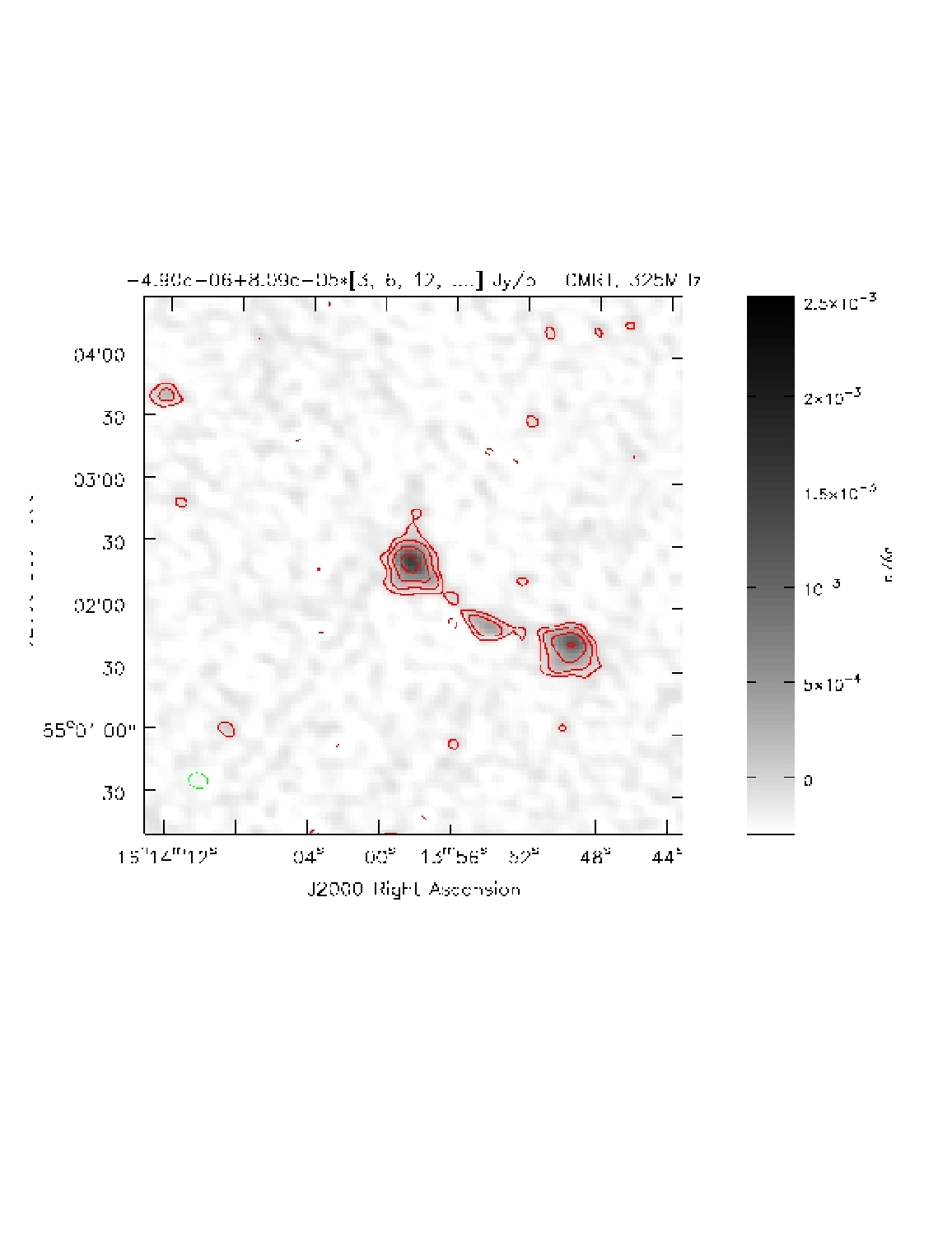}
   \includegraphics[angle=0, totalheight=1.8in, viewport=19 212 573 627, clip]{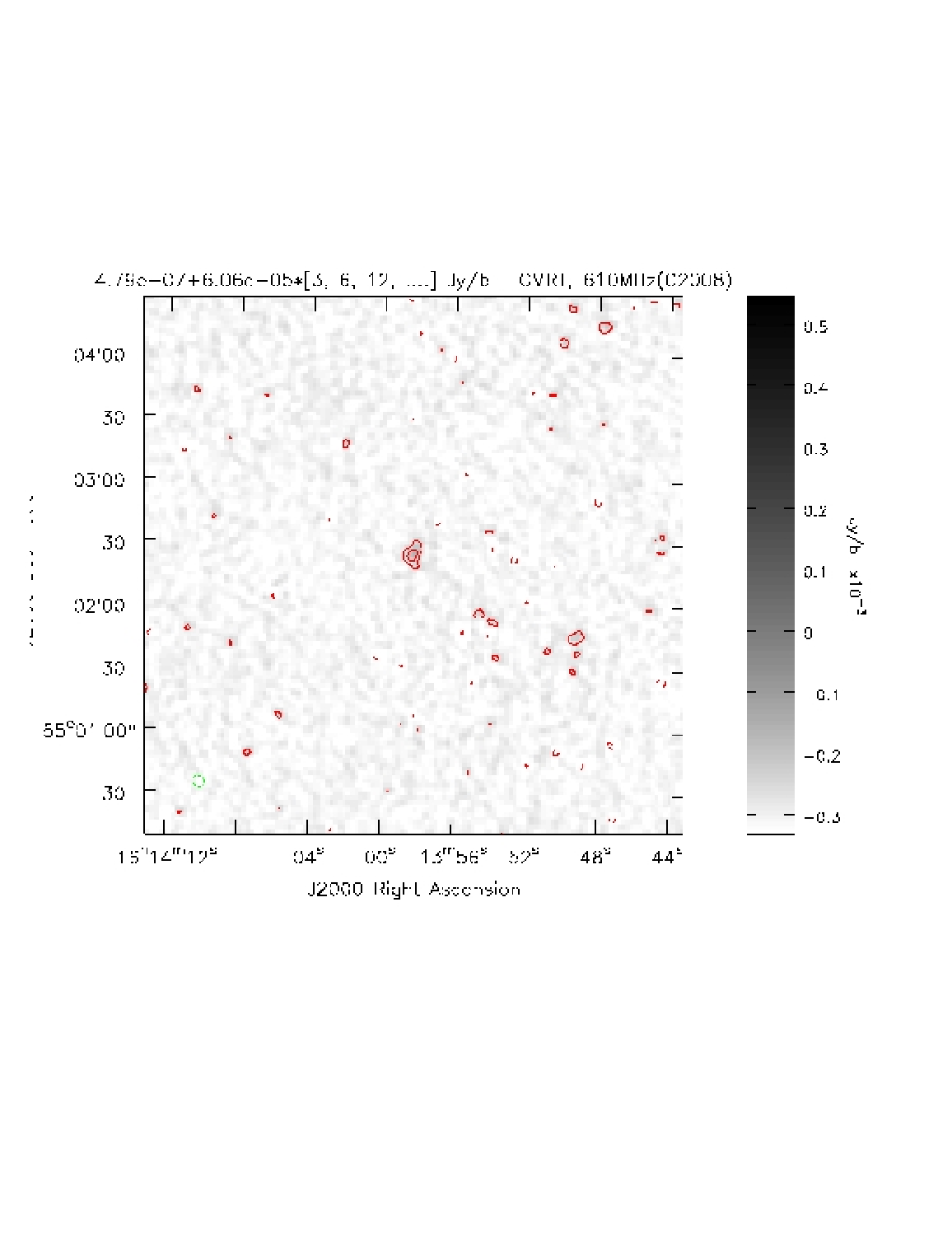}
   \includegraphics[angle=0, totalheight=1.8in, viewport=19 212 573 627, clip]{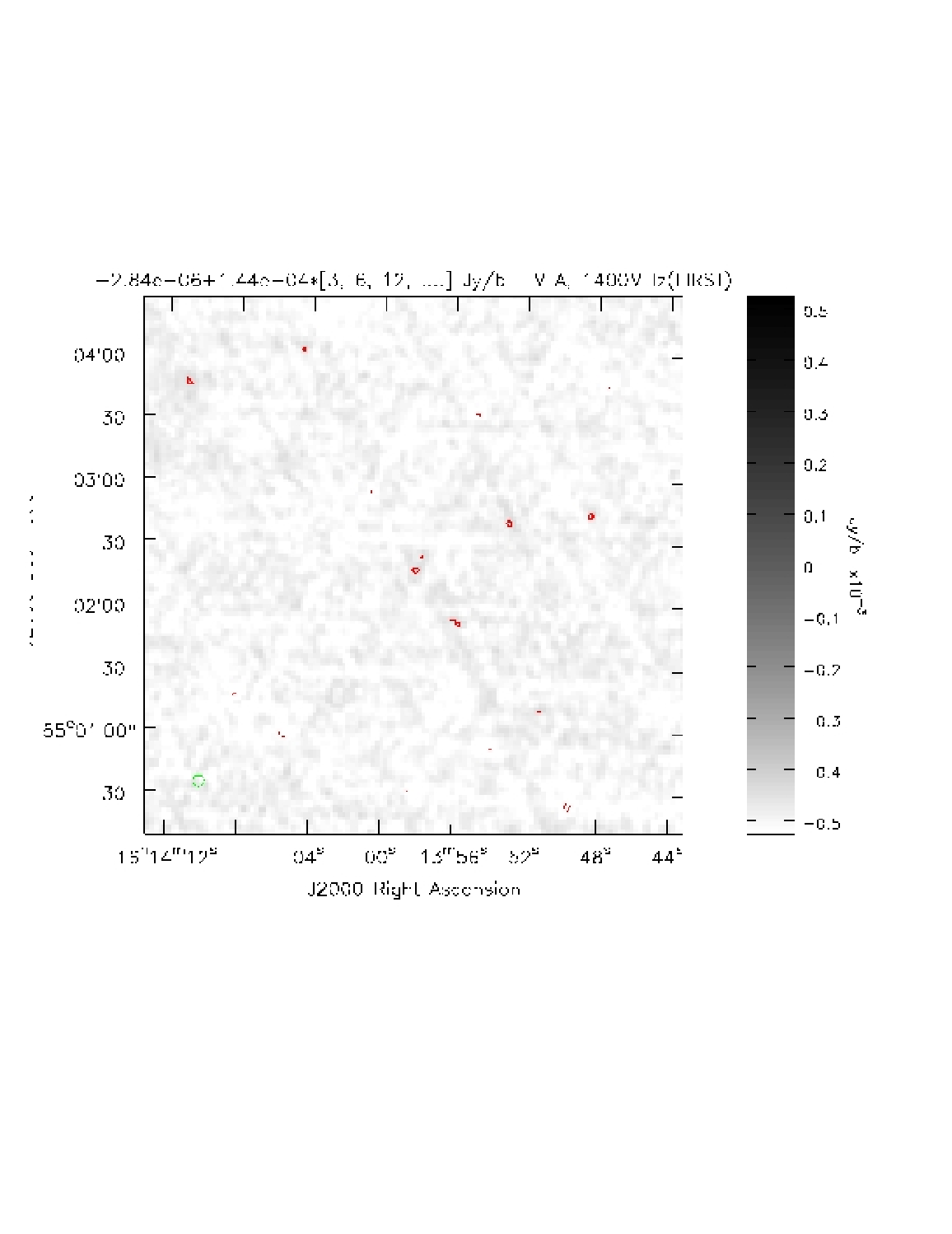}
  }
}
\caption{Images of two of the candidate very steep spectrum sources, GMRT160948+551649 (upper row) and
GMRT161357+550213 (lower row) at 325 (left), 610 (middle, G2008) and 1400 MHz (right, FIRST). } 
\label{ef0:fig:ss_limit}
\end{center}
\end{figure*}

The candidate very steep spectrum sources are listed in Table~\ref{ef0:table:ss_src}.
Images of two of the sources from our observations
and the corresponding regions from G2008 and FIRST are shown in 
Fig.~\ref{ef0:fig:ss_limit}. Although the sources are visible in G2008, 
not all the extended emission seen at 325 MHz is visible in the 610-MHz image.
Also, there is no significant emission seen in the FIRST images. 
Since these candidates have been identified from 
observations which differ in resolution by a factor of two, it is important
to confirm their steep spectral indices from observations with the same
resolution at the different frequencies. 

\begin{table*}
\caption{Candidate very steep spectrum sources}
\label{ef0:table:ss_src}
\begin{tabular}{l r r r r r r r r r}
\hline

\multicolumn{1}{c}{Source name} & \multicolumn{1}{c}{RA} & \multicolumn{1}{c}{DEC} & \multicolumn{1}{c}{Dist} & \multicolumn{1}{c}{$\sigma_{rms}$} & \multicolumn{1}{c}{S$_{peak}$} & \multicolumn{1}{c}{S$_{total~3\sigma}$} & \multicolumn{1}{c}{Size} & \multicolumn{1}{c}{Notes} \\ 
                    & \multicolumn{1}{c}{hh:mm:ss.s} & \multicolumn{1}{c}{dd:mm:ss.s} & \multicolumn{1}{c}{deg} & \multicolumn{1}{c}{$\mu$Jy b$^{-1}$}  & \multicolumn{1}{c}{mJy b$^{-1}$} &  \multicolumn{1}{c}{mJy}   & \multicolumn{1}{c}{$^{\prime\prime}$} &        \\ 
\multicolumn{1}{c}{(1)} & \multicolumn{1}{c}{(2)} & \multicolumn{1}{c}{(3)} & \multicolumn{1}{c}{(4)} &  \multicolumn{1}{c}{(5)} & \multicolumn{1}{c}{(6)} & \multicolumn{1}{c}{(7)} & \multicolumn{1}{c}{(8)} & \multicolumn{1}{c}{(9)} & \\
\hline
GMRT160551+543842  & 16:05:51.9 & 54:38:42.7 &     0.60 &      478 &     4.96 &    11.57 &     12.4 & $a$ \\
GMRT160613+545322  & 16:06:13.3 & 54:53:22.6 &     0.59 &      398 &     4.45 &    14.50 &     14.8 & $a$ \\
GMRT160613+550149  & 16:06:13.6 & 55:01:49.1 &     0.65 &      200 &     2.70 &     7.90 &     22.4 \\
GMRT160621+545636  & 16:06:21.2 & 54:56:36.1 &     0.59 &      202 &     4.10 &    22.44 &     48.5 \\
GMRT160626+544906  & 16:06:26.3 & 54:49:06.1 &     0.54 &      178 &     2.27 &     6.09 &     15.0 & $b$ \\
GMRT160636+543247  & 16:06:36.3 & 54:32:47.6 &     0.51 &      137 &    17.34 &    45.50 &     18.0 \\
GMRT160725+544720  & 16:07:25.7 & 54:47:20.6 &     0.39 &       91 &     1.83 &     5.90 &     41.1 \\
GMRT160732+545106  & 16:07:32.0 & 54:51:06.3 &     0.40 &       76 &     2.87 &     6.85 &     29.9 \\
GMRT160733+545639  & 16:07:33.4 & 54:56:39.3 &     0.45 &       94 &     1.24 &    13.70 &     58.9 \\
GMRT160746+541159  & 16:07:46.1 & 54:11:59.9 &     0.57 &       71 &     4.93 &     7.38 &     11.5 \\
GMRT160757+551013  & 16:07:57.6 & 55:10:13.0 &     0.58 &       82 &     2.85 &     8.33 &     15.2 \\
GMRT160800+542144  & 16:08:00.9 & 54:21:44.8 &     0.42 &       63 &     3.54 &     9.02 &     34.9 & $c$ \\
GMRT160801+542025  & 16:08:01.8 & 54:20:25.2 &     0.43 &       76 &     8.51 &    25.69 &     23.4 \\
GMRT160825+551755  & 16:08:25.5 & 55:17:55.4 &     0.67 &       83 &     2.74 &     7.54 &     26.8 \\
GMRT160850+545348  & 16:08:50.4 & 54:53:48.1 &     0.28 &       58 &     2.04 &     6.21 &     15.9 \\
GMRT160931+541915  & 16:09:31.5 & 54:19:15.4 &     0.35 &       58 &     3.07 &     5.93 &      9.3 \\
GMRT160932+542028  & 16:09:32.7 & 54:20:28.5 &     0.33 &       63 &     1.27 &     6.65 &     33.3 & $d$ \\
GMRT160948+551646  & 16:09:48.9 & 55:16:46.9 &     0.61 &      109 &     2.67 &    17.40 &     57.4 \\
GMRT160949+540826  & 16:09:49.6 & 54:08:26.2 &     0.53 &       70 &     4.87 &     8.23 &     21.3 \\
GMRT161008+540753  & 16:10:08.7 & 54:07:53.4 &     0.54 &       66 &     4.14 &    16.77 &     38.7 \\
GMRT161013+544924  & 16:10:13.4 & 54:49:24.9 &     0.16 &       55 &     1.65 &    10.09 &     93.5 \\
GMRT161016+540143  & 16:10:16.8 & 54:01:43.1 &     0.64 &       70 &     4.11 &     5.50 &     11.7 \\
GMRT161030+540249  & 16:10:30.7 & 54:02:49.4 &     0.62 &       78 &     1.80 &     7.17 &     39.4 \\
GMRT161040+540626  & 16:10:40.6 & 54:06:26.6 &     0.57 &       65 &     4.74 &    12.69 &     28.7 \\
GMRT161040+550824  & 16:10:40.7 & 55:08:24.8 &     0.48 &       76 &     2.18 &     6.10 &     35.1 \\
GMRT161230+545105  & 16:12:30.5 & 54:51:05.3 &     0.41 &       62 &     3.98 &     7.99 &     10.2 \\
GMRT161319+541033  & 16:13:19.7 & 54:10:33.1 &     0.69 &       91 &     2.85 &     6.99 &     29.1 \\
GMRT161349+550140  & 16:13:49.5 & 55:01:40.2 &     0.66 &       86 &     2.04 &     6.41 &     15.1 \\
GMRT161358+550219  & 16:13:58.2 & 55:02:19.9 &     0.68 &       94 &     2.53 &     8.06 &     15.0 & $e$ \\
GMRT161424+544314  & 16:14:24.4 & 54:43:14.4 &     0.64 &      101 &     3.23 &     6.06 &      9.1 \\
\hline \hline
\end{tabular}

Notes: $a$~strong source nearby; $b$~diffuse extended emission visible in the NVSS image; $c$~double-lobed source;   
$d$~northern component of an asymmetric double; $e$~north-western component of a triple lobed source.
\end{table*}

\subsubsection{Spectral index$-$flux density relationships}
\begin{figure}
\begin{center}
\includegraphics[angle=270, totalheight=2.5in, width=3.5in]{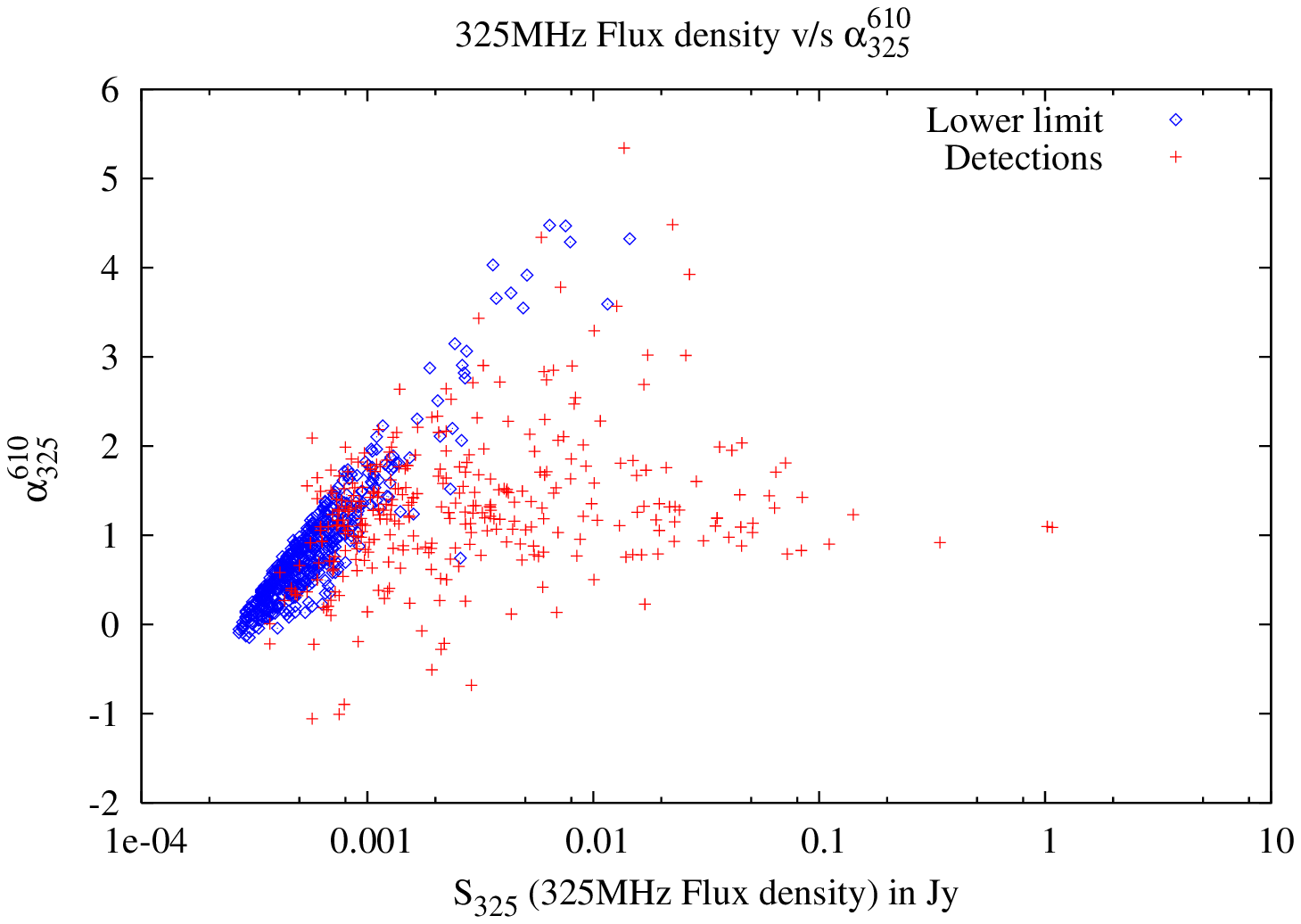}
\includegraphics[angle=270, totalheight=2.5in, width=3.5in]{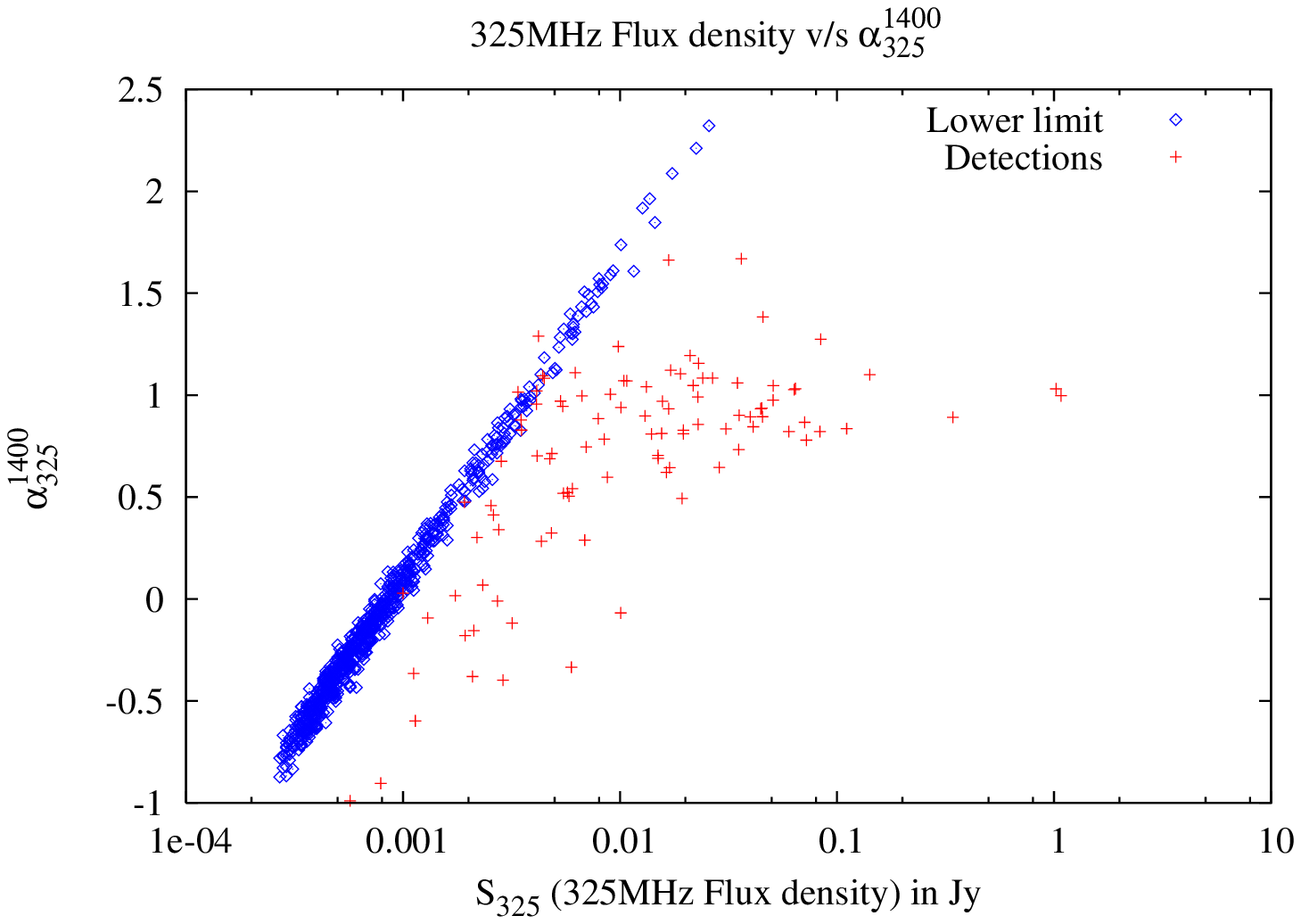}
\caption{The spectral indices as a function of flux density. The sources with lower limits 
are shown by open diamonds, while those with measured values are shown by the $+$ sign.}
\label{ef0:fig:flux_sp}
\end{center}
\end{figure}

In Fig.~\ref{ef0:fig:flux_sp} we present the plots of  $\alpha_{325}^{610}$,
and $\alpha_{325}^{1400}$ against the flux
density at 325 MHz. 
Any dependence of spectral index on flux density can be used to 
study the different populations of sources which may be present at
different flux density levels. It is well known that the normalized
source counts tend to flatten at low flux densities corresponding to
$\leq$0.5 mJy at 1400 MHz and 1 mJy at 610 MHz (\citeauthor{2007ASPC..380..205P}~\citeyear{2007ASPC..380..205P};
G2008 and references therein). At high flux densities the
radio source counts are due to double-lobed radio galaxies and quasars
with spectral indices ranging from $\sim$0.6 to 1.5 (cf. \citeauthor{1980MNRAS.190..903L}~\citeyear{1980MNRAS.190..903L}). The spectral indices of these sources are known to 
be correlated with radio luminosity and/or redshift, leading to
several searches for high-redshift galaxies amongst the very steep
spectrum radio sources 
(e.g.  \citeauthor{1990AJ....100.1014M}~\citeyear{1990AJ....100.1014M};
\citeauthor{2000A+AS..143..303D}~\citeyear{2000A+AS..143..303D}). The flattening of the source counts 
at low flux densities is generally ascribed to a population of 
starburst galaxies, low luminosity AGN and radio quiet QSOs 
\citep[e.g.][]{2004MNRAS.352..131S, 2006MNRAS.372..741S, 2007ASPC..380..205P, 2008ApJS..177...14S}. While
starburst galaxies are expected to have a radio spectral index in
the range of $\sim$0.5 to 0.8 at low radio frequencies which are
dominated by non-thermal emission, the nuclear components of low
luminosity AGN could have spectral indices $\leq$0.5. In a study 
of the VLA Deep Field, \citet{2007A+A...463..519B} find the median spectral
index to be 0.46$\pm$0.03 for sources with a flux density 0.15$\leq$S$<$0.50
mJy at 1400 MHz while those above this flux density have a median 
spectral index of 0.67$\pm$0.05. They attribute the flatter spectral
index at the lower flux density level to a population of low luminosity AGN.
Although we need to determine the spectral indices from similar-resolution data,
Fig.~\ref{ef0:fig:flux_sp} shows $\alpha_{325}^{610}$,
and $\alpha_{325}^{1400}$ against the flux density at 325 MHz. For $\alpha_{325}^{610}$,
510 sources have limits while for $\alpha_{325}^{1400}$ 742 sources have limits to their
spectral indices. Given the large fraction of sources with limits to their spectral
indices, it is difficult to determine reliably any dependence of spectral index on 
flux density. It is required to  
observe this field with deeper sensitivity at higher frequencies
to determine their spectral indices and investigate any such dependence.

\subsection{Source counts}
\begin{figure}
\begin{center}
\includegraphics[angle=270, totalheight=2.5in, width=3.5in]{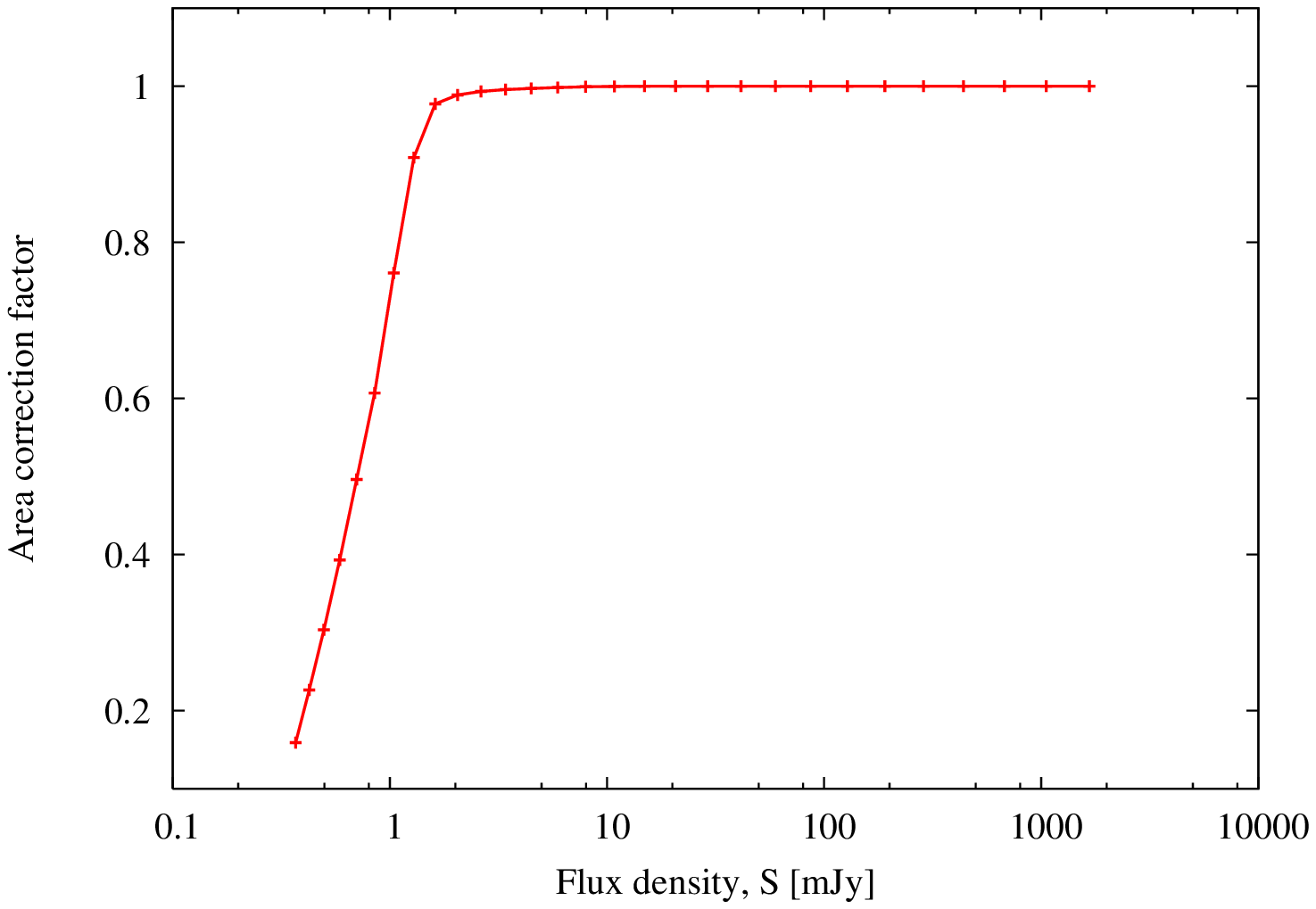}
\caption{Visibility area as a function of radio flux density}
\label{ef0:fig:gmrt_carea}
\end{center}
\end{figure}

\begin{figure}
\begin{center}
\includegraphics[angle=270, totalheight=2.5in, width=3.5in]{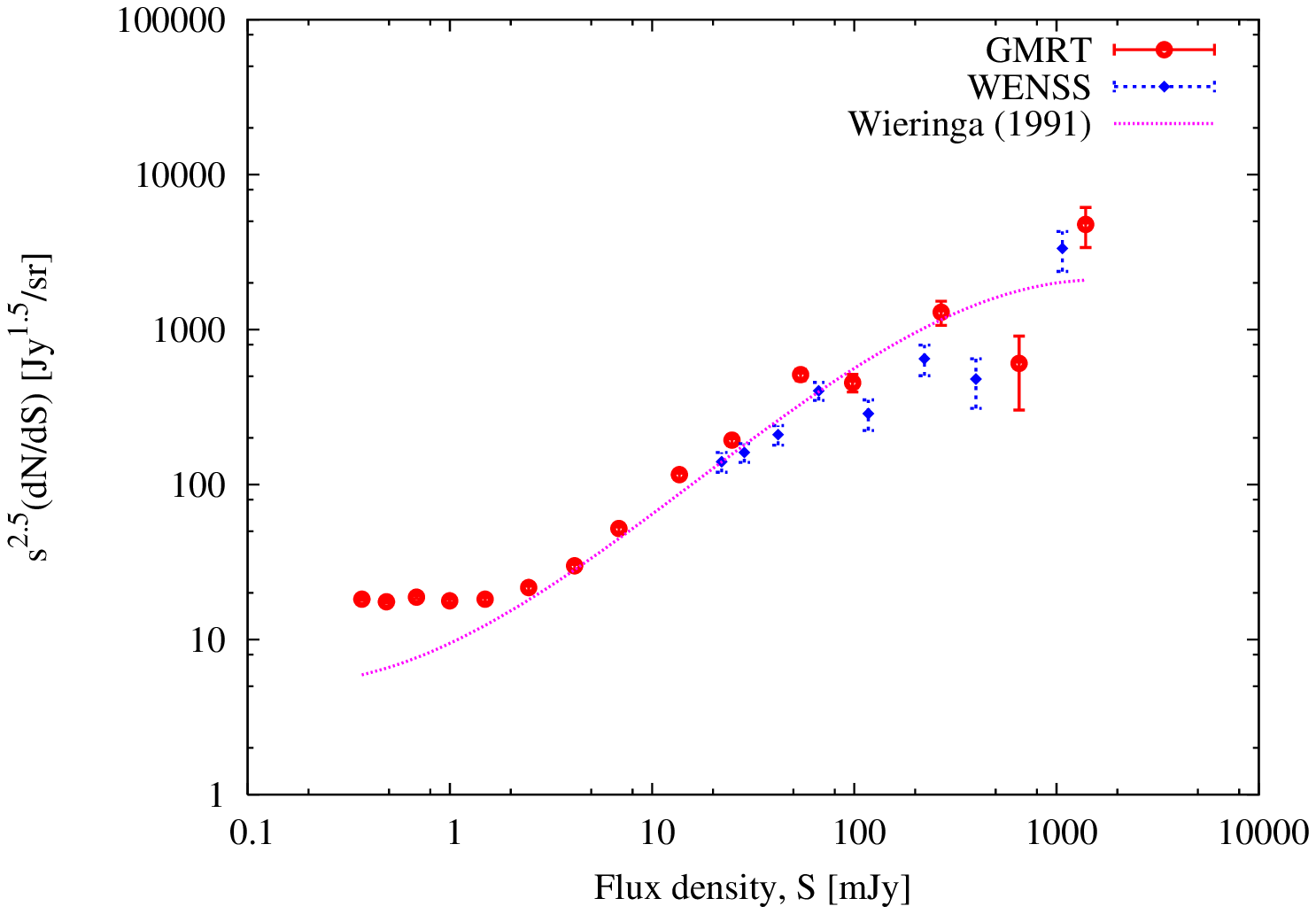}
\caption{The normalized differential source counts at 325 MHz for sources from our 
present observations (circles) and WENSS at 330 MHz (diamonds)
for all sources located within 1.1$^\circ$ of the phase centre of our 
observations. This is $\sim$18.3 per cent of the beamwidth of the GMRT primary beam at 325 MHz.
The continuous curve represents the functional form of the fit to the source counts by Wierenga (1991)
for sources in the flux density range of $\sim$4~mJy~to~1~Jy.}
\label{ef0:fig:gmrt_wenss_ns}
\end{center}
\end{figure}

We constructed the source counts at 325 MHz for our sample of sources by binning them in
different ranges of flux density starting from 315 $\mu$Jy, so that even the sources
in the lowest-flux density bin have a typical peak flux density to rms ratio of about 8. The
corrections due to source incompleteness is negligible as can also be seen in the results
from Fig. 5 of \cite{2006A+A...456..791T}.  The source counts were 
corrected for the fraction of the area (Fig.~\ref{ef0:fig:gmrt_carea}) over which the 
source could be detected because
of the increase in noise near the bright sources. The flux density bins, the average
flux density for each bin, the number of sources in each bin, the noise corrected number 
of sources, the differential source counts and the normalized source counts are listed 
in Table~\ref{ef0:table:dns}. The differential source counts, which have been estimated by dividing the noise
corrected number of sources by the area of the image in steradians and the width of the
flux density range, is shown in Fig.~\ref{ef0:fig:gmrt_wenss_ns}. We have also plotted the
differential source counts for the WENSS sources for the same area. The WENSS sources plotted
here are stronger than about 18 mJy, while the sources from our survey are much weaker,
extending to about 270 $\mu$Jy, although in present study for source counts we consider
only those above 315 $\mu$Jy. There is evidence of the source counts 
flattening at about a mJy, consistent with higher frequency studies, although more data
at low flux density levels are required at this frequency.  Since the earlier
surveys in the literature at 325 MHz are limited to higher flux density levels, the
flattening in the source counts at this frequency has not been reported earlier
\citep{1991PhDT.......241W, 2006A+A...456..791T}. 
The functional form fitted by \cite{1991PhDT.......241W} to the differential source counts ranging 
from about 4 mJy to 1 Jy from his deep 325-MHz Westerbork survey, shown in 
Fig.~\ref{ef0:fig:gmrt_wenss_ns}, also illustrates clearly the flattening of the source
counts at low flux densities. This functional form is also consistent with the 
measurements of \cite{2006A+A...456..791T} (see their Fig.~7) whose sources range from 
about 3 to 500 mJy.  A more
detailed modeling of the source counts will be presented in a subsequent paper, after 
combining with data from other fields observed with the GMRT at the same frequency.

We have integrated the number counts from our survey, yielding a surface brightness 
of $4.04\times10^{-5}$~nW~m$^{-2}$~sr$^{-1}$, which gives a lower
limit to the contribution of extragalactic sources to the 
cosmic radio background at 325 MHz \citep{2002ApJ...575....7D}.

\begin{table}
\caption{Differential source counts at 325 MHz}
\label{ef0:table:dns}
\begin{tabular}{r @{-} r r r r r @{$\pm$} l}
\hline
\multicolumn{2}{c}{Flux bin} & \multicolumn{1}{c}{$\langle$S$\rangle$} & \multicolumn{1}{c}{N} & \multicolumn{1}{c}{N$_c$} & \multicolumn{2}{c}{$\frac{dN}{dS}\langle$S$\rangle^{2.5}$}\\
\multicolumn{2}{c}{mJy} & \multicolumn{1}{c}{mJy} & \multicolumn{1}{c}{} & \multicolumn{1}{c}{} & \multicolumn{2}{c}{sr$^{-1}$Jy$^{1.5}$}\\
\hline
  0.315 &    0.413 &     0.367  & 120  & 800.4 &   18.25 &    0.65  \\
  0.413 &    0.566 &     0.485  & 174  & 601.4 &   17.55 &    0.72  \\
  0.566 &    0.813 &     0.684  & 209  & 438.8 &   18.80 &    0.90  \\
  0.813 &    1.221 &     0.998  & 194  & 267.0 &   17.78 &    1.09  \\
  1.221 &    1.917 &     1.495  & 165  & 170.1 &   18.22 &    1.40  \\
  1.917 &    3.151 &     2.453  & 103  & 103.8 &   21.65 &    2.13  \\
  3.151 &    5.416 &     4.139  &  71  &  71.2 &   29.93 &    3.55  \\
  5.416 &    9.742 &     6.851  &  67  &  67.1 &   52.03 &    6.35  \\
  9.742 &    18.33 &     13.64  &  53  &  53.0 &   115.9 &   15.92  \\
  18.33 &    36.08 &     24.83  &  41  &  41.0 &   193.8 &   30.27  \\
  36.08 &    74.32 &     54.31  &  33  &  33.0 &   512.3 &   89.19  \\
  74.32 &    160.1 &     98.06  &  15  &  15.0 &   454.5 &   117.4  \\
  160.1 &    360.9 &     269.3  &   8  &   8.0 &   1294  &   457.6  \\  
  360.9 &    851.2 &     652.4  &   1  &   1.0 &   605.5 &   605.5  \\
  851.2 &    2100  &     1395   &   3  &   3.0 &   4764  &   2751   \\
\hline 
\end{tabular}
\end{table}

\section{Concluding remarks}
We have presented deep observations of the ELAIS-N1 field at 325 MHz using the GMRT. 
This is part of an ongoing program to examine the source counts and the nature of 
radio sources identified from sensitive observations at low radio frequencies 
with the objective of identifying AGN and starburst galaxies and examining their evolution
with cosmic epoch. For this data set we have achieved a median rms noise of  
$\approx 40~\mu$Jy beam$^{-1}$ towards the phase centre by combining data from two different days,
which is about the lowest that has been achieved at this frequency.  We detect 1286 sources
with a total flux density above $\approx270 \mu$Jy within a radius of 1.1$^\circ$ of the
phase centre. By comparing our results with those of G2008 and the FIRST survey, whose
resolutions are within a factor of two we have identified candidate very steep spectrum 
sources with a $\alpha\geq$1.3 and candidate GPS objects for further investigations.
Considering only those sources with a flux density $>$315~$\mu$Jy, so that any effects of
incompleteness are negligible,
our data show evidence of a flattening of the source counts at low flux densities at 325 MHz, 
which has been reported earlier at higher frequencies and attributed to both starburst 
galaxies and low luminosity AGN. This needs to be investigated further by combining it with
sensitive observations from other fields as well at this frequency.

\section*{Acknowledgments}
We thank an anonymous referee and the staff of the GMRT who have made these observations possible. 
GMRT is run by the National Centre for Radio Astrophysics of the Tata Institute of Fundamental Research. 
This research has made use of the NASA/IPAC Extragalactic Database (NED) which is operated by 
the Jet Propulsion Laboratory, California Institute of Technology, under contract with the National 
Aeronautics and Space Administration.

\bibliography{../bib/ref.bib} 

\begin{thebibliography}{}

\bibitem[\protect\citeauthoryear{{Adelman-McCarthy}, {et al.,}, {} \&
  {}}{{Adelman-McCarthy} et~al.}{2008}]{2008ApJS..175..297A}
{Adelman-McCarthy} J.~K.,  {et al.,} {}   {} 2008, \apjs, 175, 297

\bibitem[\protect\citeauthoryear{{Aharonian}, {Akhperjanian}, {Bazer-Bachi} \&
  {HESS~Collaboration}}{{Aharonian} et~al.}{2006}]{2006Natur.440.1018A}
{Aharonian} F.,  {Akhperjanian} A.~G.,  {Bazer-Bachi} A.~R.,
  {HESS~Collaboration} 2006, \nat, 440, 1018

\bibitem[\protect\citeauthoryear{{Becker}, {White} \& {Helfand}}{{Becker}
  et~al.}{1995}]{1995ApJ...450..559B}
{Becker} R.~H.,  {White} R.~L.,    {Helfand} D.~J.,  1995, \apj, 450, 559

\bibitem[\protect\citeauthoryear{{Bondi}, {et al.,} \& {}}{{Bondi}
  et~al.}{2007}]{2007A+A...463..519B}
{Bondi} M.,  {et al.,}   {} 2007, \aap, 463, 519

\bibitem[\protect\citeauthoryear{{Chapman}, {Blain}, {Smail} \&
  {Ivison}}{{Chapman} et~al.}{2005}]{2005ApJ...622..772C}
{Chapman} S.~C.,  {Blain} A.~W.,  {Smail} I.,    {Ivison} R.~J.,  2005, \apj,
  622, 772

\bibitem[\protect\citeauthoryear{{Ciliegi}, {et al.,} \& {}}{{Ciliegi}
  et~al.}{1999}]{1999MNRAS.302..222C}
{Ciliegi} P.,  {et al.,}   {} 1999, \mnras, 302, 222

\bibitem[\protect\citeauthoryear{{Condon}}{{Condon}}{1989}]{1989ApJ...338...13%
C}
{Condon} J.~J.,  1989, \apj, 338, 13

\bibitem[\protect\citeauthoryear{{Condon}}{{Condon}}{1992}]{1992ARA+A..30..575%
C}
{Condon} J.~J.,  1992, \araa, 30, 575

\bibitem[\protect\citeauthoryear{{Condon}, {Cotton}, {Greisen}, {Yin},
  {Perley}, {Taylor} \& {Broderick}}{{Condon}
  et~al.}{1998}]{1998AJ....115.1693C}
{Condon} J.~J.,  {Cotton} W.~D.,  {Greisen} E.~W.,  {Yin} Q.~F.,  {Perley}
  R.~A.,  {Taylor} G.~B.,    {Broderick} J.~J.,  1998, \aj, 115, 1693

\bibitem[\protect\citeauthoryear{{De Breuck}, {van Breugel}, {R{\"o}ttgering}
  \& {Miley}}{{De Breuck} et~al.}{2000}]{2000A+AS..143..303D}
{De Breuck} C.,  {van Breugel} W.,  {R{\"o}ttgering} H.~J.~A.,    {Miley} G.,
  2000, \aaps, 143, 303

\bibitem[\protect\citeauthoryear{{de Bruyn}, {et al.,} \& {}}{{de Bruyn}
  et~al.}{2000}]{2000yCat.8062....0D}
{de Bruyn} G.,  {et al.,}   {} 2000, VizieR Online Data Catalog, 8062, 0

\bibitem[\protect\citeauthoryear{{Dennefeld}, {Lagache}, {Mei}, {Ciliegi},
  {Dole}, {Mann}, {Taylor} \& {Vaccari}}{{Dennefeld}
  et~al.}{2005}]{2005A+A...440....5D}
{Dennefeld} M.,  {Lagache} G.,  {Mei} S.,  {Ciliegi} P.,  {Dole} H.,  {Mann}
  R.~G.,  {Taylor} E.~L.,    {Vaccari} M.,  2005, \aap, 440, 5

\bibitem[\protect\citeauthoryear{{Dole}, {et al.,} \& {}}{{Dole}
  et~al.}{2001}]{2001A+A...372..364D}
{Dole} H.,  {et al.,}   {} 2001, \aap, 372, 364

\bibitem[\protect\citeauthoryear{{Dole}, {et al.,} \& {}}{{Dole}
  et~al.}{2006}]{2006A+A...451..417D}
{Dole} H.,  {et al.,}   {} 2006, \aap, 451, 417

\bibitem[\protect\citeauthoryear{{Dwek} \& {Barker}}{{Dwek} \&
  {Barker}}{2002}]{2002ApJ...575....7D}
{Dwek} E.,  {Barker} M.~K.,  2002, \apj, 575, 7

\bibitem[\protect\citeauthoryear{{Garn}, {Green}, {Riley} \&
  {Alexander}}{{Garn} et~al.}{2008}]{2008MNRAS.383...75G}
{Garn} T.,  {Green} D.~A.,  {Riley} J.~M.,    {Alexander} P.,  2008, \mnras,
  383, 75~(G2008)

\bibitem[\protect\citeauthoryear{{Garrett}}{{Garrett}}{2002}]{2002A+A...384L..%
19G}
{Garrett} M.~A.,  2002, \aap, 384, L19

\bibitem[\protect\citeauthoryear{{Gispert}, {Lagache} \& {Puget}}{{Gispert}
  et~al.}{2000}]{2000A+A...360....1G}
{Gispert} R.,  {Lagache} G.,    {Puget} J.~L.,  2000, \aap, 360, 1

\bibitem[\protect\citeauthoryear{{Hauser} \& {Dwek}}{{Hauser} \&
  {Dwek}}{2001}]{2001ARA+A..39..249H}
{Hauser} M.~G.,  {Dwek} E.,  2001, \araa, 39, 249

\bibitem[\protect\citeauthoryear{{Hauser}, {et al.,} \& {}}{{Hauser}
  et~al.}{1998}]{1998ApJ...508...25H}
{Hauser} M.~G.,  {et al.,}   {} 1998, \apj, 508, 25

\bibitem[\protect\citeauthoryear{{Helou}, {Soifer} \& {Rowan-Robinson}}{{Helou}
  et~al.}{1985}]{1985ApJ...298L...7H}
{Helou} G.,  {Soifer} B.~T.,    {Rowan-Robinson} M.,  1985, \apjl, 298, L7

\bibitem[\protect\citeauthoryear{{Hopkins} \& {Beacom}}{{Hopkins} \&
  {Beacom}}{2006}]{2006ApJ...651..142H}
{Hopkins} A.~M.,  {Beacom} J.~F.,  2006, \apj, 651, 142

\bibitem[\protect\citeauthoryear{{Hughes}, {Wong}, {Ekers}, {Staveley-Smith},
  {Filipovic}, {Maddison}, {Fukui} \& {Mizuno}}{{Hughes}
  et~al.}{2006}]{2006MNRAS.370..363H}
{Hughes} A.,  {Wong} T.,  {Ekers} R.,  {Staveley-Smith} L.,  {Filipovic} M.,
  {Maddison} S.,  {Fukui} Y.,    {Mizuno} N.,  2006, \mnras, 370, 363

\bibitem[\protect\citeauthoryear{{Kashlinsky}}{{Kashlinsky}}{2005}]{2005PhR...%
409..361K}
{Kashlinsky} A.,  2005, \physrep, 409, 361

\bibitem[\protect\citeauthoryear{{Lagache}, {Abergel}, {Boulanger},
  {D{\'e}sert} \& {Puget}}{{Lagache} et~al.}{1999}]{1999A+A...344..322L}
{Lagache} G.,  {Abergel} A.,  {Boulanger} F.,  {D{\'e}sert} F.~X.,    {Puget}
  J.~L.,  1999, \aap, 344, 322

\bibitem[\protect\citeauthoryear{{Lagache}, {Puget} \& {Dole}}{{Lagache}
  et~al.}{2005}]{2005ARA+A..43..727L}
{Lagache} G.,  {Puget} J.~L.,    {Dole} H.,  2005, \araa, 43, 727

\bibitem[\protect\citeauthoryear{{Laing} \& {Peacock}}{{Laing} \&
  {Peacock}}{1980}]{1980MNRAS.190..903L}
{Laing} R.~A.,  {Peacock} J.~A.,  1980, \mnras, 190, 903

\bibitem[\protect\citeauthoryear{{Longair}}{{Longair}}{1966}]{1966MNRAS.133..4%
21L}
{Longair} M.~S.,  1966, \mnras, 133, 421

\bibitem[\protect\citeauthoryear{{Lonsdale}, {et al.,} \& {}}{{Lonsdale}
  et~al.}{2004}]{2004ApJS..154...54L}
{Lonsdale} C.,  {et al.,}   {} 2004, \apjs, 154, 54

\bibitem[\protect\citeauthoryear{{Luo} \& {Wu}}{{Luo} \&
  {Wu}}{2005}]{2005ChJAA...5..448L}
{Luo} S.~G.,  {Wu} X.~B.,  2005, Chinese Journal of Astronomy and Astrophysics,
  5, 448

\bibitem[\protect\citeauthoryear{{Madau}, {Ferguson}, {Dickinson},
  {Giavalisco}, {Steidel} \& {Fruchter}}{{Madau}
  et~al.}{1996}]{1996MNRAS.283.1388M}
{Madau} P.,  {Ferguson} H.~C.,  {Dickinson} M.~E.,  {Giavalisco} M.,  {Steidel}
  C.~C.,    {Fruchter} A.,  1996, \mnras, 283, 1388

\bibitem[\protect\citeauthoryear{{Madau}, {Pozzetti} \& {Dickinson}}{{Madau}
  et~al.}{1998}]{1998ApJ...498..106M}
{Madau} P.,  {Pozzetti} L.,    {Dickinson} M.,  1998, \apj, 498, 106

\bibitem[\protect\citeauthoryear{{McCarthy}, {Kapahi}, {van Breugel} \&
  {Subrahmanya}}{{McCarthy} et~al.}{1990}]{1990AJ....100.1014M}
{McCarthy} P.~J.,  {Kapahi} V.~K.,  {van Breugel} W.,    {Subrahmanya} C.~R.,
  1990, \aj, 100, 1014

\bibitem[\protect\citeauthoryear{{McMahon}, {Walton}, {Irwin}, {Lewis},
  {Bunclark} \& {Jones}}{{McMahon} et~al.}{2001}]{2001NewAR..45...97M}
{McMahon} R.~G.,  {Walton} N.~A.,  {Irwin} M.~J.,  {Lewis} J.~R.,  {Bunclark}
  P.~S.,    {Jones} D.~H.,  2001, New Astronomy Review, 45, 97

\bibitem[\protect\citeauthoryear{{Moss}, {Seymour}, {McHardy}, {Dwelly}, {Page}
  \& {Loaring}}{{Moss} et~al.}{2007}]{2007MNRAS.378..995M}
{Moss} D.,  {Seymour} N.,  {McHardy} I.~M.,  {Dwelly} T.,  {Page} M.~J.,
  {Loaring} N.~S.,  2007, \mnras, 378, 995

\bibitem[\protect\citeauthoryear{{Murphy}, {et al.,} \& {}}{{Murphy}
  et~al.}{2006}]{2006ApJ...638..157M}
{Murphy} E.~J.,  {et al.,}   {} 2006, \apj, 638, 157

\bibitem[\protect\citeauthoryear{{Ochsenbein}, {Bauer} \&
  {Marcout}}{{Ochsenbein} et~al.}{2000}]{2000A+AS..143...23O}
{Ochsenbein} F.,  {Bauer} P.,    {Marcout} J.,  2000, \aaps, 143, 23

\bibitem[\protect\citeauthoryear{{Padovani}, {Mainieri}, {Tozzi}, {Kellermann},
  {Fomalont}, {Miller}, {Rosati} \& {Shaver}}{{Padovani}
  et~al.}{2007}]{2007ASPC..380..205P}
{Padovani} P.,  {Mainieri} V.,  {Tozzi} P.,  {Kellermann} K.~I.,  {Fomalont}
  E.~B.,  {Miller} N.,  {Rosati} P.,    {Shaver} P.,  2007, in {Afonso} J.,
  {Ferguson} H.~C.,  {Mobasher} B.,   {Norris} R.,  eds, Deepest Astronomical
  Surveys. ASP Conf. series, 380, 205

\bibitem[\protect\citeauthoryear{{Prandoni}, {Parma}, {Wieringa}, {de Ruiter},
  {Gregorini}, {Mignano}, {Vettolani} \& {Ekers}}{{Prandoni}
  et~al.}{2006}]{2006A+A...457..517P}
{Prandoni} I.,  {Parma} P.,  {Wieringa} M.~H.,  {de Ruiter} H.~R.,  {Gregorini}
  L.,  {Mignano} A.,  {Vettolani} G.,    {Ekers} R.~D.,  2006, \aap, 457, 517

\bibitem[\protect\citeauthoryear{{Puget}, {Abergel}, {Bernard}, {Boulanger},
  {Burton}, {Desert} \& {Hartmann}}{{Puget} et~al.}{1996}]{1996A+A...308L...5P}
{Puget} J.~L.,  {Abergel} A.,  {Bernard} J.~P.,  {Boulanger} F.,  {Burton}
  W.~B.,  {Desert} F.~X.,    {Hartmann} D.,  1996, \aap, 308, L5

\bibitem[\protect\citeauthoryear{{Rengelink}, {Tang}, {de Bruyn}, {Miley},
  {Bremer}, {Roettgering} \& {Bremer}}{{Rengelink}
  et~al.}{1997}]{1997A+AS..124..259R}
{Rengelink} R.~B.,  {Tang} Y.,  {de Bruyn} A.~G.,  {Miley} G.~K.,  {Bremer}
  M.~N.,  {Roettgering} H.~J.~A.,    {Bremer} M.~A.~R.,  1997, \aaps, 124, 259

\bibitem[\protect\citeauthoryear{{Rowan-Robinson}, {Benn}, {Lawrence},
  {McMahon} \& {Broadhurst}}{{Rowan-Robinson}
  et~al.}{1993}]{1993MNRAS.263..123R}
{Rowan-Robinson} M.,  {Benn} C.~R.,  {Lawrence} A.,  {McMahon} R.~G.,
  {Broadhurst} T.~J.,  1993, \mnras, 263, 123

\bibitem[\protect\citeauthoryear{{Rowan-Robinson}, {et al.,}, {} \&
  {}}{{Rowan-Robinson} et~al.}{2008}]{2008MNRAS.386..697R}
{Rowan-Robinson} M.,  {et al.,} {}   {} 2008, \mnras, 386, 697

\bibitem[\protect\citeauthoryear{{Rowan-Robinson}, {et al.,} \&
  {}}{{Rowan-Robinson} et~al.}{2004}]{2004yCat..73511290R}
{Rowan-Robinson} M.,  {et al.,}   {} 2004, VizieR Online Data Catalog, 735,
  11290

\bibitem[\protect\citeauthoryear{{Seymour}, {McHardy} \& {Gunn}}{{Seymour}
  et~al.}{2004}]{2004MNRAS.352..131S}
{Seymour} N.,  {McHardy} I.~M.,    {Gunn} K.~F.,  2004, \mnras, 352, 131

\bibitem[\protect\citeauthoryear{{Simpson}, {Mart{\'{\i}}nez-Sansigre},
  {Rawlings}, {Ivison}, {Akiyama}, {Sekiguchi}, {Takata}, {Ueda} \&
  {Watson}}{{Simpson} et~al.}{2006}]{2006MNRAS.372..741S}
{Simpson} C.,  {Mart{\'{\i}}nez-Sansigre} A.,  {Rawlings} S.,  {Ivison} R.,
  {Akiyama} M.,  {Sekiguchi} K.,  {Takata} T.,  {Ueda} Y.,    {Watson} M.,
  2006, \mnras, 372, 741

\bibitem[\protect\citeauthoryear{{Sirothia}}{{Sirothia}}{2008}]{sks2008}
{Sirothia} S.~K.,  2008, \mnras, submitted

\bibitem[\protect\citeauthoryear{{Sirothia}, {Saikia}, {Ishwara-Chandra} \&
  {Kantharia}}{{Sirothia} et~al.}{2008}]{2008arXiv0809.4565S}
{Sirothia} S.~K.,  {Saikia} D.~J.,  {Ishwara-Chandra} C.~H.,    {Kantharia}
  N.~G.,  2008, \mnras, in press, arXiv:0809.4565

\bibitem[\protect\citeauthoryear{{Smol{\v c}i{\'c}}, {et. al.}, {} \&
  {}}{{Smol{\v c}i{\'c}} et~al.}{2008}]{2008ApJS..177...14S}
{Smol{\v c}i{\'c}} V.,  {et. al.} {}   {} 2008, \apjs, 177, 14

\bibitem[\protect\citeauthoryear{{Tasse}, {et al.,} \& {}}{{Tasse}
  et~al.}{2006}]{2006A+A...456..791T}
{Tasse} C.,  {et al.,}   {} 2006, \aap, 456, 791

\bibitem[\protect\citeauthoryear{{Taylor}, {et al.,} \& {}}{{Taylor}
  et~al.}{2007}]{2007ApJ...666..201T}
{Taylor} A.~R.,  {et al.,}   {} 2007, \apj, 666, 201

\bibitem[\protect\citeauthoryear{{Wieringa}}{{Wieringa}}{1991}]{1991PhDT......%
.241W}
{Wieringa} M.~H.,  1991, PhD thesis, Rijksuniversiteit Leiden, (1991)

\bibitem[\protect\citeauthoryear{{Wieringa}, {de Bruyn} \&
  {Katgert}}{{Wieringa} et~al.}{1992}]{1992A+A...256..331W}
{Wieringa} M.~H.,  {de Bruyn} A.~G.,    {Katgert} P.,  1992, \aap, 256, 331

\bibitem[\protect\citeauthoryear{{Windhorst}, {Mathis} \&
  {Neuschaefer}}{{Windhorst} et~al.}{1990}]{1990ASPC...10..389W}
{Windhorst} R.,  {Mathis} D.,    {Neuschaefer} L.,  1990, in {Kron} R.~G.,
  ed., Evolution of the Universe of Galaxies. ASP Conf. series, 10, 389

\bibitem[\protect\citeauthoryear{{Wright}}{{Wright}}{2004}]{2004NewAR..48..465%
W}
{Wright} E.~L.,  2004, New Astronomy Review, 48, 465

\end{thebibliography}
\bibliographystyle{./mn2e.bst}

\label{lastpage}

\begin{table*}
\setcounter{table}{2}
\caption{The source catalogue (available in online version only).}
\label{ef0:table:ot}
\begin{tabular}{l r r r r r r r r r r r}
\hline
\multicolumn{1}{c}{Source name} & \multicolumn{1}{c}{RA} & \multicolumn{1}{c}{DEC} & \multicolumn{1}{c}{Dist} & \multicolumn{1}{c}{$\sigma_{rms}$} & \multicolumn{1}{c}{S$_{peak}$} & \multicolumn{1}{c}{S$_{total~3\sigma}$} & \multicolumn{1}{c}{S$_{total}$} & \multicolumn{1}{c}{Size} & \multicolumn{1}{c}{Notes} \\ 
                    & \multicolumn{1}{c}{hh:mm:ss.s} & \multicolumn{1}{c}{dd:mm:ss.s} & \multicolumn{1}{c}{deg} & \multicolumn{1}{c}{$\mu$Jy b$^{-1}$}  & \multicolumn{1}{c}{mJy b$^{-1}$} &  \multicolumn{1}{c}{mJy} &  \multicolumn{1}{c}{mJy}  & \multicolumn{1}{c}{$^{\prime\prime}$} &        \\ 
\multicolumn{1}{c}{(1)} & \multicolumn{1}{c}{(2)} & \multicolumn{1}{c}{(3)} & \multicolumn{1}{c}{(4)} &  \multicolumn{1}{c}{(5)} & \multicolumn{1}{c}{(6)} & \multicolumn{1}{c}{(7)} & \multicolumn{1}{c}{(8)} & \multicolumn{1}{c}{(9)} & \multicolumn{1}{c}{(10)} \\
\hline
GMRT160232+0545404  & 16:02:32.8 & +54:54:06.9 &     1.10 &      267 &    22.92 &    48.57 &    48.57 &     16.6 \\
GMRT160234+0544706  & 16:02:34.7 & +54:47:10.0 &     1.08 &      212 &     1.56 &     1.59 &     1.56 &        U \\
GMRT160248+0544735  & 16:02:48.5 & +54:47:39.3 &     1.05 &      223 &     8.46 &    12.47 &    12.47 &      9.9 \\
GMRT160253+0545951  & 16:02:53.0 & +54:59:51.6 &     1.08 &      205 &     2.23 &     6.17 &     6.17 &     19.1 \\
GMRT160255+0550533  & 16:02:55.1 & +55:05:34.7 &     1.10 &      244 &     2.20 &     3.62 &     3.62 &     10.8 \\
\hline \hline
\end{tabular}

\end{table*}
\begin{table*}
\contcaption{}
\begin{tabular}{l r r r r r r r r r r r}
\hline
\multicolumn{1}{c}{Source name} & \multicolumn{1}{c}{RA} & \multicolumn{1}{c}{DEC} & \multicolumn{1}{c}{Dist} & \multicolumn{1}{c}{$\sigma_{rms}$} & \multicolumn{1}{c}{S$_{peak}$} & \multicolumn{1}{c}{S$_{total~3\sigma}$} & \multicolumn{1}{c}{S$_{total}$} & \multicolumn{1}{c}{Size} & \multicolumn{1}{c}{Notes} \\ 
                    & \multicolumn{1}{c}{hh:mm:ss.s} & \multicolumn{1}{c}{dd:mm:ss.s} & \multicolumn{1}{c}{deg} & \multicolumn{1}{c}{$\mu$Jy b$^{-1}$}  & \multicolumn{1}{c}{mJy b$^{-1}$} &  \multicolumn{1}{c}{mJy} &  \multicolumn{1}{c}{mJy}  & \multicolumn{1}{c}{$^{\prime\prime}$} &        \\ 
\multicolumn{1}{c}{(1)} & \multicolumn{1}{c}{(2)} & \multicolumn{1}{c}{(3)} & \multicolumn{1}{c}{(4)} &  \multicolumn{1}{c}{(5)} & \multicolumn{1}{c}{(6)} & \multicolumn{1}{c}{(7)} & \multicolumn{1}{c}{(8)} & \multicolumn{1}{c}{(9)} & \multicolumn{1}{c}{(10)} \\
\hline
GMRT160440+0545725  & 16:04:40.7 & +54:57:35.0 &     0.82 &      145 &     1.57 &     1.60 &     1.57 &        U \\
GMRT160441+0543828  & 16:04:41.2 & +54:38:28.0 &     0.77 &      291 &    45.93 &   104.51 &   104.51 &     30.5 \\
GMRT160441+0545047  & 16:04:41.0 & +54:50:48.5 &     0.79 &      145 &     1.67 &     1.50 &     1.67 &        U \\
GMRT160442+0543059  & 16:04:42.1 & +54:30:60.1 &     0.78 &      288 &     1.86 &     6.83 &     6.83 &     12.2 \\
GMRT160442+0544319  & 16:04:42.1 & +54:43:19.9 &     0.77 &      170 &     2.08 &     2.98 &     2.08 &        U \\
\hline \hline
\end{tabular}

\end{table*}
\begin{table*}
\contcaption{}
\begin{tabular}{l r r r r r r r r r r r}
\hline
\multicolumn{1}{c}{Source name} & \multicolumn{1}{c}{RA} & \multicolumn{1}{c}{DEC} & \multicolumn{1}{c}{Dist} & \multicolumn{1}{c}{$\sigma_{rms}$} & \multicolumn{1}{c}{S$_{peak}$} & \multicolumn{1}{c}{S$_{total~3\sigma}$} & \multicolumn{1}{c}{S$_{total}$} & \multicolumn{1}{c}{Size} & \multicolumn{1}{c}{Notes} \\ 
                    & \multicolumn{1}{c}{hh:mm:ss.s} & \multicolumn{1}{c}{dd:mm:ss.s} & \multicolumn{1}{c}{deg} & \multicolumn{1}{c}{$\mu$Jy b$^{-1}$}  & \multicolumn{1}{c}{mJy b$^{-1}$} &  \multicolumn{1}{c}{mJy} &  \multicolumn{1}{c}{mJy}  & \multicolumn{1}{c}{$^{\prime\prime}$} &        \\ 
\multicolumn{1}{c}{(1)} & \multicolumn{1}{c}{(2)} & \multicolumn{1}{c}{(3)} & \multicolumn{1}{c}{(4)} &  \multicolumn{1}{c}{(5)} & \multicolumn{1}{c}{(6)} & \multicolumn{1}{c}{(7)} & \multicolumn{1}{c}{(8)} & \multicolumn{1}{c}{(9)} & \multicolumn{1}{c}{(10)} \\
\hline
GMRT160553+0544537  & 16:05:53.3 & +54:45:42.0 &     0.60 &      268 &     3.62 &    26.68 &    26.68 &     57.5 \\
GMRT160553+0551030  & 16:05:53.3 & +55:10:37.9 &     0.78 &      149 &    23.33 &    30.29 &    30.29 &     29.8 \\
GMRT160555+0551311  & 16:05:55.3 & +55:13:11.7 &     0.81 &      151 &     1.12 &     1.32 &     1.12 &        U \\
GMRT160556+0543043  & 16:05:56.8 & +54:30:52.1 &     0.61 &      106 &     0.95 &     0.92 &     0.95 &        U \\
GMRT160556+0550832  & 16:05:56.3 & +55:08:39.2 &     0.75 &      130 &     1.15 &     0.90 &     1.15 &        U \\
\hline \hline
\end{tabular}

\end{table*}
\begin{table*}
\contcaption{}
\begin{tabular}{l r r r r r r r r r r r}
\hline
\multicolumn{1}{c}{Source name} & \multicolumn{1}{c}{RA} & \multicolumn{1}{c}{DEC} & \multicolumn{1}{c}{Dist} & \multicolumn{1}{c}{$\sigma_{rms}$} & \multicolumn{1}{c}{S$_{peak}$} & \multicolumn{1}{c}{S$_{total~3\sigma}$} & \multicolumn{1}{c}{S$_{total}$} & \multicolumn{1}{c}{Size} & \multicolumn{1}{c}{Notes} \\ 
                    & \multicolumn{1}{c}{hh:mm:ss.s} & \multicolumn{1}{c}{dd:mm:ss.s} & \multicolumn{1}{c}{deg} & \multicolumn{1}{c}{$\mu$Jy b$^{-1}$}  & \multicolumn{1}{c}{mJy b$^{-1}$} &  \multicolumn{1}{c}{mJy} &  \multicolumn{1}{c}{mJy}  & \multicolumn{1}{c}{$^{\prime\prime}$} &        \\ 
\multicolumn{1}{c}{(1)} & \multicolumn{1}{c}{(2)} & \multicolumn{1}{c}{(3)} & \multicolumn{1}{c}{(4)} &  \multicolumn{1}{c}{(5)} & \multicolumn{1}{c}{(6)} & \multicolumn{1}{c}{(7)} & \multicolumn{1}{c}{(8)} & \multicolumn{1}{c}{(9)} & \multicolumn{1}{c}{(10)} \\
\hline
GMRT160649+0535021  & 16:06:49.0 & +53:50:25.6 &     0.95 &      152 &     2.01 &     2.44 &     2.01 &        U \\
GMRT160650+0542511  & 16:06:50.7 & +54:25:16.1 &     0.52 &       75 &     0.68 &     1.32 &     0.68 &        U \\
GMRT160651+0541449  & 16:06:51.3 & +54:14:49.1 &     0.62 &      103 &     1.50 &     4.90 &     4.90 &     24.4 \\
GMRT160655+0542753  & 16:06:55.2 & +54:27:60.0 &     0.49 &       74 &     0.55 &     0.52 &     0.55 &        U \\
GMRT160656+0541457  & 16:06:56.1 & +54:14:63.0 &     0.61 &       91 &     1.14 &     1.11 &     1.14 &        U \\
\hline \hline
\end{tabular}

\end{table*}
\begin{table*}
\contcaption{}
\begin{tabular}{l r r r r r r r r r r r}
\hline
\multicolumn{1}{c}{Source name} & \multicolumn{1}{c}{RA} & \multicolumn{1}{c}{DEC} & \multicolumn{1}{c}{Dist} & \multicolumn{1}{c}{$\sigma_{rms}$} & \multicolumn{1}{c}{S$_{peak}$} & \multicolumn{1}{c}{S$_{total~3\sigma}$} & \multicolumn{1}{c}{S$_{total}$} & \multicolumn{1}{c}{Size} & \multicolumn{1}{c}{Notes} \\ 
                    & \multicolumn{1}{c}{hh:mm:ss.s} & \multicolumn{1}{c}{dd:mm:ss.s} & \multicolumn{1}{c}{deg} & \multicolumn{1}{c}{$\mu$Jy b$^{-1}$}  & \multicolumn{1}{c}{mJy b$^{-1}$} &  \multicolumn{1}{c}{mJy} &  \multicolumn{1}{c}{mJy}  & \multicolumn{1}{c}{$^{\prime\prime}$} &        \\ 
\multicolumn{1}{c}{(1)} & \multicolumn{1}{c}{(2)} & \multicolumn{1}{c}{(3)} & \multicolumn{1}{c}{(4)} &  \multicolumn{1}{c}{(5)} & \multicolumn{1}{c}{(6)} & \multicolumn{1}{c}{(7)} & \multicolumn{1}{c}{(8)} & \multicolumn{1}{c}{(9)} & \multicolumn{1}{c}{(10)} \\
\hline
GMRT160724+0550128  & 16:07:24.3 & +55:01:30.7 &     0.52 &       77 &     0.61 &     0.86 &     0.61 &        U \\
GMRT160725+0535351  & 16:07:25.7 & +53:53:53.1 &     0.86 &      132 &     4.08 &     4.66 &     4.66 &      9.4 \\
GMRT160725+0544720  & 16:07:25.7 & +54:47:26.8 &     0.39 &       91 &     1.83 &     5.90 &     5.90 &     41.1 \\
GMRT160726+0553522  & 16:07:26.1 & +55:35:30.6 &     0.99 &      181 &    16.32 &    20.83 &    20.83 &      6.2 \\
GMRT160727+0541833  & 16:07:27.6 & +54:18:38.8 &     0.51 &       64 &     0.40 &     0.36 &     0.40 &        U \\
\hline \hline
\end{tabular}

\end{table*}
\begin{table*}
\contcaption{}
\begin{tabular}{l r r r r r r r r r r r}
\hline
\multicolumn{1}{c}{Source name} & \multicolumn{1}{c}{RA} & \multicolumn{1}{c}{DEC} & \multicolumn{1}{c}{Dist} & \multicolumn{1}{c}{$\sigma_{rms}$} & \multicolumn{1}{c}{S$_{peak}$} & \multicolumn{1}{c}{S$_{total~3\sigma}$} & \multicolumn{1}{c}{S$_{total}$} & \multicolumn{1}{c}{Size} & \multicolumn{1}{c}{Notes} \\ 
                    & \multicolumn{1}{c}{hh:mm:ss.s} & \multicolumn{1}{c}{dd:mm:ss.s} & \multicolumn{1}{c}{deg} & \multicolumn{1}{c}{$\mu$Jy b$^{-1}$}  & \multicolumn{1}{c}{mJy b$^{-1}$} &  \multicolumn{1}{c}{mJy} &  \multicolumn{1}{c}{mJy}  & \multicolumn{1}{c}{$^{\prime\prime}$} &        \\ 
\multicolumn{1}{c}{(1)} & \multicolumn{1}{c}{(2)} & \multicolumn{1}{c}{(3)} & \multicolumn{1}{c}{(4)} &  \multicolumn{1}{c}{(5)} & \multicolumn{1}{c}{(6)} & \multicolumn{1}{c}{(7)} & \multicolumn{1}{c}{(8)} & \multicolumn{1}{c}{(9)} & \multicolumn{1}{c}{(10)} \\
\hline
GMRT160751+0553154  & 16:07:52.0 & +55:31:62.2 &     0.92 &      142 &     1.49 &     1.30 &     1.49 &        U \\
GMRT160752+0540832  & 16:07:52.2 & +54:08:35.5 &     0.61 &       83 &     4.14 &     5.45 &     5.45 &      3.6 \\
GMRT160752+0542432  & 16:07:52.2 & +54:24:39.6 &     0.40 &       60 &     0.58 &     0.85 &     0.58 &        U \\
GMRT160752+0544231  & 16:07:52.5 & +54:42:36.0 &     0.31 &       59 &     1.03 &     1.08 &     1.03 &        U \\
GMRT160752+0552022  & 16:07:52.1 & +55:20:31.8 &     0.74 &      100 &     0.77 &     0.75 &     0.77 &        U \\
\hline \hline
\end{tabular}

\end{table*}
\begin{table*}
\contcaption{}
\begin{tabular}{l r r r r r r r r r r r}
\hline
\multicolumn{1}{c}{Source name} & \multicolumn{1}{c}{RA} & \multicolumn{1}{c}{DEC} & \multicolumn{1}{c}{Dist} & \multicolumn{1}{c}{$\sigma_{rms}$} & \multicolumn{1}{c}{S$_{peak}$} & \multicolumn{1}{c}{S$_{total~3\sigma}$} & \multicolumn{1}{c}{S$_{total}$} & \multicolumn{1}{c}{Size} & \multicolumn{1}{c}{Notes} \\ 
                    & \multicolumn{1}{c}{hh:mm:ss.s} & \multicolumn{1}{c}{dd:mm:ss.s} & \multicolumn{1}{c}{deg} & \multicolumn{1}{c}{$\mu$Jy b$^{-1}$}  & \multicolumn{1}{c}{mJy b$^{-1}$} &  \multicolumn{1}{c}{mJy} &  \multicolumn{1}{c}{mJy}  & \multicolumn{1}{c}{$^{\prime\prime}$} &        \\ 
\multicolumn{1}{c}{(1)} & \multicolumn{1}{c}{(2)} & \multicolumn{1}{c}{(3)} & \multicolumn{1}{c}{(4)} &  \multicolumn{1}{c}{(5)} & \multicolumn{1}{c}{(6)} & \multicolumn{1}{c}{(7)} & \multicolumn{1}{c}{(8)} & \multicolumn{1}{c}{(9)} & \multicolumn{1}{c}{(10)} \\
\hline
GMRT160812+0545523  & 16:08:12.5 & +54:55:29.0 &     0.36 &       51 &     1.05 &     1.29 &     1.05 &        U \\
GMRT160812+0550012  & 16:08:12.2 & +55:00:21.4 &     0.42 &       62 &     2.12 &     1.99 &     2.12 &        U \\
GMRT160814+0541939  & 16:08:14.0 & +54:19:42.4 &     0.43 &       57 &     0.93 &     0.83 &     0.93 &        U \\
GMRT160814+0542832  & 16:08:14.3 & +54:28:32.7 &     0.32 &       57 &     1.86 &     2.72 &     2.72 &      7.7 \\
GMRT160814+0552659  & 16:08:15.0 & +55:26:63.5 &     0.82 &      135 &     5.22 &    13.90 &    13.90 &     38.2 \\
\hline \hline
\end{tabular}

\end{table*}
\begin{table*}
\contcaption{}
\begin{tabular}{l r r r r r r r r r r r}
\hline
\multicolumn{1}{c}{Source name} & \multicolumn{1}{c}{RA} & \multicolumn{1}{c}{DEC} & \multicolumn{1}{c}{Dist} & \multicolumn{1}{c}{$\sigma_{rms}$} & \multicolumn{1}{c}{S$_{peak}$} & \multicolumn{1}{c}{S$_{total~3\sigma}$} & \multicolumn{1}{c}{S$_{total}$} & \multicolumn{1}{c}{Size} & \multicolumn{1}{c}{Notes} \\ 
                    & \multicolumn{1}{c}{hh:mm:ss.s} & \multicolumn{1}{c}{dd:mm:ss.s} & \multicolumn{1}{c}{deg} & \multicolumn{1}{c}{$\mu$Jy b$^{-1}$}  & \multicolumn{1}{c}{mJy b$^{-1}$} &  \multicolumn{1}{c}{mJy} &  \multicolumn{1}{c}{mJy}  & \multicolumn{1}{c}{$^{\prime\prime}$} &        \\ 
\multicolumn{1}{c}{(1)} & \multicolumn{1}{c}{(2)} & \multicolumn{1}{c}{(3)} & \multicolumn{1}{c}{(4)} &  \multicolumn{1}{c}{(5)} & \multicolumn{1}{c}{(6)} & \multicolumn{1}{c}{(7)} & \multicolumn{1}{c}{(8)} & \multicolumn{1}{c}{(9)} & \multicolumn{1}{c}{(10)} \\
\hline
GMRT160835+0535932  & 16:08:35.5 & +53:59:40.0 &     0.70 &       86 &     4.06 &     6.26 &     6.26 &     12.2 \\
GMRT160835+0541630  & 16:08:35.5 & +54:16:39.4 &     0.44 &       71 &     1.81 &     2.09 &     2.09 &      3.8 \\
GMRT160835+0544944  & 16:08:35.9 & +54:49:49.6 &     0.26 &       55 &     0.70 &     0.83 &     0.70 &        U \\
GMRT160836+0541415  & 16:08:36.4 & +54:14:21.0 &     0.47 &       66 &     7.39 &    14.99 &    14.99 &     18.2 \\
GMRT160836+0551414  & 16:08:36.5 & +55:14:20.1 &     0.60 &       80 &     2.44 &     2.94 &     2.94 &      9.4 \\
\hline \hline
\end{tabular}

\end{table*}
\begin{table*}
\contcaption{}
\begin{tabular}{l r r r r r r r r r r r}
\hline
\multicolumn{1}{c}{Source name} & \multicolumn{1}{c}{RA} & \multicolumn{1}{c}{DEC} & \multicolumn{1}{c}{Dist} & \multicolumn{1}{c}{$\sigma_{rms}$} & \multicolumn{1}{c}{S$_{peak}$} & \multicolumn{1}{c}{S$_{total~3\sigma}$} & \multicolumn{1}{c}{S$_{total}$} & \multicolumn{1}{c}{Size} & \multicolumn{1}{c}{Notes} \\ 
                    & \multicolumn{1}{c}{hh:mm:ss.s} & \multicolumn{1}{c}{dd:mm:ss.s} & \multicolumn{1}{c}{deg} & \multicolumn{1}{c}{$\mu$Jy b$^{-1}$}  & \multicolumn{1}{c}{mJy b$^{-1}$} &  \multicolumn{1}{c}{mJy} &  \multicolumn{1}{c}{mJy}  & \multicolumn{1}{c}{$^{\prime\prime}$} &        \\ 
\multicolumn{1}{c}{(1)} & \multicolumn{1}{c}{(2)} & \multicolumn{1}{c}{(3)} & \multicolumn{1}{c}{(4)} &  \multicolumn{1}{c}{(5)} & \multicolumn{1}{c}{(6)} & \multicolumn{1}{c}{(7)} & \multicolumn{1}{c}{(8)} & \multicolumn{1}{c}{(9)} & \multicolumn{1}{c}{(10)} \\
\hline
GMRT160857+0541026  & 16:08:57.9 & +54:10:27.2 &     0.52 &       63 &     0.89 &     0.89 &     0.89 &        U \\
GMRT160858+0541815  & 16:08:58.3 & +54:18:16.8 &     0.39 &       53 &     4.70 &     5.48 &     5.48 &      9.4 \\
GMRT160858+0543941  & 16:08:58.3 & +54:39:46.6 &     0.15 &       46 &     2.96 &     3.31 &     3.31 &      4.7 \\
GMRT160859+0540226  & 16:08:59.1 & +54:02:32.5 &     0.64 &       83 &     5.15 &     5.97 &     5.97 &      9.4 \\
GMRT160900+0552839  & 16:09:00.9 & +55:28:42.0 &     0.82 &      116 &     1.53 &     1.86 &     1.53 &        U \\
\hline \hline
\end{tabular}

\end{table*}
\begin{table*}
\contcaption{}
\begin{tabular}{l r r r r r r r r r r r}
\hline
\multicolumn{1}{c}{Source name} & \multicolumn{1}{c}{RA} & \multicolumn{1}{c}{DEC} & \multicolumn{1}{c}{Dist} & \multicolumn{1}{c}{$\sigma_{rms}$} & \multicolumn{1}{c}{S$_{peak}$} & \multicolumn{1}{c}{S$_{total~3\sigma}$} & \multicolumn{1}{c}{S$_{total}$} & \multicolumn{1}{c}{Size} & \multicolumn{1}{c}{Notes} \\ 
                    & \multicolumn{1}{c}{hh:mm:ss.s} & \multicolumn{1}{c}{dd:mm:ss.s} & \multicolumn{1}{c}{deg} & \multicolumn{1}{c}{$\mu$Jy b$^{-1}$}  & \multicolumn{1}{c}{mJy b$^{-1}$} &  \multicolumn{1}{c}{mJy} &  \multicolumn{1}{c}{mJy}  & \multicolumn{1}{c}{$^{\prime\prime}$} &        \\ 
\multicolumn{1}{c}{(1)} & \multicolumn{1}{c}{(2)} & \multicolumn{1}{c}{(3)} & \multicolumn{1}{c}{(4)} &  \multicolumn{1}{c}{(5)} & \multicolumn{1}{c}{(6)} & \multicolumn{1}{c}{(7)} & \multicolumn{1}{c}{(8)} & \multicolumn{1}{c}{(9)} & \multicolumn{1}{c}{(10)} \\
\hline
GMRT160918+0550719  & 16:09:18.6 & +55:07:28.8 &     0.47 &       61 &     0.40 &     0.46 &     0.40 &        U \\
GMRT160918+0552114  & 16:09:18.8 & +55:21:18.1 &     0.69 &       98 &     2.67 &     3.06 &     3.06 &      4.3 \\
GMRT160919+0542237  & 16:09:19.1 & +54:22:41.0 &     0.31 &       49 &     0.97 &     1.02 &     0.97 &        U \\
GMRT160919+0544004  & 16:09:19.2 & +54:40:07.2 &     0.10 &       46 &     0.48 &     0.41 &     0.48 &        U \\
GMRT160919+0544033  & 16:09:19.6 & +54:40:35.1 &     0.10 &       45 &     0.94 &     0.77 &     0.94 &        U \\
\hline \hline
\end{tabular}

\end{table*}
\begin{table*}
\contcaption{}
\begin{tabular}{l r r r r r r r r r r r}
\hline
\multicolumn{1}{c}{Source name} & \multicolumn{1}{c}{RA} & \multicolumn{1}{c}{DEC} & \multicolumn{1}{c}{Dist} & \multicolumn{1}{c}{$\sigma_{rms}$} & \multicolumn{1}{c}{S$_{peak}$} & \multicolumn{1}{c}{S$_{total~3\sigma}$} & \multicolumn{1}{c}{S$_{total}$} & \multicolumn{1}{c}{Size} & \multicolumn{1}{c}{Notes} \\ 
                    & \multicolumn{1}{c}{hh:mm:ss.s} & \multicolumn{1}{c}{dd:mm:ss.s} & \multicolumn{1}{c}{deg} & \multicolumn{1}{c}{$\mu$Jy b$^{-1}$}  & \multicolumn{1}{c}{mJy b$^{-1}$} &  \multicolumn{1}{c}{mJy} &  \multicolumn{1}{c}{mJy}  & \multicolumn{1}{c}{$^{\prime\prime}$} &        \\ 
\multicolumn{1}{c}{(1)} & \multicolumn{1}{c}{(2)} & \multicolumn{1}{c}{(3)} & \multicolumn{1}{c}{(4)} &  \multicolumn{1}{c}{(5)} & \multicolumn{1}{c}{(6)} & \multicolumn{1}{c}{(7)} & \multicolumn{1}{c}{(8)} & \multicolumn{1}{c}{(9)} & \multicolumn{1}{c}{(10)} \\
\hline
GMRT160933+0535718  & 16:09:33.9 & +53:57:24.9 &     0.71 &       99 &     0.76 &     0.65 &     0.76 &        U \\
GMRT160933+0540140  & 16:09:33.2 & +54:01:43.3 &     0.64 &       76 &     0.80 &     0.94 &     0.80 &        U \\
GMRT160933+0545215  & 16:09:33.6 & +54:52:17.5 &     0.21 &       45 &     0.38 &     0.36 &     0.38 &        U \\
GMRT160933+0551947  & 16:09:33.1 & +55:19:50.6 &     0.67 &       83 &     1.23 &     2.05 &     2.05 &      5.4 \\
GMRT160934+0535217  & 16:09:34.8 & +53:52:25.6 &     0.80 &      108 &     0.98 &     1.21 &     0.98 &        U \\
\hline \hline
\end{tabular}

\end{table*}
\begin{table*}
\contcaption{}
\begin{tabular}{l r r r r r r r r r r r}
\hline
\multicolumn{1}{c}{Source name} & \multicolumn{1}{c}{RA} & \multicolumn{1}{c}{DEC} & \multicolumn{1}{c}{Dist} & \multicolumn{1}{c}{$\sigma_{rms}$} & \multicolumn{1}{c}{S$_{peak}$} & \multicolumn{1}{c}{S$_{total~3\sigma}$} & \multicolumn{1}{c}{S$_{total}$} & \multicolumn{1}{c}{Size} & \multicolumn{1}{c}{Notes} \\ 
                    & \multicolumn{1}{c}{hh:mm:ss.s} & \multicolumn{1}{c}{dd:mm:ss.s} & \multicolumn{1}{c}{deg} & \multicolumn{1}{c}{$\mu$Jy b$^{-1}$}  & \multicolumn{1}{c}{mJy b$^{-1}$} &  \multicolumn{1}{c}{mJy} &  \multicolumn{1}{c}{mJy}  & \multicolumn{1}{c}{$^{\prime\prime}$} &        \\ 
\multicolumn{1}{c}{(1)} & \multicolumn{1}{c}{(2)} & \multicolumn{1}{c}{(3)} & \multicolumn{1}{c}{(4)} &  \multicolumn{1}{c}{(5)} & \multicolumn{1}{c}{(6)} & \multicolumn{1}{c}{(7)} & \multicolumn{1}{c}{(8)} & \multicolumn{1}{c}{(9)} & \multicolumn{1}{c}{(10)} \\
\hline
GMRT160949+0540826  & 16:09:49.7 & +54:08:28.3 &     0.53 &       70 &     4.87 &     8.23 &     8.23 &     21.3 \\
GMRT160949+0542744  & 16:09:49.6 & +54:27:47.3 &     0.21 &       42 &     0.43 &     0.31 &     0.43 &        U \\
GMRT160949+0543809  & 16:09:49.3 & +54:38:10.0 &     0.04 &       46 &     0.36 &     0.31 &     0.36 &        U \\
GMRT160949+0550347  & 16:09:49.8 & +55:03:50.7 &     0.40 &       64 &     1.14 &     1.13 &     1.14 &        U \\
GMRT160950+0541811  & 16:09:50.4 & +54:18:15.9 &     0.36 &       54 &     0.36 &     0.30 &     0.36 &        U \\
\hline \hline
\end{tabular}

\end{table*}
\begin{table*}
\contcaption{}
\begin{tabular}{l r r r r r r r r r r r}
\hline
\multicolumn{1}{c}{Source name} & \multicolumn{1}{c}{RA} & \multicolumn{1}{c}{DEC} & \multicolumn{1}{c}{Dist} & \multicolumn{1}{c}{$\sigma_{rms}$} & \multicolumn{1}{c}{S$_{peak}$} & \multicolumn{1}{c}{S$_{total~3\sigma}$} & \multicolumn{1}{c}{S$_{total}$} & \multicolumn{1}{c}{Size} & \multicolumn{1}{c}{Notes} \\ 
                    & \multicolumn{1}{c}{hh:mm:ss.s} & \multicolumn{1}{c}{dd:mm:ss.s} & \multicolumn{1}{c}{deg} & \multicolumn{1}{c}{$\mu$Jy b$^{-1}$}  & \multicolumn{1}{c}{mJy b$^{-1}$} &  \multicolumn{1}{c}{mJy} &  \multicolumn{1}{c}{mJy}  & \multicolumn{1}{c}{$^{\prime\prime}$} &        \\ 
\multicolumn{1}{c}{(1)} & \multicolumn{1}{c}{(2)} & \multicolumn{1}{c}{(3)} & \multicolumn{1}{c}{(4)} &  \multicolumn{1}{c}{(5)} & \multicolumn{1}{c}{(6)} & \multicolumn{1}{c}{(7)} & \multicolumn{1}{c}{(8)} & \multicolumn{1}{c}{(9)} & \multicolumn{1}{c}{(10)} \\
\hline
GMRT161007+0544140  & 16:10:07.6 & +54:41:40.1 &     0.03 &       41 &     0.57 &     0.75 &     0.57 &        U \\
GMRT161007+0545033  & 16:10:07.9 & +54:50:39.6 &     0.18 &       48 &     0.44 &     0.33 &     0.44 &        U \\
GMRT161008+0535027  & 16:10:08.8 & +53:50:27.7 &     0.83 &      115 &     0.77 &     0.58 &     0.77 &        U \\
GMRT161008+0540753  & 16:10:08.7 & +54:07:57.1 &     0.54 &       66 &     4.14 &    16.77 &    16.77 &     38.7 \\
GMRT161008+0553611  & 16:10:08.0 & +55:36:14.8 &     0.94 &      147 &     1.26 &     1.48 &     1.26 &        U \\
\hline \hline
\end{tabular}

\end{table*}
\begin{table*}
\contcaption{}
\begin{tabular}{l r r r r r r r r r r r}
\hline
\multicolumn{1}{c}{Source name} & \multicolumn{1}{c}{RA} & \multicolumn{1}{c}{DEC} & \multicolumn{1}{c}{Dist} & \multicolumn{1}{c}{$\sigma_{rms}$} & \multicolumn{1}{c}{S$_{peak}$} & \multicolumn{1}{c}{S$_{total~3\sigma}$} & \multicolumn{1}{c}{S$_{total}$} & \multicolumn{1}{c}{Size} & \multicolumn{1}{c}{Notes} \\ 
                    & \multicolumn{1}{c}{hh:mm:ss.s} & \multicolumn{1}{c}{dd:mm:ss.s} & \multicolumn{1}{c}{deg} & \multicolumn{1}{c}{$\mu$Jy b$^{-1}$}  & \multicolumn{1}{c}{mJy b$^{-1}$} &  \multicolumn{1}{c}{mJy} &  \multicolumn{1}{c}{mJy}  & \multicolumn{1}{c}{$^{\prime\prime}$} &        \\ 
\multicolumn{1}{c}{(1)} & \multicolumn{1}{c}{(2)} & \multicolumn{1}{c}{(3)} & \multicolumn{1}{c}{(4)} &  \multicolumn{1}{c}{(5)} & \multicolumn{1}{c}{(6)} & \multicolumn{1}{c}{(7)} & \multicolumn{1}{c}{(8)} & \multicolumn{1}{c}{(9)} & \multicolumn{1}{c}{(10)} \\
\hline
GMRT161022+0545404  & 16:10:22.1 & +54:54:06.9 &     0.24 &       49 &     0.45 &     0.35 &     0.45 &        U \\
GMRT161023+0540325  & 16:10:23.5 & +54:03:33.4 &     0.61 &       68 &     3.17 &     4.12 &     4.12 &      8.2 \\
GMRT161023+0543007  & 16:10:23.1 & +54:30:09.9 &     0.17 &       57 &    17.12 &    34.76 &    34.76 &     30.5 \\
GMRT161023+0545622  & 16:10:23.3 & +54:56:30.0 &     0.28 &       48 &     0.46 &     0.55 &     0.46 &        U \\
GMRT161023+0550056  & 16:10:23.7 & +55:00:65.9 &     0.35 &       55 &     1.38 &     1.50 &     1.38 &        U \\
\hline \hline
\end{tabular}

\end{table*}
\begin{table*}
\contcaption{}
\begin{tabular}{l r r r r r r r r r r r}
\hline
\multicolumn{1}{c}{Source name} & \multicolumn{1}{c}{RA} & \multicolumn{1}{c}{DEC} & \multicolumn{1}{c}{Dist} & \multicolumn{1}{c}{$\sigma_{rms}$} & \multicolumn{1}{c}{S$_{peak}$} & \multicolumn{1}{c}{S$_{total~3\sigma}$} & \multicolumn{1}{c}{S$_{total}$} & \multicolumn{1}{c}{Size} & \multicolumn{1}{c}{Notes} \\ 
                    & \multicolumn{1}{c}{hh:mm:ss.s} & \multicolumn{1}{c}{dd:mm:ss.s} & \multicolumn{1}{c}{deg} & \multicolumn{1}{c}{$\mu$Jy b$^{-1}$}  & \multicolumn{1}{c}{mJy b$^{-1}$} &  \multicolumn{1}{c}{mJy} &  \multicolumn{1}{c}{mJy}  & \multicolumn{1}{c}{$^{\prime\prime}$} &        \\ 
\multicolumn{1}{c}{(1)} & \multicolumn{1}{c}{(2)} & \multicolumn{1}{c}{(3)} & \multicolumn{1}{c}{(4)} &  \multicolumn{1}{c}{(5)} & \multicolumn{1}{c}{(6)} & \multicolumn{1}{c}{(7)} & \multicolumn{1}{c}{(8)} & \multicolumn{1}{c}{(9)} & \multicolumn{1}{c}{(10)} \\
\hline
GMRT161040+0540626  & 16:10:40.6 & +54:06:32.2 &     0.57 &       65 &     4.74 &    12.69 &    12.69 &     28.7 \\
GMRT161040+0542253  & 16:10:40.7 & +54:22:57.8 &     0.30 &       49 &     0.30 &     0.56 &     0.30 &        U \\
GMRT161040+0542638  & 16:10:40.6 & +54:26:43.3 &     0.24 &       48 &     0.44 &     0.32 &     0.44 &        U \\
GMRT161040+0550824  & 16:10:40.8 & +55:08:32.8 &     0.48 &       76 &     2.18 &     6.10 &     6.10 &     35.1 \\
GMRT161041+0541029  & 16:10:41.5 & +54:10:33.8 &     0.50 &       58 &     1.32 &     1.68 &     1.32 &        U \\
\hline \hline
\end{tabular}

\end{table*}
\begin{table*}
\contcaption{}
\begin{tabular}{l r r r r r r r r r r r}
\hline
\multicolumn{1}{c}{Source name} & \multicolumn{1}{c}{RA} & \multicolumn{1}{c}{DEC} & \multicolumn{1}{c}{Dist} & \multicolumn{1}{c}{$\sigma_{rms}$} & \multicolumn{1}{c}{S$_{peak}$} & \multicolumn{1}{c}{S$_{total~3\sigma}$} & \multicolumn{1}{c}{S$_{total}$} & \multicolumn{1}{c}{Size} & \multicolumn{1}{c}{Notes} \\ 
                    & \multicolumn{1}{c}{hh:mm:ss.s} & \multicolumn{1}{c}{dd:mm:ss.s} & \multicolumn{1}{c}{deg} & \multicolumn{1}{c}{$\mu$Jy b$^{-1}$}  & \multicolumn{1}{c}{mJy b$^{-1}$} &  \multicolumn{1}{c}{mJy} &  \multicolumn{1}{c}{mJy}  & \multicolumn{1}{c}{$^{\prime\prime}$} &        \\ 
\multicolumn{1}{c}{(1)} & \multicolumn{1}{c}{(2)} & \multicolumn{1}{c}{(3)} & \multicolumn{1}{c}{(4)} &  \multicolumn{1}{c}{(5)} & \multicolumn{1}{c}{(6)} & \multicolumn{1}{c}{(7)} & \multicolumn{1}{c}{(8)} & \multicolumn{1}{c}{(9)} & \multicolumn{1}{c}{(10)} \\
\hline
GMRT161058+0540525  & 16:10:58.2 & +54:05:26.6 &     0.59 &       62 &     0.45 &     0.43 &     0.45 &        U \\
GMRT161058+0543521  & 16:10:58.2 & +54:35:27.6 &     0.16 &       42 &     0.48 &     0.40 &     0.48 &        U \\
GMRT161059+0550706  & 16:10:59.7 & +55:07:06.0 &     0.47 &       69 &     3.37 &     4.13 &     4.13 &      4.0 \\
GMRT161100+0535658  & 16:11:00.0 & +53:56:58.9 &     0.73 &       78 &     0.53 &     0.40 &     0.53 &        U \\
GMRT161100+0544200  & 16:11:00.5 & +54:42:08.2 &     0.15 &      114 &    58.81 &   110.86 &   110.86 &     60.8 \\
\hline \hline
\end{tabular}

\end{table*}
\begin{table*}
\contcaption{}
\begin{tabular}{l r r r r r r r r r r r}
\hline
\multicolumn{1}{c}{Source name} & \multicolumn{1}{c}{RA} & \multicolumn{1}{c}{DEC} & \multicolumn{1}{c}{Dist} & \multicolumn{1}{c}{$\sigma_{rms}$} & \multicolumn{1}{c}{S$_{peak}$} & \multicolumn{1}{c}{S$_{total~3\sigma}$} & \multicolumn{1}{c}{S$_{total}$} & \multicolumn{1}{c}{Size} & \multicolumn{1}{c}{Notes} \\ 
                    & \multicolumn{1}{c}{hh:mm:ss.s} & \multicolumn{1}{c}{dd:mm:ss.s} & \multicolumn{1}{c}{deg} & \multicolumn{1}{c}{$\mu$Jy b$^{-1}$}  & \multicolumn{1}{c}{mJy b$^{-1}$} &  \multicolumn{1}{c}{mJy} &  \multicolumn{1}{c}{mJy}  & \multicolumn{1}{c}{$^{\prime\prime}$} &        \\ 
\multicolumn{1}{c}{(1)} & \multicolumn{1}{c}{(2)} & \multicolumn{1}{c}{(3)} & \multicolumn{1}{c}{(4)} &  \multicolumn{1}{c}{(5)} & \multicolumn{1}{c}{(6)} & \multicolumn{1}{c}{(7)} & \multicolumn{1}{c}{(8)} & \multicolumn{1}{c}{(9)} & \multicolumn{1}{c}{(10)} \\
\hline
GMRT161117+0540216  & 16:11:17.5 & +54:02:22.6 &     0.66 &       80 &     0.79 &     0.79 &     0.79 &        U \\
GMRT161117+0541625  & 16:11:17.6 & +54:16:34.8 &     0.44 &       63 &     2.11 &     2.82 &     2.82 &      4.8 \\
GMRT161118+0553345  & 16:11:18.1 & +55:33:53.9 &     0.92 &      272 &    24.18 &    88.84 &    88.84 &     32.5 \\
GMRT161119+0541253  & 16:11:19.5 & +54:12:58.5 &     0.49 &       58 &     1.01 &     1.21 &     1.01 &        U \\
GMRT161119+0542801  & 16:11:19.9 & +54:28:03.3 &     0.28 &       46 &     0.75 &     0.80 &     0.75 &        U \\
\hline \hline
\end{tabular}

\end{table*}
\begin{table*}
\contcaption{}
\begin{tabular}{l r r r r r r r r r r r}
\hline
\multicolumn{1}{c}{Source name} & \multicolumn{1}{c}{RA} & \multicolumn{1}{c}{DEC} & \multicolumn{1}{c}{Dist} & \multicolumn{1}{c}{$\sigma_{rms}$} & \multicolumn{1}{c}{S$_{peak}$} & \multicolumn{1}{c}{S$_{total~3\sigma}$} & \multicolumn{1}{c}{S$_{total}$} & \multicolumn{1}{c}{Size} & \multicolumn{1}{c}{Notes} \\ 
                    & \multicolumn{1}{c}{hh:mm:ss.s} & \multicolumn{1}{c}{dd:mm:ss.s} & \multicolumn{1}{c}{deg} & \multicolumn{1}{c}{$\mu$Jy b$^{-1}$}  & \multicolumn{1}{c}{mJy b$^{-1}$} &  \multicolumn{1}{c}{mJy} &  \multicolumn{1}{c}{mJy}  & \multicolumn{1}{c}{$^{\prime\prime}$} &        \\ 
\multicolumn{1}{c}{(1)} & \multicolumn{1}{c}{(2)} & \multicolumn{1}{c}{(3)} & \multicolumn{1}{c}{(4)} &  \multicolumn{1}{c}{(5)} & \multicolumn{1}{c}{(6)} & \multicolumn{1}{c}{(7)} & \multicolumn{1}{c}{(8)} & \multicolumn{1}{c}{(9)} & \multicolumn{1}{c}{(10)} \\
\hline
GMRT161140+0541438  & 16:11:40.6 & +54:14:38.9 &     0.49 &       63 &     1.44 &     1.36 &     1.44 &        U \\
GMRT161140+0544423  & 16:11:40.2 & +54:44:24.3 &     0.25 &       49 &     0.83 &     0.87 &     0.83 &        U \\
GMRT161141+0535418  & 16:11:41.9 & +53:54:27.6 &     0.80 &      101 &     0.88 &     0.82 &     0.88 &        U \\
GMRT161143+0535311  & 16:11:43.8 & +53:53:13.8 &     0.82 &       96 &     1.63 &     1.77 &     1.63 &        U \\
GMRT161143+0540703  & 16:11:43.7 & +54:07:07.4 &     0.60 &       62 &     2.10 &     2.24 &     2.10 &        U \\
\hline \hline
\end{tabular}

\end{table*}

~~~~~
\newpage

\begin{table*}
\contcaption{}
\begin{tabular}{l r r r r r r r r r r r}
\hline
\multicolumn{1}{c}{Source name} & \multicolumn{1}{c}{RA} & \multicolumn{1}{c}{DEC} & \multicolumn{1}{c}{Dist} & \multicolumn{1}{c}{$\sigma_{rms}$} & \multicolumn{1}{c}{S$_{peak}$} & \multicolumn{1}{c}{S$_{total~3\sigma}$} & \multicolumn{1}{c}{S$_{total}$} & \multicolumn{1}{c}{Size} & \multicolumn{1}{c}{Notes} \\ 
                    & \multicolumn{1}{c}{hh:mm:ss.s} & \multicolumn{1}{c}{dd:mm:ss.s} & \multicolumn{1}{c}{deg} & \multicolumn{1}{c}{$\mu$Jy b$^{-1}$}  & \multicolumn{1}{c}{mJy b$^{-1}$} &  \multicolumn{1}{c}{mJy} &  \multicolumn{1}{c}{mJy}  & \multicolumn{1}{c}{$^{\prime\prime}$} &        \\ 
\multicolumn{1}{c}{(1)} & \multicolumn{1}{c}{(2)} & \multicolumn{1}{c}{(3)} & \multicolumn{1}{c}{(4)} &  \multicolumn{1}{c}{(5)} & \multicolumn{1}{c}{(6)} & \multicolumn{1}{c}{(7)} & \multicolumn{1}{c}{(8)} & \multicolumn{1}{c}{(9)} & \multicolumn{1}{c}{(10)} \\
\hline
GMRT161204+0540933  & 16:12:04.5 & +54:09:41.6 &     0.59 &       71 &     0.47 &     0.44 &     0.47 &        U \\
GMRT161204+0543446  & 16:12:04.7 & +54:34:55.0 &     0.31 &       49 &     0.79 &     0.81 &     0.79 &        U \\
GMRT161204+0545505  & 16:12:04.9 & +54:55:08.3 &     0.39 &       68 &     0.60 &     0.51 &     0.60 &        U \\
GMRT161205+0535954  & 16:12:05.3 & +53:59:59.4 &     0.73 &       90 &     0.83 &     0.73 &     0.83 &        U \\
GMRT161205+0544146  & 16:12:05.0 & +54:41:47.1 &     0.30 &       47 &     1.25 &     1.18 &     1.25 &        U \\
\hline \hline
\end{tabular}

\end{table*}
\begin{table*}
\contcaption{}
\begin{tabular}{l r r r r r r r r r r r}
\hline
\multicolumn{1}{c}{Source name} & \multicolumn{1}{c}{RA} & \multicolumn{1}{c}{DEC} & \multicolumn{1}{c}{Dist} & \multicolumn{1}{c}{$\sigma_{rms}$} & \multicolumn{1}{c}{S$_{peak}$} & \multicolumn{1}{c}{S$_{total~3\sigma}$} & \multicolumn{1}{c}{S$_{total}$} & \multicolumn{1}{c}{Size} & \multicolumn{1}{c}{Notes} \\ 
                    & \multicolumn{1}{c}{hh:mm:ss.s} & \multicolumn{1}{c}{dd:mm:ss.s} & \multicolumn{1}{c}{deg} & \multicolumn{1}{c}{$\mu$Jy b$^{-1}$}  & \multicolumn{1}{c}{mJy b$^{-1}$} &  \multicolumn{1}{c}{mJy} &  \multicolumn{1}{c}{mJy}  & \multicolumn{1}{c}{$^{\prime\prime}$} &        \\ 
\multicolumn{1}{c}{(1)} & \multicolumn{1}{c}{(2)} & \multicolumn{1}{c}{(3)} & \multicolumn{1}{c}{(4)} &  \multicolumn{1}{c}{(5)} & \multicolumn{1}{c}{(6)} & \multicolumn{1}{c}{(7)} & \multicolumn{1}{c}{(8)} & \multicolumn{1}{c}{(9)} & \multicolumn{1}{c}{(10)} \\
\hline
GMRT161224+0543901  & 16:12:24.4 & +54:39:10.8 &     0.35 &       54 &     0.98 &     1.11 &     0.98 &        U \\
GMRT161225+0542319  & 16:12:25.4 & +54:23:22.5 &     0.45 &       66 &     0.73 &     0.67 &     0.73 &        U \\
GMRT161225+0545143  & 16:12:25.6 & +54:51:52.0 &     0.40 &       63 &     0.54 &     0.35 &     0.54 &        U \\
GMRT161225+0545459  & 16:12:25.6 & +54:54:67.4 &     0.43 &       57 &     3.51 &     3.54 &     3.51 &        U \\
GMRT161226+0540229  & 16:12:26.9 & +54:02:36.8 &     0.72 &       84 &     1.66 &     1.67 &     1.66 &        U \\
\hline \hline
\end{tabular}

\end{table*}
\begin{table*}
\contcaption{}
\begin{tabular}{l r r r r r r r r r r r}
\hline
\multicolumn{1}{c}{Source name} & \multicolumn{1}{c}{RA} & \multicolumn{1}{c}{DEC} & \multicolumn{1}{c}{Dist} & \multicolumn{1}{c}{$\sigma_{rms}$} & \multicolumn{1}{c}{S$_{peak}$} & \multicolumn{1}{c}{S$_{total~3\sigma}$} & \multicolumn{1}{c}{S$_{total}$} & \multicolumn{1}{c}{Size} & \multicolumn{1}{c}{Notes} \\ 
                    & \multicolumn{1}{c}{hh:mm:ss.s} & \multicolumn{1}{c}{dd:mm:ss.s} & \multicolumn{1}{c}{deg} & \multicolumn{1}{c}{$\mu$Jy b$^{-1}$}  & \multicolumn{1}{c}{mJy b$^{-1}$} &  \multicolumn{1}{c}{mJy} &  \multicolumn{1}{c}{mJy}  & \multicolumn{1}{c}{$^{\prime\prime}$} &        \\ 
\multicolumn{1}{c}{(1)} & \multicolumn{1}{c}{(2)} & \multicolumn{1}{c}{(3)} & \multicolumn{1}{c}{(4)} &  \multicolumn{1}{c}{(5)} & \multicolumn{1}{c}{(6)} & \multicolumn{1}{c}{(7)} & \multicolumn{1}{c}{(8)} & \multicolumn{1}{c}{(9)} & \multicolumn{1}{c}{(10)} \\
\hline
GMRT161251+0541649  & 16:12:51.2 & +54:16:53.0 &     0.57 &       74 &     0.48 &     0.85 &     0.48 &        U \\
GMRT161251+0542846  & 16:12:51.4 & +54:28:46.6 &     0.45 &       87 &     1.54 &     1.61 &     1.54 &        U \\
GMRT161252+0550659  & 16:12:53.0 & +55:06:61.3 &     0.61 &       81 &     1.52 &     1.49 &     1.52 &        U \\
GMRT161253+0544109  & 16:12:53.6 & +54:41:17.1 &     0.42 &       55 &     0.77 &     0.90 &     0.77 &        U \\
GMRT161254+0544740  & 16:12:54.7 & +54:47:44.0 &     0.44 &       59 &     0.81 &     1.39 &     1.39 &     14.1 \\
\hline \hline
\end{tabular}

\end{table*}
\begin{table*}
\contcaption{}
\begin{tabular}{l r r r r r r r r r r r}
\hline
\multicolumn{1}{c}{Source name} & \multicolumn{1}{c}{RA} & \multicolumn{1}{c}{DEC} & \multicolumn{1}{c}{Dist} & \multicolumn{1}{c}{$\sigma_{rms}$} & \multicolumn{1}{c}{S$_{peak}$} & \multicolumn{1}{c}{S$_{total~3\sigma}$} & \multicolumn{1}{c}{S$_{total}$} & \multicolumn{1}{c}{Size} & \multicolumn{1}{c}{Notes} \\ 
                    & \multicolumn{1}{c}{hh:mm:ss.s} & \multicolumn{1}{c}{dd:mm:ss.s} & \multicolumn{1}{c}{deg} & \multicolumn{1}{c}{$\mu$Jy b$^{-1}$}  & \multicolumn{1}{c}{mJy b$^{-1}$} &  \multicolumn{1}{c}{mJy} &  \multicolumn{1}{c}{mJy}  & \multicolumn{1}{c}{$^{\prime\prime}$} &        \\ 
\multicolumn{1}{c}{(1)} & \multicolumn{1}{c}{(2)} & \multicolumn{1}{c}{(3)} & \multicolumn{1}{c}{(4)} &  \multicolumn{1}{c}{(5)} & \multicolumn{1}{c}{(6)} & \multicolumn{1}{c}{(7)} & \multicolumn{1}{c}{(8)} & \multicolumn{1}{c}{(9)} & \multicolumn{1}{c}{(10)} \\
\hline
GMRT161322+0543745  & 16:13:22.6 & +54:37:48.9 &     0.49 &       68 &     0.42 &     0.36 &     0.42 &        U \\
GMRT161322+0550614  & 16:13:22.1 & +55:06:15.1 &     0.65 &       83 &     1.07 &     0.93 &     1.07 &        U \\
GMRT161323+0545725  & 16:13:23.3 & +54:57:25.4 &     0.57 &       68 &     0.58 &     0.42 &     0.58 &        U \\
GMRT161325+0553927  & 16:13:25.4 & +55:39:27.2 &     1.10 &      287 &    16.40 &    23.01 &    23.01 &      4.1 \\
GMRT161326+0544447  & 16:13:26.2 & +54:44:48.0 &     0.50 &       66 &     0.45 &     0.37 &     0.45 &        U \\
\hline \hline
\end{tabular}

\end{table*}
\begin{table*}
\contcaption{}
\begin{tabular}{l r r r r r r r r r r r}
\hline
\multicolumn{1}{c}{Source name} & \multicolumn{1}{c}{RA} & \multicolumn{1}{c}{DEC} & \multicolumn{1}{c}{Dist} & \multicolumn{1}{c}{$\sigma_{rms}$} & \multicolumn{1}{c}{S$_{peak}$} & \multicolumn{1}{c}{S$_{total~3\sigma}$} & \multicolumn{1}{c}{S$_{total}$} & \multicolumn{1}{c}{Size} & \multicolumn{1}{c}{Notes} \\ 
                    & \multicolumn{1}{c}{hh:mm:ss.s} & \multicolumn{1}{c}{dd:mm:ss.s} & \multicolumn{1}{c}{deg} & \multicolumn{1}{c}{$\mu$Jy b$^{-1}$}  & \multicolumn{1}{c}{mJy b$^{-1}$} &  \multicolumn{1}{c}{mJy} &  \multicolumn{1}{c}{mJy}  & \multicolumn{1}{c}{$^{\prime\prime}$} &        \\ 
\multicolumn{1}{c}{(1)} & \multicolumn{1}{c}{(2)} & \multicolumn{1}{c}{(3)} & \multicolumn{1}{c}{(4)} &  \multicolumn{1}{c}{(5)} & \multicolumn{1}{c}{(6)} & \multicolumn{1}{c}{(7)} & \multicolumn{1}{c}{(8)} & \multicolumn{1}{c}{(9)} & \multicolumn{1}{c}{(10)} \\
\hline
GMRT161359+0550746  & 16:13:59.5 & +55:07:46.2 &     0.74 &       88 &     1.37 &     1.37 &     1.37 &        U \\
GMRT161400+0535711  & 16:14:00.6 & +53:57:15.6 &     0.92 &      172 &    19.41 &    47.11 &    47.11 &     36.3 \\
GMRT161401+0553111  & 16:14:01.6 & +55:31:16.4 &     1.03 &      225 &     2.67 &     2.28 &     2.67 &        U \\
GMRT161402+0544417  & 16:14:02.6 & +54:44:23.6 &     0.59 &      133 &    30.44 &   141.48 &   141.48 &     48.6 \\
GMRT161403+0540157  & 16:14:03.8 & +54:01:65.6 &     0.87 &      122 &     0.75 &     0.70 &     0.75 &        U \\
\hline \hline
\end{tabular}

\end{table*}
\begin{table*}
\contcaption{}
\begin{tabular}{l r r r r r r r r r r r}
\hline
\multicolumn{1}{c}{Source name} & \multicolumn{1}{c}{RA} & \multicolumn{1}{c}{DEC} & \multicolumn{1}{c}{Dist} & \multicolumn{1}{c}{$\sigma_{rms}$} & \multicolumn{1}{c}{S$_{peak}$} & \multicolumn{1}{c}{S$_{total~3\sigma}$} & \multicolumn{1}{c}{S$_{total}$} & \multicolumn{1}{c}{Size} & \multicolumn{1}{c}{Notes} \\ 
                    & \multicolumn{1}{c}{hh:mm:ss.s} & \multicolumn{1}{c}{dd:mm:ss.s} & \multicolumn{1}{c}{deg} & \multicolumn{1}{c}{$\mu$Jy b$^{-1}$}  & \multicolumn{1}{c}{mJy b$^{-1}$} &  \multicolumn{1}{c}{mJy} &  \multicolumn{1}{c}{mJy}  & \multicolumn{1}{c}{$^{\prime\prime}$} &        \\ 
\multicolumn{1}{c}{(1)} & \multicolumn{1}{c}{(2)} & \multicolumn{1}{c}{(3)} & \multicolumn{1}{c}{(4)} &  \multicolumn{1}{c}{(5)} & \multicolumn{1}{c}{(6)} & \multicolumn{1}{c}{(7)} & \multicolumn{1}{c}{(8)} & \multicolumn{1}{c}{(9)} & \multicolumn{1}{c}{(10)} \\
\hline
GMRT161432+0542600  & 16:14:32.8 & +54:26:06.0 &     0.70 &      110 &     1.12 &     1.28 &     1.12 &        U \\
GMRT161432+0544052  & 16:14:32.9 & +54:40:54.8 &     0.66 &       83 &     2.44 &     2.72 &     2.72 &      3.1 \\
GMRT161432+0544852  & 16:14:32.7 & +54:48:61.5 &     0.67 &       77 &     0.73 &     0.65 &     0.73 &        U \\
GMRT161432+0550507  & 16:14:32.6 & +55:05:12.0 &     0.78 &       90 &     0.87 &     0.82 &     0.87 &        U \\
GMRT161433+0541421  & 16:14:33.9 & +54:14:28.8 &     0.79 &      104 &     3.81 &    12.91 &    12.91 &     23.3 \\
\hline \hline
\end{tabular}

\end{table*}
\begin{table*}
\contcaption{}
\begin{tabular}{l r r r r r r r r r r r}
\hline
\multicolumn{1}{c}{Source name} & \multicolumn{1}{c}{RA} & \multicolumn{1}{c}{DEC} & \multicolumn{1}{c}{Dist} & \multicolumn{1}{c}{$\sigma_{rms}$} & \multicolumn{1}{c}{S$_{peak}$} & \multicolumn{1}{c}{S$_{total~3\sigma}$} & \multicolumn{1}{c}{S$_{total}$} & \multicolumn{1}{c}{Size} & \multicolumn{1}{c}{Notes} \\ 
                    & \multicolumn{1}{c}{hh:mm:ss.s} & \multicolumn{1}{c}{dd:mm:ss.s} & \multicolumn{1}{c}{deg} & \multicolumn{1}{c}{$\mu$Jy b$^{-1}$}  & \multicolumn{1}{c}{mJy b$^{-1}$} &  \multicolumn{1}{c}{mJy} &  \multicolumn{1}{c}{mJy}  & \multicolumn{1}{c}{$^{\prime\prime}$} &        \\ 
\multicolumn{1}{c}{(1)} & \multicolumn{1}{c}{(2)} & \multicolumn{1}{c}{(3)} & \multicolumn{1}{c}{(4)} &  \multicolumn{1}{c}{(5)} & \multicolumn{1}{c}{(6)} & \multicolumn{1}{c}{(7)} & \multicolumn{1}{c}{(8)} & \multicolumn{1}{c}{(9)} & \multicolumn{1}{c}{(10)} \\
\hline
GMRT161521+0541812  & 16:15:21.9 & +54:18:20.5 &     0.86 &      107 &     4.13 &     4.84 &     4.84 &      3.5 \\
GMRT161521+0543145  & 16:15:22.0 & +54:31:46.1 &     0.79 &       98 &     2.45 &     2.77 &     2.45 &        U \\
GMRT161521+0543515  & 16:15:21.9 & +54:35:18.2 &     0.78 &       94 &     5.06 &     5.48 &     5.48 &      4.3 \\
GMRT161521+0552038  & 16:15:21.9 & +55:20:45.4 &     1.02 &      180 &     3.53 &     4.07 &     3.53 &        U \\
GMRT161522+0545822  & 16:15:22.5 & +54:58:24.7 &     0.83 &      109 &     5.27 &     5.75 &     5.75 &      3.2 \\
\hline \hline
\end{tabular}

\end{table*}
\begin{table*}
\contcaption{}
\begin{tabular}{l r r r r r r r r r r r}
\hline
\multicolumn{1}{c}{Source name} & \multicolumn{1}{c}{RA} & \multicolumn{1}{c}{DEC} & \multicolumn{1}{c}{Dist} & \multicolumn{1}{c}{$\sigma_{rms}$} & \multicolumn{1}{c}{S$_{peak}$} & \multicolumn{1}{c}{S$_{total~3\sigma}$} & \multicolumn{1}{c}{S$_{total}$} & \multicolumn{1}{c}{Size} & \multicolumn{1}{c}{Notes} \\ 
                    & \multicolumn{1}{c}{hh:mm:ss.s} & \multicolumn{1}{c}{dd:mm:ss.s} & \multicolumn{1}{c}{deg} & \multicolumn{1}{c}{$\mu$Jy b$^{-1}$}  & \multicolumn{1}{c}{mJy b$^{-1}$} &  \multicolumn{1}{c}{mJy} &  \multicolumn{1}{c}{mJy}  & \multicolumn{1}{c}{$^{\prime\prime}$} &        \\ 
\multicolumn{1}{c}{(1)} & \multicolumn{1}{c}{(2)} & \multicolumn{1}{c}{(3)} & \multicolumn{1}{c}{(4)} &  \multicolumn{1}{c}{(5)} & \multicolumn{1}{c}{(6)} & \multicolumn{1}{c}{(7)} & \multicolumn{1}{c}{(8)} & \multicolumn{1}{c}{(9)} & \multicolumn{1}{c}{(10)} \\
\hline
GMRT161611+0551448  & 16:16:11.6 & +55:14:51.6 &     1.06 &      186 &     2.70 &     3.22 &     2.70 &        U \\
GMRT161615+0543719  & 16:16:15.6 & +54:37:25.7 &     0.91 &      114 &     2.31 &     2.67 &     2.31 &        U \\
GMRT161616+0541720  & 16:16:16.4 & +54:17:21.3 &     0.99 &      182 &    16.31 &    41.92 &    41.92 &     28.6 \\
GMRT161616+0551006  & 16:16:17.0 & +55:10:06.1 &     1.03 &      168 &     2.93 &     3.16 &     2.93 &        U \\
GMRT161623+0544526  & 16:16:23.4 & +54:45:35.1 &     0.93 &      125 &     1.33 &     1.75 &     1.33 &        U \\
\hline \hline
\end{tabular}

\end{table*}

\end{document}